\documentclass[
    aps,
    prb,
    twocolumn,
    superscriptaddress,
    nobibnotes,
    floatfix,
    longbibliography
]{revtex4-2}

\usepackage{color}
\usepackage{graphicx}
\usepackage{epstopdf}
\usepackage{physics}
\usepackage{xcolor}
\usepackage{bm} 
\usepackage{amsmath, amsthm, amssymb, amsfonts}
\usepackage{mathtools}
\usepackage{array}
\usepackage{feynmp-auto}
\usepackage[compat=1.1.0]{tikz-feynman}
\usepackage{float}

\usepackage[final]{changes}
\definechangesauthor[color=green]{aitor}
\definechangesauthor[color=red]{diego}
\definechangesauthor[color=purple]{manish}
\definechangesauthor[color=blue]{pavel}
\definechangesauthor[color=brown]{maria}
\definechangesauthor[color=orange]{asier}

\newcommand{\be}{\begin{equation}}
\newcommand{\ee}{\end{equation}}

\newcommand{\dD}{\delta D}

\newcommand{\meV}{{\text{meV}}}
\newcommand{\eV}{{\text{eV}}}
\newcommand{\K}{\text{K}}
\newcommand{\HWHM}{\text{HWHM}}
\newcommand{\FWHM}{\text{FWHM}}
\newcommand{\NRG}{\text{NRG}}
\newcommand{\ex}{\text{ex}}
\newcommand{\ZP}{\textbf{PorA}_{\bf 2}}
\newcommand{\FAT}{\textbf{4''}^+}

\begin{document}
\title{Theoretical model for multi-orbital Kondo screening in strongly correlated molecules with several unpaired electrons}

\newcommand{\FISI}[0]{{Departamento de F\'{\i}sica, Universidad del Pa\'{\i}s Vasco UPV-EHU, 48080 Leioa, Spain}}
\newcommand{\QUIMI}[0]{{Departamento de Polímeros y Materiales Avanzados: F\'{\i}sica, Qu\'{\i}mica y Tecnolog\'{\i}a, Universidad del Pa\'{\i}s Vasco UPV-EHU, 20018 Donostia-San Sebasti\'an, Spain}}
\newcommand{\DIPC}[0]{{Donostia International Physics Center (DIPC),
20018 Donostia-San Sebasti\'an, Spain}}
\newcommand{\EHUQC}[0]{{EHU Quantum Center, Universidad del Pa\'{\i}s Vasco UPV-EHU, 48080 Leioa, Spain}}
\newcommand{\CFM}[0]{{Centro de F\'{\i}sica de Materiales CFM/MPC (CSIC-UPV/EHU), 20018 Donostia-San Sebasti\'an, Spain}}

\author{Aitor Calvo-Fern{\'a}ndez}
\thanks{These authors contributed equally to this work.}
\affiliation{\FISI}
\affiliation{\DIPC} 

\author{Manish Kumar}
\thanks{These authors contributed equally to this work.}
\affiliation{Institute of Physics, Academy of Sciences of the Czech Republic, Cukrovarnicka 10, Prague 6, CZ 16200, Czech Republic}
\affiliation{Department of Condensed Matter Physics, Faculty of Mathematics and Physics, Charles University, Prague, Czech Republic}

\author{Diego Soler-Polo}
\affiliation{Institute of Physics, Academy of Sciences of the Czech Republic, Cukrovarnicka 10, Prague 6, CZ 16200, Czech Republic}

\author{Asier Eiguren}
\affiliation{\FISI}
\affiliation{\DIPC}
\affiliation{\EHUQC}

\author{Mar{\'i}a Blanco-Rey}
\thanks{Corresponding author}
\affiliation{\QUIMI}
\affiliation{\DIPC}
\affiliation{\CFM}

\author{Pavel Jel\'{\i}nek}
\thanks{Corresponding author}
\affiliation{Institute of Physics, Academy of Sciences of the Czech Republic, Cukrovarnicka 10, Prague 6, CZ 16200, Czech Republic}

\affiliation{Regional Centre of Advanced Technologies and Materials, Czech Advanced Technology and Research Institute (CATRIN), Palacký University Olomouc,Olomouc 78371, Czech Republic}

\date{\today}

\begin{abstract}
The mechanism of Kondo screening in strongly correlated molecules with several unpaired electrons on a metal surface is still under debate. Here, we provide a theoretical framework that rationalizes the emergence of Kondo screening involving several extended molecular orbitals with unpaired electrons. We introduce a~perturbative model, which provides simple rules to identify the presence of antiferromagnetic spin-flip channels involving charged molecular multiplets responsible for Kondo screening. The Kondo regime is confirmed by numerical renormalization group calculations. In addition, we introduce the concept of Kondo orbitals as molecular orbitals associated with the Kondo screening process, which provide a direct interpretation of experimental d$I$/d$V$ maps of Kondo resonances.  We demonstrate that this theoretical framework can be applied to different strongly correlated open-shell molecules on metal surfaces, obtaining good agreement with previously published experimental data.
\end{abstract}

\maketitle

\section{Introduction}
The Kondo effect \cite{bib:hewson,kondo1964resistance} is a widely studied many-body phenomenon in solid state physics, which has received a lot of attention. The origin of the Kondo effect is related to the exchange interaction between a single magnetic impurity and conduction electrons in a non-magnetic metal. This scattering process may establish a many-body singlet ground state, which is manifested as a strong zero-energy resonance in the spectral function. Originally, the Kondo effect was observed in transport measurements in non-magnetic metals with a diluted concentration of magnetic impurities \cite{deHaas1934,NP2014-Kondo}. Later, it was also observed in single atoms with partially occupied $d$-electrons on metallic substrates by scanning tunneling microscopy \cite{bib:madhavan98,bib:li98,bib:knorr02,bib:otte08}. More recently, it has been demonstrated that unpaired electrons in extended molecular orbitals (MO) can also act as magnetic impurities and scatter conduction electrons from the underlying metal surface, leading to the molecular Kondo effect \cite{Esat2015,bib:scott10,bib:komeda11,bib:minamitani12,bib:hiraoka17,bib:zitko21,bib:fernandez08,bib:choi10,bib:perera10,bib:requist14,bib:minamitani15,bib:patera19,bib:koshida21,bib:lu21,Li2019,bib:li20,bib:mishra20,bib:jacob21,bib:turco23}.

 While the conventional $S=1/2$ Kondo effect for magnetic impurities has been extensively studied and is very well understood, the mechanism of the Kondo effect in magnetic impurities with various strongly correlated unpaired electrons is barely explored. In particular, here the so-called underscreened Kondo effect can take place if there are less than $2S$ screening channels \cite{bib:nozieres80}, whereby the ground state shows a residual magnetic moment. This effect was extensively studied for quantum dot systems both experimentally \cite{bib:roch09,bib:parks10} and theoretically \cite{pustilnik2001,mitchell2017}. Recently, the Kondo effect in molecular systems with several unpaired electrons was reported for open-shell triplet ($S=1$) nanographenes \cite{bib:li20,bib:turco23} and porphyrins \cite{bib:girovsky2017,yo8,Sun2020} on metallic surfaces. The former case  was theoretically rationalized within the one-crossing approximation \cite{bib:jacob21}. 

Nevertheless, a broader picture of the Kondo effect in open-shell molecular
systems with several unpaired electrons is still missing. This is a
multi-orbital Kondo problem where the involved orbitals have an extended
non-trivial spatial distribution. Therefore, in order to describe it we need to
consider, on the same footing, aspects such as the number and symmetry of the
screening channels defined by the MOs hosting the unpaired electrons, as well as
the strongly correlated multiplet (i.e., many-electron) nature of the
molecular electronic configuration, which cannot be properly described by
mean-field theories such as density functional theory (DFT).

In this work, we present a theoretical framework for the molecular multi-orbital Kondo effect and apply it to strongly correlated molecules deposited on metal surfaces. The theoretical framework consists of three steps: (i) Complete Active Space Configuration Interaction (CASCI) calculations \cite{CAS} accurately provide the electronic states of neutral and charged molecular multiplets; (ii) a generalized perturbative analysis of the Anderson Hamiltonian, which accounts for the multiplet structure, reveals the antiferromagnetically coupled channels responsible for Kondo screening by the metal substrate; and finally (iii) numerical renormalization group (NRG) calculations \cite{krishna-murthy1980a} provide observable quantities. 
In addition, we introduce the concept of \emph{Kondo orbitals} (KO) as unitary transformations of the MOs inside the active space that show pure antiferromagnetic or ferromagnetic coupling to conduction electrons. Upon critical comparison with the well-established quantum-chemical concept of natural transition orbitals (NTO), we show that the KOs can be directly compared to experimental d$I$/d$V$ maps of the Kondo signal.

To benchmark the proposed theoretical framework, we select examples of open-shell molecules with several unpaired spins deposited on Au(111) surfaces, where Kondo resonances were experimentally observed. In particular, we try to reproduce Kondo features observed in extended porphyrins \cite{Sun2020} and graphene nano-sized flakes \cite{bib:li20,bib:calupitan23,bib:vilas23}. By the combination of the Ovchinnikov rules \cite{bib:ovchinnikov78}, charge transfer to the substrate, and {\it ad hoc} radical formation by partial dehydrogenation of edge carbon sites using STM manipulation, high-spin states states can be realized in these systems, resulting in the presence of coexisting underscreened Kondo and spin excitation features in the d$I$/d$V$ signals \cite{bib:ternes15}.

\section{Rationalization of the Kondo peak}

\subsection{Multi-channel Anderson model}

To model a molecule adsorbed on a metallic surface we use the ionic Anderson impurity model \cite{anderson1970,hirst1978} given by the Hamiltonian
\begin{equation}
    \hat{\mathcal H}_\text{A}
    = 
    \hat{H}_\text{mol} + \hat{H}_\text{con} + \hat{H}_\text{hyb}.
\label{eq:AndersonHamiltonian}
\end{equation}
\par
The molecular part of the Hamiltonian is given by the term 
\begin{equation}
    \hat H_\text{mol}
    =
    \sum_{M,m} \epsilon_M
    \ket{Mm}\bra{Mm},
\label{eq:Hmol}
\end{equation}
which contains the energies $\epsilon_M$ of the multiplet
states $\ket{Mm}$ obtained from CASCI calculations, which
will be described in Section~\ref{elecstr}. Here $M$
denotes the order of the multiplet state and $m$ is the
magnetic quantum number. Since we are mainly interested in
capturing the Kondo physics of the system, we keep only the
neutral $N_0$-particle ground multiplet and charged
multiplets with $N_0\pm 1$  particles involved in virtual
spin excitations, as discussed later. We incorporate the
effect of the substrate on the molecule by shifting the
charged multiplet energies as $\epsilon_M=E_M+(N_M-N_0)\mu$,
where $N_M$ is the number of particles in the multiplet $M$,
$E_{M}$ is the energy of the multiplet in gas phase, and
$\mu$ is the chemical potential induced by the substrate.  
\par
The substrate conduction electrons are described by the term
\begin{equation}
    \hat H_\text{con}
    =
    \sum_{\alpha,\sigma}
    \int_{-D}^{D}
    d\epsilon 
    \hat c^\dagger_{\epsilon\alpha\sigma}
    \hat c_{\epsilon\alpha\sigma}
    \epsilon,
\label{eq:Hcon}
\end{equation}
where $\hat c^\dagger_{\epsilon\alpha\sigma}$
($\hat c_{\epsilon\alpha\sigma}$) creates (annihilates) a conduction electron with spin $\sigma$ and energy $\epsilon$ 
covering a band-width from $-D$ to $D$, belonging to channel $\alpha$. The electronic states in channel
$\alpha$ couple to a single one-electron MO, which can be labeled the same as $\alpha$, and are constructed as
\begin{equation}
    \ket{\epsilon\alpha\sigma} 
    = 
    \frac{1}{[\rho_\alpha(\epsilon)]^\frac{1}{2}
    V_\alpha(\epsilon)} 
    \sum_{\mathbf k} 
    V_{\alpha \mathbf k}
    \ket{\mathbf k\sigma}
    \delta(\epsilon-\epsilon_{\mathbf k}),
\label{eq:ChannelElectrons}
\end{equation}
where $V_{\alpha\mathbf k}$ is the tunneling amplitude for an electron in MO $\alpha$ into 
a conduction state with momentum $\mathbf k$, and $[\rho_\alpha (\epsilon)]^\frac{1}{2} V_\alpha(\epsilon)$ is the integrated tunneling amplitude
between electrons in the MO and channel with the same label $\alpha$, given by
$\rho_\alpha(\epsilon) V_\alpha^2(\epsilon)= \sum_{\mathbf k} |V_{\alpha\mathbf k}|^2\delta(\epsilon-\epsilon_{\mathbf k})$. The integrated tunneling amplitude contains the density of states $\rho_\alpha(\epsilon)$ and the tunneling amplitude $V_\alpha(\epsilon)$ in the channel representation.
We take the usual wide-band limit where the hybridization is energy independent in the bandwidth of interest and also equal for all MOs, i.e. $[\rho_\alpha(\epsilon)]^\frac{1}{2} V_\alpha(\epsilon)=\rho^\frac{1}{2} V$. 
We assume the constant coupling approximation to all MOs, as they consist of $\pi$-orbitals. Moreover, molecules considered here are planar molecules physisorbed on a metal surface. Thus, MOs have a very similar electronic overlap with the substrate, and the constant coupling approximation seems to be justified.
\par 
The coupling of the molecule to the substrate is described by the hybridization term
\begin{equation}
\begin{aligned}
\hat{H}_\text{hyb}
    &=
    \sum_{M_i m_i}
    \sum_{M_f m_f}
    \sum_{\alpha \sigma}
    \int_{-D}^{D}
    d\epsilon 
    \rho^\frac{1}{2}
    V
    \\ &
    \times\bra{M_f m_f}
    \hat f^\dagger_{\alpha \sigma} \ket{M_i m_i}
   \ket{M_f m_f}\bra{M_i m_i}
    \hat{c}_{\epsilon \alpha \sigma}
    + \text{h.c.},
\label{AndersonHamiltonian}
\end{aligned}
\end{equation}
which introduces processes where the impurity gains (loses) an electron in MO $\alpha$ with spin  $\sigma$, which is created (annihilated) by the operator $\hat f^\dagger_{\alpha\sigma}$ ($\hat  f_{\alpha\sigma}$), and goes from the initial state $\ket{M_i m_i}$ to the final state $\ket{M_f m_f}$. 

\subsection{Effective Kondo model} \label{kondo_model}
At energy scales much smaller than the molecular excitation energies, the
charged multiplets can be integrated out using the Schrieffer-Wolff
transformation \cite{schrieffer1966}. This transformation can
be generalized to ionic and multichannel Kondo systems
\cite{flores2017,pustilnik2001,mitchell2023}.
We perform the transformation to second order in perturbation theory by summing over the virtual processes depicted in Fig. \ref{fig:feynman_schrieffer-wolff}, which involve virtual fluctuations of the molecule into the charged multiplets (see Appendix~\ref{appendix_schrieffer_wolff}). The resulting effective interaction, ignoring the potential term, is given by the Kondo Hamiltonian
\begin{equation}
    \hat{\mathcal{H}}_\text{K}
    =
    \sum_{\alpha\alpha'}
    J_{\alpha\alpha'}
    \hat{\mathbf{S}} 
    \cdot
    \hat{\mathbf{s}}_{\alpha\alpha'},
\label{eq:KondoHamiltonian}
\end{equation}
where $J$ is the matrix containing the effective coupling constants $J_{\alpha\alpha'}$, $\hat{\mathbf S}$ is the molecular spin  operator, and $\hat{\mathbf{s}}_{\alpha\alpha'}=\sum_{ii'}\frac{1}{2}\hat \psi^\dagger_{\alpha i} \vec \sigma_{ii'} \hat \psi_{\alpha'i'}$ is the local spin density operator \cite{pustilnik2001}. In this Hamiltonian, the molecule is reduced to the spin degree of freedom of the ground multiplet. The conduction electrons scatter off this spin by interchanging spin, $i'\rightarrow i$, and hopping between channels, $\alpha'\rightarrow\alpha$. The strength of the coupling between channels is given by the constants
\begin{figure}[t]
\scalebox{1.3}{
\begin{tikzpicture}
\begin{feynman}
\vertex (i1) {\tiny$M_0,m_0$};
\vertex [above=of i1] (e1) {\tiny$\epsilon\alpha i$};
\vertex [dot, right=of i1] (v1) {};
\vertex [dot, right=of v1] (v2) {};
\vertex [right=of v2] (i2) {\tiny$M_0,m_0'$};
\vertex [above=of i2] (e2) {\tiny$\epsilon'\alpha' i'$};
\diagram* {
(e2) -- [charged scalar, bend right] (v2) ,
(v1) -- [charged scalar, bend right] (e1),
(i2) -- [fermion] (v2) -- [fermion, edge
label'=\tiny{$M_c,m_c$}] (v1) -- [fermion] (i1),
};
\end{feynman}
\end{tikzpicture} 
}
\scalebox{1.3}{
\begin{tikzpicture}
\begin{feynman}
\vertex (i1) {\tiny{$M_0,m_0$}};
\vertex [above=of i1] (e1) {\tiny$\epsilon\alpha i$};
\vertex [dot, right=of i1] (v1) {};
\vertex [dot, right=of v1] (v2) {};
\vertex [right=of v2] (i2){\tiny{$M_0,m_0'$}};
\vertex [above=of i2] (e2){\tiny$\epsilon'\alpha' i'$};
\diagram* {
(e2) -- [charged scalar, bend right] (v1),
(v2) -- [charged scalar, bend right] (e1),
(i2) -- [fermion] (v2) -- [fermion, edge
label'=\tiny{$M_c,m_c$}] (v1) -- [fermion] (i1),
};
\end{feynman}
\end{tikzpicture}
}
\caption{
    Feynman diagrams for virtual scattering process through
    the intermediate molecular multiplet $M_c$ with
    occupation $N_0+1$ (top) and $N_0-1$
    (bottom). 
}
\label{fig:feynman_schrieffer-wolff}
\end{figure}
\begin{equation}
    J_{\alpha\alpha'}
    = 
    V^2
    \sum_{M_c}
    \frac{C(M_c)_{\alpha\alpha'}}{\epsilon_{M_c}-\epsilon_{M_0}},
\label{eq:J}
\end{equation}
which contain contributions from all the excited multiplets $M_c$ weighted by their excitation energy and a coefficient $C(M)_{\alpha\alpha'}$ determined by the many-body structure of the multiplets (see Appendix~\ref{appendix_schrieffer_wolff} and~\ref{appendix_lehmann}),
\begin{widetext}
\begin{equation}
    C(M_c)_{\alpha\alpha'}
    =
    \begin{cases}
    \frac{
        \bra{M_c}|\hat{f}^\dagger_{\alpha}|\ket{M_0}^*
        \bra{M_c}|\hat{f}^\dagger_{\alpha^\prime}|\ket{M_0}
    }{
        (S_0+\frac{1}{2})
    }, & \text{if}\quad N_c=N_0+1,\;S_c=S_0-\frac{1}{2},
    \\
    \frac{
        \bra{M_0}|\hat{f}^{\dagger}_{\alpha^\prime}|\ket{M_c}
        \bra{M_0}|\hat{f}^\dagger_{\alpha}|\ket{M_c}^*
    }{
        S_0
    }, & \text{if}\quad N_c=N_0-1,\;S_c=S_0-\frac{1}{2},
    \\
    - \frac{
        \bra{M_c}|\hat{f}^\dagger_{\alpha}|\ket{M_0}^*
        \bra{M_c}|\hat{f}^\dagger_{\alpha^\prime}|\ket{M_0}
    }{
        (S_0+\frac{1}{2})
    }, & \text{if}\quad N_c=N_0+1,\;S_c=S_0+\frac{1}{2},
    \\
    - \frac{
        \bra{M_0}|\hat{f}^{\dagger}_{\alpha^\prime}|\ket{M_c}
        \bra{M_0}|\hat{f}^\dagger_{\alpha}|\ket{M_c}^*
    }{
        (S_0+1)
    }, & \text{if}\quad N_c=N_0-1,\;S_c=S_0+\frac{1}{2},
    \end{cases}
\label{eq:C}
\end{equation}
\end{widetext}
where $\langle \dots || f^\dagger_{\alpha} || \dots \rangle$
are reduced matrix elements of the creation operator and
$N_0$ ($N_c$) and $S_0$ ($S_c$) are the occupation and spin
of the ground (excited) multiplet, respectively.

\subsection{Kondo orbitals}\label{kondo_orbitals}
Using the information contained in the coupling constants $J_{\alpha\alpha'}$ characterizing the scattering between \textit{channels}, we can identify the \textit{orbitals} involved in the screening of the molecular spin through the virtual processes depicted in Fig. \ref{fig:feynman_schrieffer-wolff}. The poor man's scaling \cite{anderson1970} equation for the coupling constant matrix, given to  second order in  perturbation theory, is
\begin{equation}
    \delta J
    =
    \rho J^2\frac{|\delta D|}{D},
\label{eq:Jscaling}
\end{equation}
which is an extension of the familiar scaling of the magnetic coupling constant to a matrix form (see Appendix~\ref{appendix_scaling}). In general, the scaling of each specific term $J_{\alpha\alpha'}$ depends quadratically on the values of other matrix elements, which hinders the solution of the equation and therefore the identification of the channels as ferromagnetic (FM) or antiferromagnetic (AFM). To circumvent this difficulty, we diagonalize the hermitian matrix $J$ to rewrite the effective interaction as $H_\K =\sum_a j_a \hat{\mathbf{S}} \cdot \hat{\mathbf{s}}_{aa}$, where $j_a$ are the eigenvalues and $\ket a = \sum_\alpha U_{\alpha a}\ket{\alpha}$ the corresponding rotated MOs, which we denote {\it  Kondo orbitals} (KO) in the following. 
\par 
In the KO basis, the scaling equations become decoupled and we can conclude that for $j_a>0$ the coupling is AFM and will increase as the energy scale is lowered, while for $j_a<0$ it is FM and will decrease with the lowering of the energy scale. Thus, the $2S_0$ AFM channels with the largest $j_a$ will screen the impurity spin, while the other AFM channels (if any) and the FM channels will decouple at low temperatures. Hence, we expect the KOs coupled to the screening AFM channels to host spectral peaks at sufficiently low temperatures, which allows us to predict the experimental d$I$/d$V$ maps to match the spatial distribution of those KOs.

\subsection{NRG and spectral functions}
We use the NRG method as implemented in the \verb|PointGroupNRG| code \cite{calvo-fernandez2023,PointGroupNRG} to solve the Anderson Hamiltonian for each molecule and compute zero-temperature orbital-resolved spectral functions that we can compare with experiment. We work in the KO basis, which allows us to leverage the Kondo orbital analysis by identifying the channels with the smallest contribution to the Kondo screening and discard them to keep a maximum of two, thereby making the systems tractable with the NRG. 
\par
Following the previous scaling arguments, the AFM channels
with largest $j_a$ should have the largest Kondo temperature
and hence the largest half-width at half-maximum (HWHM) in
the spectral peaks. To prove the validity of our model, we
show that our free parameters $\Gamma=\rho\pi V^2$ and $\mu$
can be adjusted to match the experimental observations: we
require the zero-bias peaks of each KO to have the
appropriate full-width at half-maximum $\FWHM=2\cdot\HWHM$,
which we estimate by subtracting the contribution from the
temperature $T_\ex$ \cite{gruber2018kondo}, 
\begin{equation}
    \FWHM_\NRG
    =
    \sqrt{\FWHM_\ex^2-(2 \pi k_B T_\ex)^2},
\end{equation}
where $\FWHM_\NRG$ is the calculated value and $\FWHM_\ex$
is the experimental value, usually obtained by fitting
lineshapes to the d$I$/d$V$ data. In the cases where an
underscreened ground state is observed, we also require the
rest of the zero-bias peaks to have a $\HWHM_\NRG$ lower than the
experimental temperature. For comparison with experiment, we
have to take into account that the measurements are made at
a non-zero temperature, causing the peaks to progressively
flatten as their associated HWHM goes past the Kondo
temperature \cite{costi1992}, which we take to be equal to
the zero-temperature half-width obtained via NRG,
$T_\K=\HWHM_\NRG$. This is an effect that we
cannot reproduce in our calculations. 
\par
We choose the half-bandwidth $D=10\eV$ value for all systems, which is larger than all excitation energies, as we have found this to yield smoother peaks with heights closer to $1/(\pi\Gamma)$. We use a discretization parameter $\Lambda=2$ and we keep 3000 multiplets at each NRG step. To obtain smoother spectral curves, we average the results over two interleaved discretization grids with twisting parameters $z=0.0$ and $0.5$ \cite{campo2005}.

\section{Electronic structure}
\label{elecstr}
We have characterized the electronic structures of the studied molecules by
means of DFT + CASCI calculations. We optimized the geometries at the PBE0 level
of theory \cite{pbe0} with the DFT code FHI-aims \cite{AIMS}. With this
optimized geometry, we perform an {\it ab initio} CASCI calculation starting
from the restricted DFT with the PBE functional~\cite{pbe} from which we obtain
the many-body multiplets, their energies, the natural orbitals and their
occupations \cite{NaturalOrbs}. We perform the latter calculation with
the quantum chemistry code ORCA \cite{orca,orca4.0} in order to
construct the integrals that define the molecular Hamiltonian in the basis of
MOs around Fermi energy:
\begin{equation}
\label{integrals1body}
    t_{ij} =  \int \phi_i(\mathbf{r}) \left( -\frac{\hbar}{2m} \nabla^2 + V(\mathbf{r})
 \right) \phi_j(\mathbf{r}) d^3\mathbf{r}
\end{equation}
and
\begin{equation}
\label{integrals 2body}
    \mathcal{V}_{ijkl} = \dfrac{1}{4\pi \epsilon_0} \int
    \dfrac{\phi_i(\mathbf{r})\phi_{j}(\mathbf{r}^\prime)\phi_{\replaced{k}{j}}(\mathbf{r}^\prime)\phi_{l}(\mathbf{r})}{\vert
    \mathbf{r} - \mathbf{r}^\prime   \vert}d^3\mathbf{r}
    d^3\mathbf{r}^\prime,
\end{equation}
where the indices $i,j,k,l$ label molecular orbitals,  ranging over the set of such orbitals selected as the active space. The one-electron potential $V(\mathbf{r})$ in Eq. \ref{integrals1body} includes the potentials of the ions and the potential from the electrons in the occupied inactive orbitals, that is,
\begin{equation*}
     V(\mathbf{r}) 
     = 
     \dfrac{1}{4\pi \epsilon_0}
     \sum_\gamma 
     \frac{e Z_\gamma}{\vert \mathbf{R}_\gamma - \mathbf{r} \vert }
     + 
     \sum_\lambda \int 
     \dfrac{\vert \phi_\lambda(\mathbf{r}^\prime) \vert ^2 }{\vert \mathbf{r} - \mathbf{r}^\prime \vert}  
     d^3\mathbf{r},
\end{equation*}
where $\gamma$ runs over the nuclei, $Z_\gamma$ is the
charge of the nuclei and the index $\lambda$ runs over
occupied inactive molecular orbitals given by
$\phi_{\lambda}(\mathbf{r})$, which are
always doubly-occupied in the possible Slater determinants
of our calculation. We then build up the many-body {\it ab
initio} molecular Hamiltonian
$\hat{\mathcal{H}}_{\text{CAS}}$ with those coefficients:
\begin{equation}
\label{molecular hamiltonian}
    \hat{\mathcal{H}}_{\text{CAS}} = \sum_{i,j, \sigma} t_{ij}\hat{C}^\dagger_{i \sigma}\hat{C}_{j \sigma} + \sum_{i,j,k,l, \sigma,\sigma^\prime} \mathcal{V}_{ijkl}\hat{C}^\dagger_{i \sigma}\hat{C}^\dagger_{j \sigma^\prime}\hat{C}_{k \sigma^\prime}\hat{C}_{l \sigma}.
\end{equation}
This many-body Hamiltonian in Eq. \ref{molecular
hamiltonian} can finally be cast into a matrix and solved by
exact diagonalization in a selected active space. The active
space is composed by selected frontier DFT orbitals around
the Fermi level (see Section \ref{Results}), which capture
the relevant electronic configurations contributing to the
ground state and low-energy excited states. The expansion in
Slater determinants, which represents the many-body
multiplets, is employed to analyze the presence of Kondo
screening processes as well as for the NRG calculation.
These multiplets for the neutral and charged molecules
studied in section IV are presented in the Appendix D.

\section{Results} \label{Results}
In this section, we apply the proposed theoretical model to rationalize previous
experimental data reporting the Kondo effect in molecules with several unpaired
electrons on the metallic surface. For each system, we provide the calculated
CASCI orbitals and the electronic configuration of the ground multiplet. We also
give the calculated $j_a$ coupling constants as a function of $\mu$ and, for the
chosen $\mu$ value, the KOs together with their d$I$/d$V$ maps. 
\added{The NTOs
and KOs are linear combinations of the orbitals of the
active space for the CASCI calculation, and therefore can be expanded in the basis of atomic orbitals. Using this atomic orbital basis, we} 
\replaced{s}{S}imulated \added{the} d$I$/d$V$ maps of NTO{s} and KO\added{s} presented in the manuscript by applying the PP-STM code \cite{Krej2017}, which takes into account the selection rules of the tunneling processes between tip and molecule. When using the NTOs, each eigenvalue of the transition density matrix provides the relative weight of a particular transition to the d$I$/d$V$ map. This analysis is then compared with experimental results using spectral functions calculated with the NRG for a value of the hybridization $\Gamma$. 

\subsection{Zinc porphyrin ($\ZP$)} \label{ZincPorphyrin}

\begin{figure}[b]
\centering
\includegraphics[width=1.0\linewidth]{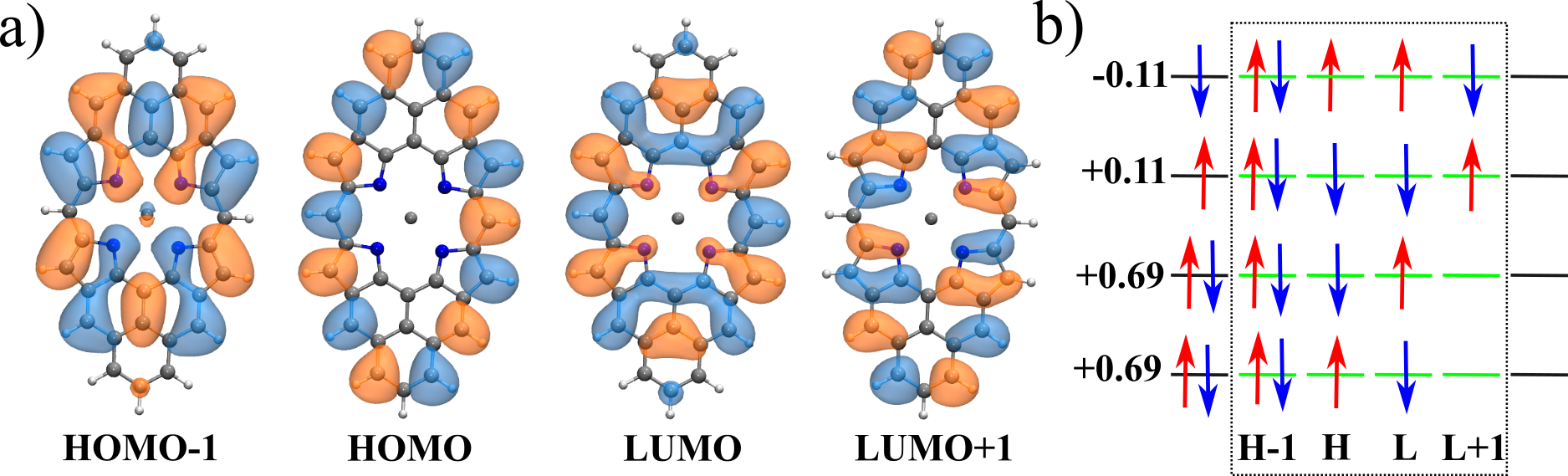}
\caption{
    a) The four main DFT orbitals employed in the active
    space calculation for the $\ZP$ molecule. b) Major
    Slater determinants for the ground states $\ZP$. The
    dashed box contains the four frontier orbitals: HOMO-1,
    HOMO, LUMO, and LUMO+1.  The values are the coefficients
    of those determinants in the orbital wave function.
} 
\label{fig: active space Zinc Porphyrin}
\end{figure}

First we discuss a~porphyrin molecule with extended $\pi$-system $\ZP$, synthesised and studied in Ref. ~\cite{Sun2020}. It hosts two radical centers at opposite pyrrol sites, created by C-H bond breaking with STM manipulation. Its open-shell configuration is 0.46\,eV more stable than the closed-shell one and the ground state is a triplet $S=1$. \deleted{When deposited on Au(111), the molecular levels undergo significant renormalization and there is a charge transfer of 0.73 electrons from the molecule to the substrate. }
The d$I$/d$V$ spectroscopy obtained at $T_\ex=4.5\,\K\approx 0.39\,\meV$ shows steps associated to inelastic spin excitation at $\pm 22$\,mV and a zero-bias feature, interpreted as a Kondo resonance.
\par

\begin{figure}[t]
\centering
\includegraphics[width=\linewidth]{"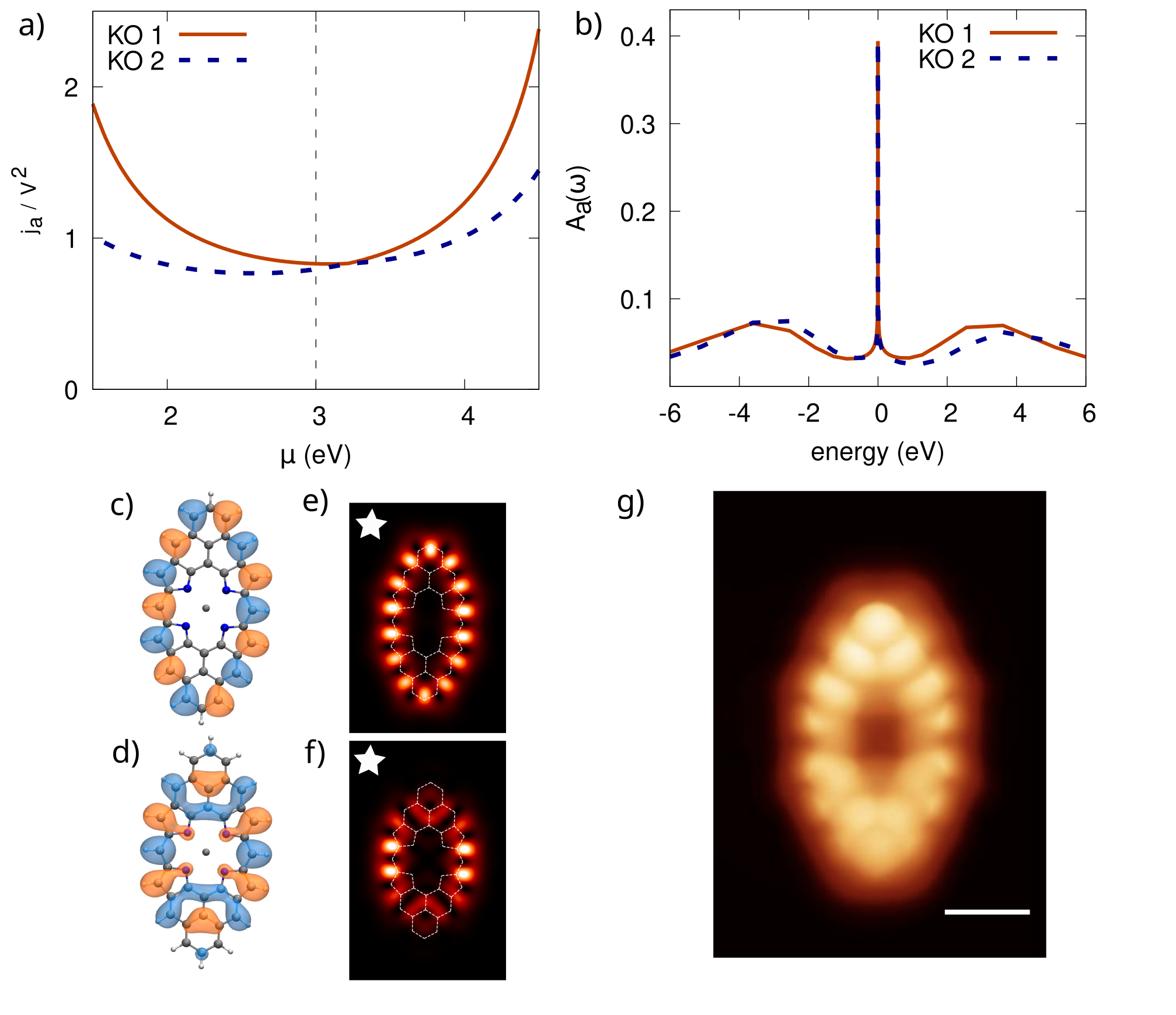"}
\caption{
    Kondo screening results for $\ZP$. (a) Rescaled coupling
    constants $j_a/V^2$ as a function of the chemical
    potential $\mu$. The vertical dashed line indicates the
    value $\mu=3\,\eV$ used in the KO and NRG
    calculations. b) Zero-temperature spectral
        functions $A_a(\omega)$ obtained with NRG with the
        hybridization set to $\Gamma=0.8$\,eV, yielding HWHM
        ($T_\K$) values of 0.8\,meV (9.3\,K) and 0.5\,meV
        (5.8\,K) for the KOs 1 and 2, respectively. Orbital
        shapes for KOs c) 1 and d) 2, and e,f) their
        respective d$I$/d$V$ maps. d$I$/d$V$ maps of KOs
        with $T_\K>T_\ex$ are indicated with white stars.
        g) Experimental low-bias constant current STM image
    acquired at 5 meV adopted from Ref. \cite{Sun2020}.
}
\label{fig:ZP}
\end{figure}

In Fig. \ref{fig: active space Zinc Porphyrin} we show the
main MOs obtained by DFT and the main Slater determinants
that make up the ground state of the molecule in gas phase.
Applying the Kondo orbital analysis from Sections
\ref{kondo_model} and \ref{kondo_orbitals}, we find two main
AFM channels in the allowed $\mu$ range, as shown in Fig.
\ref{fig:ZP} a). We choose $\mu=3$~eV, for which we compute
the associated KOs shown in Fig. \ref{fig:ZP} c,d), which
correspond very well with the HOMO and LUMO orbitals,
respectively. Their d$I$/d$V$ maps, shown in Fig.
\ref{fig:ZP} e,f), are compatible with the measured STM
maps shown in Fig. \ref{fig:ZP} g), so that the two channels coupled to the KOs can
participate in the screening. The calculated spectral
functions are shown in Fig. \ref{fig:ZP} b). They feature
zero-bias peaks with $\HWHM_\NRG=$~0.8\,meV 
and 0.5\,meV for
a hybridization $\Gamma=$~0.8\,eV. In Ref. \cite{Sun2020} the
authors use Ternes' model \cite{bib:ternes15} to fit the
d$I$/d$V$ spectrum with an AFM coupling of $|\rho J|=0.14$.
We obtain an equivalent coupling $\rho j_1\approx\rho
j_2\approx 0.14$ using $\Gamma=0.55$\,eV, but this results
in $T_\K=\HWHM_\NRG\approx$~0.8\,K and
0.5\,K for
KOs 1 and 2, respectively, both below the experimental
temperature. Note, however, that in the fitting used in Ref.
\cite{Sun2020} the spin-excitation cusps are considered as
well as the zero-bias peak.

\subsection{Fused Aza-[3]-Triangulene ($\FAT$)} \label{FusedAzaTriangulene}

The spin $S$ of [$N$]-Triangulene  increases linearly with
the number $N$ of units on the sides of the molecule
according to Ovchinnikov's rule \cite{bib:ovchinnikov78}.
The value of $S$ is decreased by 1/2 when a N atom
substitutional is introduced at the center
(Aza-[$N$]-Triangulene). Further alterations of the spin
state can be achieved by charge transfer upon adsorption on
a metal, dehydrogenation, and also by fusing Triangulenes at
different C sites. In Ref.~\cite{bib:calupitan23} the fused
Aza-[3]-Triangulene is reported. This molecule has
open-shell character, with a $S=1$ ground state and +1 net
charge when adsorbed on Au(111). The conductance spectrum
shows a zero-bias peak consistent with a Kondo effect at
$T_\ex=4.3$\,K~$\approx 0.37$\,meV. A fit to the Frota
function \cite{frota1992} yields $\FWHM_\ex=15.0$\,meV.

\begin{figure}[b]
\centering
\includegraphics[width=1.0\linewidth]{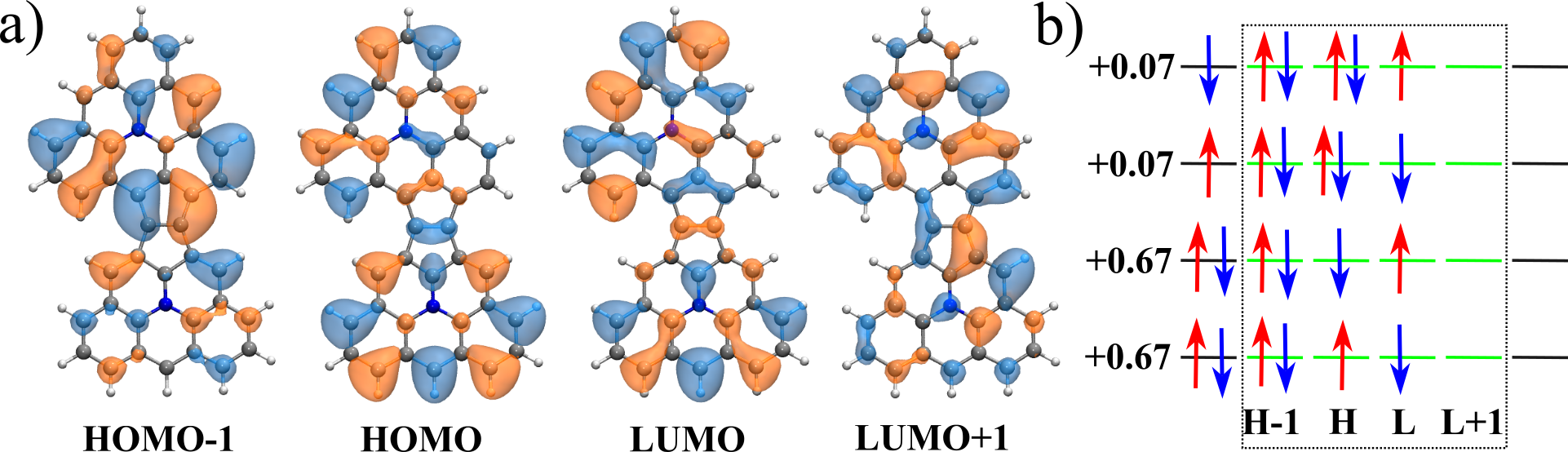}
\caption{Same information as in Fig. \ref{fig: active space Zinc Porphyrin} for the $\FAT$ molecule.}
\label{fig: active space Fused Aza Triangulene}
\end{figure}

\par
\begin{figure}
\centering
\includegraphics[width=\linewidth]{"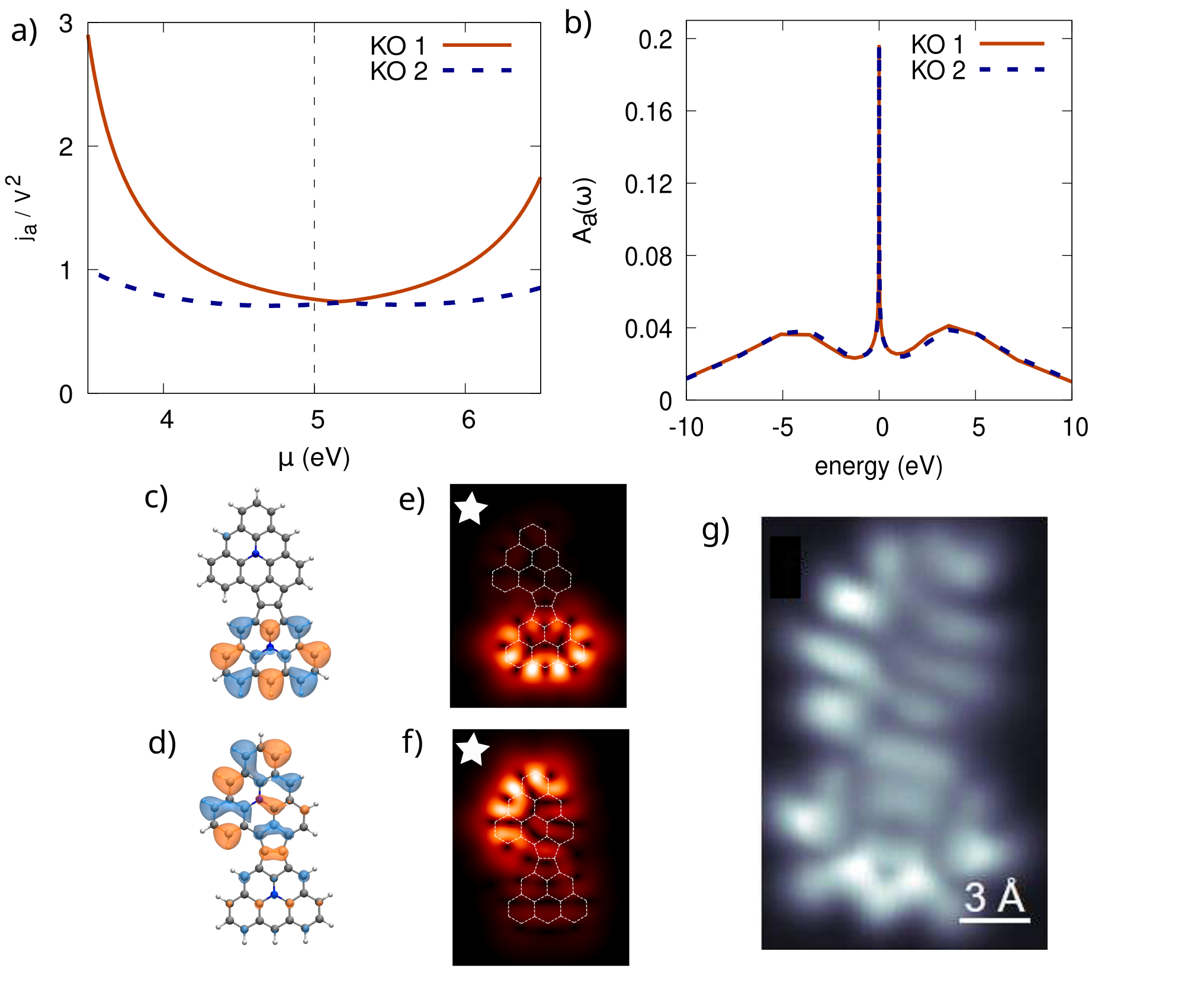"}
\caption{
    Same information as in Fig. \ref{fig:ZP} for the $\FAT$
    molecule with parameters $\mu=5$\,eV and
    $\Gamma=$~1.6~eV, and spectral
    functions with FWHM ($T_\K$) values
    14.6\,meV (84.7~K) and
    13.4\,meV (77.8~K) for KOs 1
    and 2, respectively. The experimental constant
    current STM image acquired at 5 meV adopted from
    Ref. \cite{bib:calupitan23}.
}
\label{fig:FAT}
\end{figure}

 
Our analysis shows that the KOs for this system are mostly combinations of the
HOMO and LUMO orbitals, given in Fig. \ref{fig: active space Fused Aza
Triangulene}, across the whole $\mu$ range. For our calculations we use
$\mu=5$\,eV, which gives us similar coupling constants for the two KOs located
in the lower (KO 1) and upper (KO 2) triangulenes, see Fig. \ref{fig:FAT}
a,c-f). With this choice of $\mu$, we expect STM signals in the
lower and upper triangulenes, in accordance with the experimental measurements
shown in Fig. \ref{fig:FAT} g). In particular, we accurately reproduce
the brighter sections in the lower part of the lower triangulene and in the
outside left edge of the upper triangulene. The computed spectral functions,
shown in Fig. \ref{fig:FAT} b), feature Kondo peaks with
$\FWHM_\NRG=$~14.6\,meV and 13.4\,meV close to
the temperature-corrected experimental value 14.8\,meV for
$\Gamma=$~1.6\,eV. The corresponding Kondo temperatures are
$T_\K=$~84.7\,K and 77.8\,K, well above the
experimental temperature.

\subsection{Rocket-shaped extended triangulene (\textbf{ETRI})} \label{Rocket} 
The \textbf{ETRI} molecule is a graphene nanoribbon studied
experimentally in Ref.~\cite{li2020}. Upon removal of two H
atoms by STM tip, it acquires a $S=1$ state. Here an
underscreened Kondo effect is probed by Zeeman splitting of
the zero-bias peak under the Kondo temperature, which is
estimated to be $T_\K\approx6$\,K from a Frota fitting of
the d$I$/d$V$ spectra, which has peak width of
$\HWHM_\ex\approx 0.75$\,meV measured at
$T_\ex=1.3$\,K~$\approx 0.11$~meV.

\begin{figure}[b]
\centering
\includegraphics[width=1.0\linewidth]{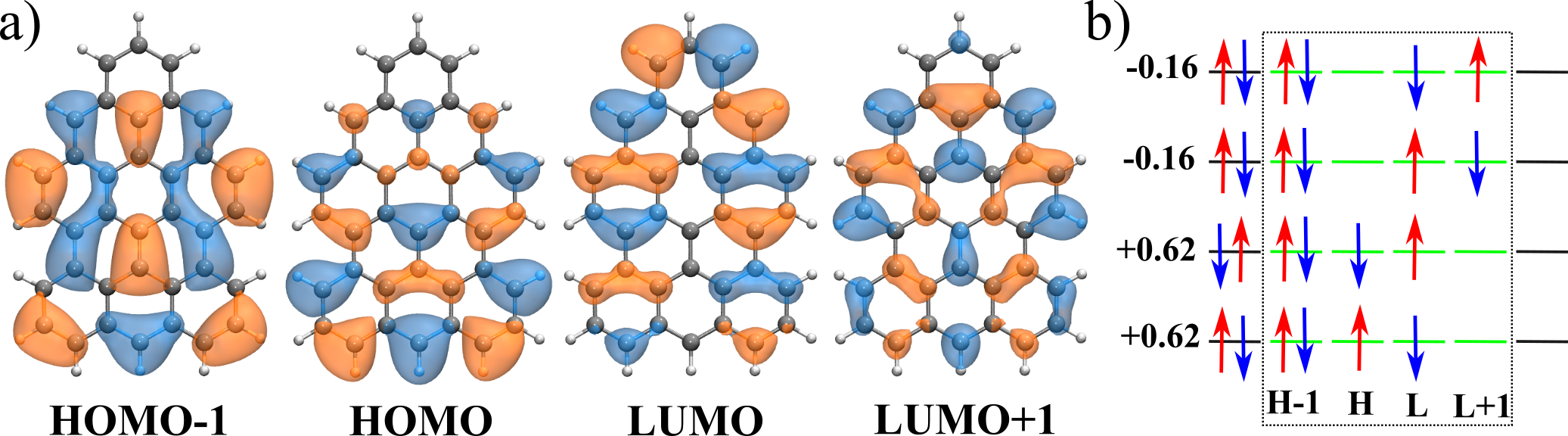}
\caption{Same information as in Fig. \ref{fig: active space Zinc Porphyrin} for the \textbf{ETRI} molecule.}
\label{fig: active space Rocket}
\end{figure}

\par
\begin{figure}
\centering
\includegraphics[width=\linewidth]{"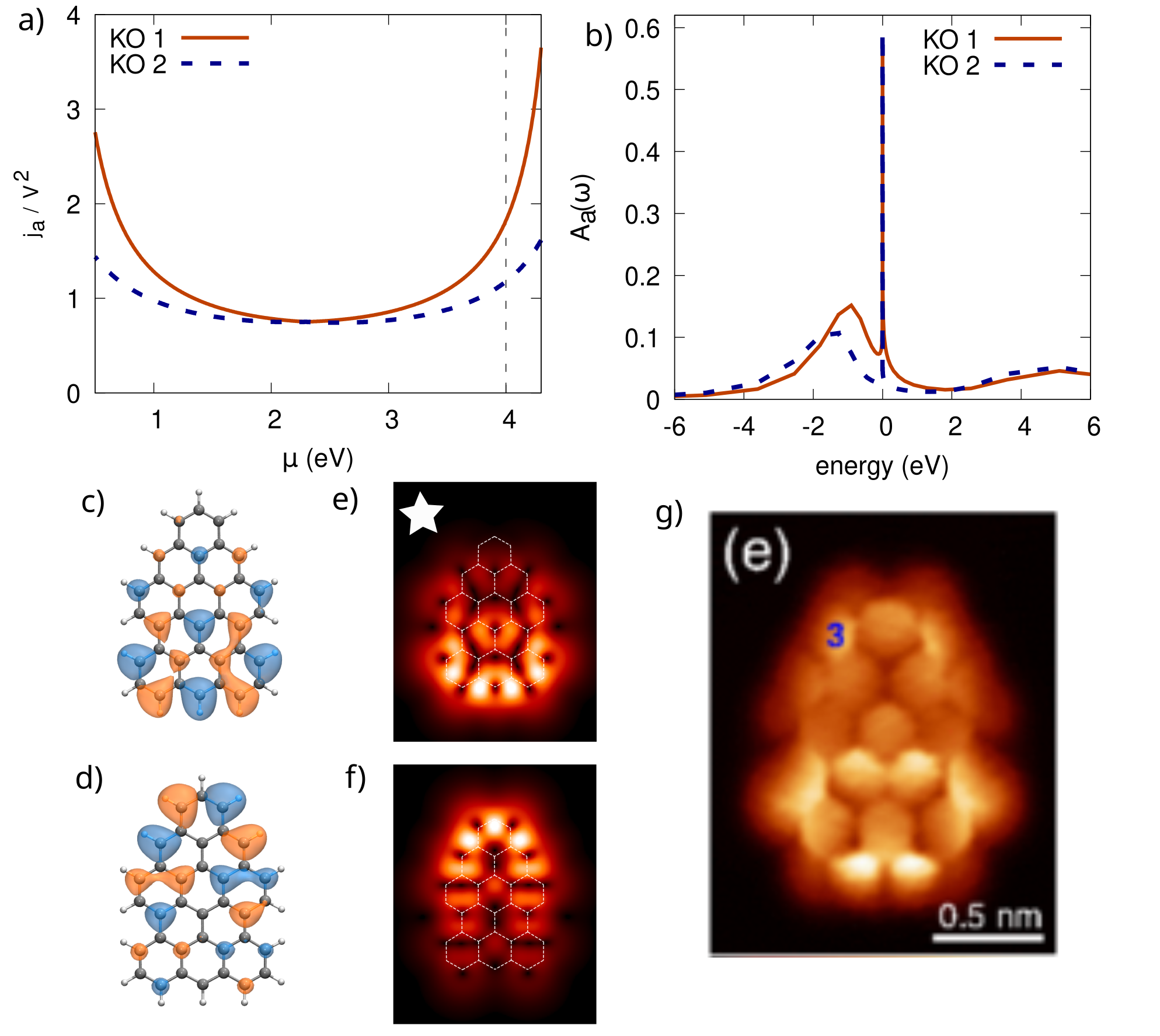"}
\caption{
    Same information as in Fig. \ref{fig:ZP} for
    \textbf{ETRI} with parameters $\mu=4\deleted{.0}$\,eV and
    $\Gamma=$~0.53\,eV, and spectral
    functions with HWHM ($T_\K$) values
    0.64\,meV (7.4\,K) and
    0.007\,meV (0.08\,K) for KOs 1 and
    2, respectively. The experimental constant current
    bond-resolved STM image acquired at 2 meV adopted from Ref.
    \cite{li2020}.
}
\label{fig:Rocket}
\end{figure}

To reproduce the experimental results, we chose a chemical
potential $\mu=4$\,meV that gives us a stronger AFM coupling
for the HOMO-shaped KO 1 located primarily in the lower half
of the molecule, where the zero-bias peaks are most intense,
see Fig. \ref{fig: active space Rocket} and Fig.
\ref{fig:Rocket} a-e). Moreover, that value of $\mu$ is
large enough to produce a sufficient separation in the
coupling strengths of both KOs, so that for
$\Gamma=$~0.53\,meV the KO 1 has an
associated
$\HWHM_\NRG=$~0.64~meV
($T_\K=7.4\,\K>T_\ex)$, close to the
temperature-corrected experimental value
0.67~meV, while the KO 2 has
$T_\K=$~0.08\,K~$\ll T_\ex$ and thus we
expect no Kondo signal from it. With these parameters we
obtain the experimentally observed underscreened ground
state as a result of an uneven magnetic coupling favoring
the HOMO-like KO, which gives us the d$I$/d$V$ map in Fig.
\ref{fig:Rocket} e). Here direct
    comparison between theory and experiment is not
straightforward as the experimental STM image, shown in Fig.
\ref{fig:Rocket} g), was taken in a~close tip-sample
distance, where lateral relaxation of CO-tip substantially
modifies the contrast \cite{Krej2017}. Nevertheless, we
still can find some similarities in the spatial distribution
with predominant Kondo signal in lower part of the
molecule.

\subsection{Aza-[5]-Triangulene (A5T)} \label{Aza5Triangulene}
The molecule $\textbf{A5T}$ is experimentally studied in
Ref.~\cite{bib:vilas23}. It has $S=3/2$ in the neutral
state, but due to charge transfer of an electron to the
Au(111) substrate it shows a $S=2$ ground state. Based on
the splitting of the Kondo peaks upon the application of a
magnetic field and the d$I$/d$V$ maps, the authors conclude
that this spin is partially screened at
$T_\ex=1.2$\,K~$\approx 0.1$\,meV by a single channel
associated with one of the molecular orbitals. The maximum
Kondo temperature estimated from fits to a Frota function is
$T_\K = 11$\,K~$\approx 0.9$\,meV and the average is
9\,K~$\approx 0.8$\,meV.

\begin{figure}
\centering
\includegraphics[width=1.0\linewidth]{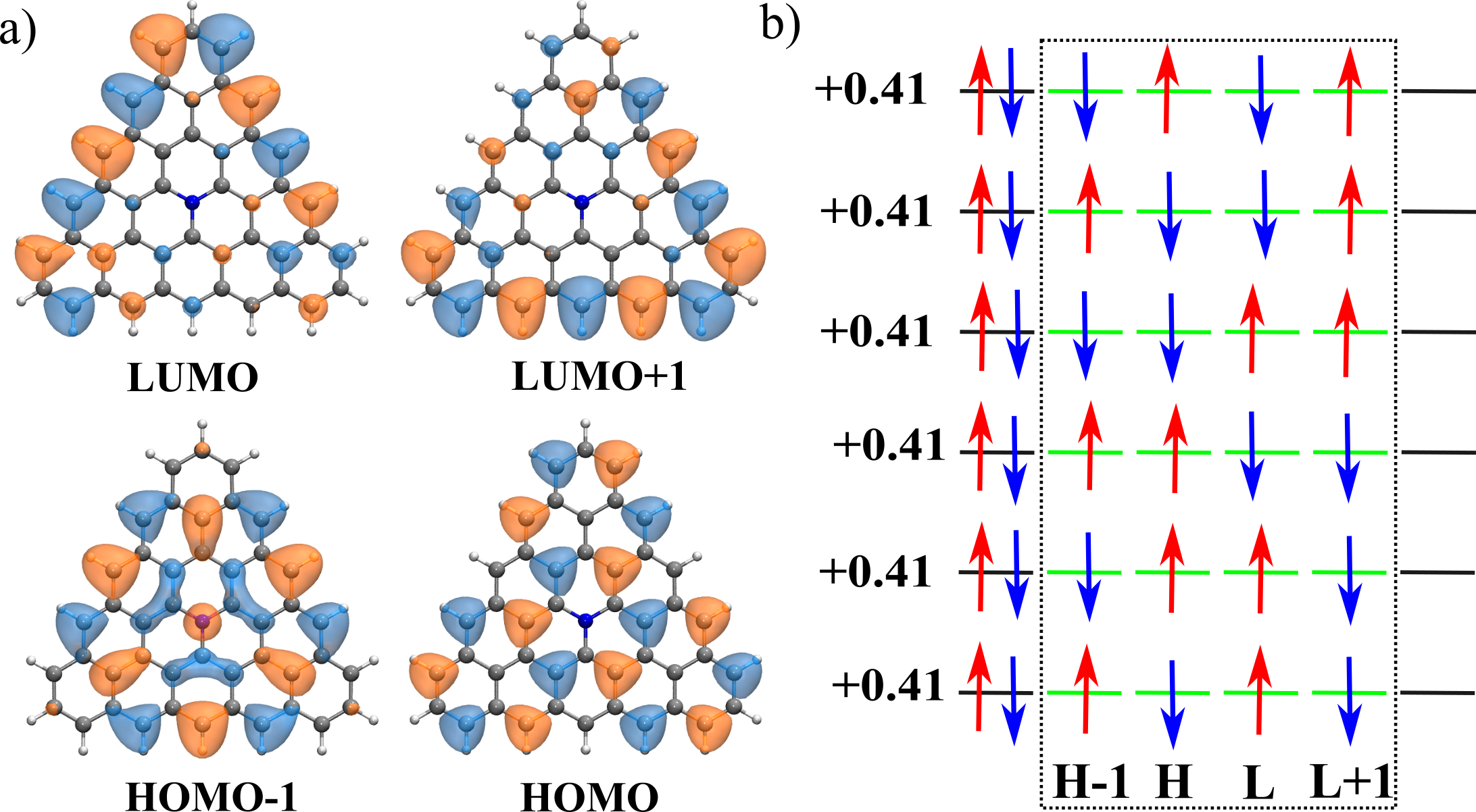}
\caption{Same information as in Fig. \ref{fig: active space Zinc Porphyrin} for \textbf{A5T}.}
\label{fig: Aza-[5]-Triangulene (A5T)}
\end{figure}

\par
\begin{figure}
\centering
\includegraphics[width=\linewidth]{"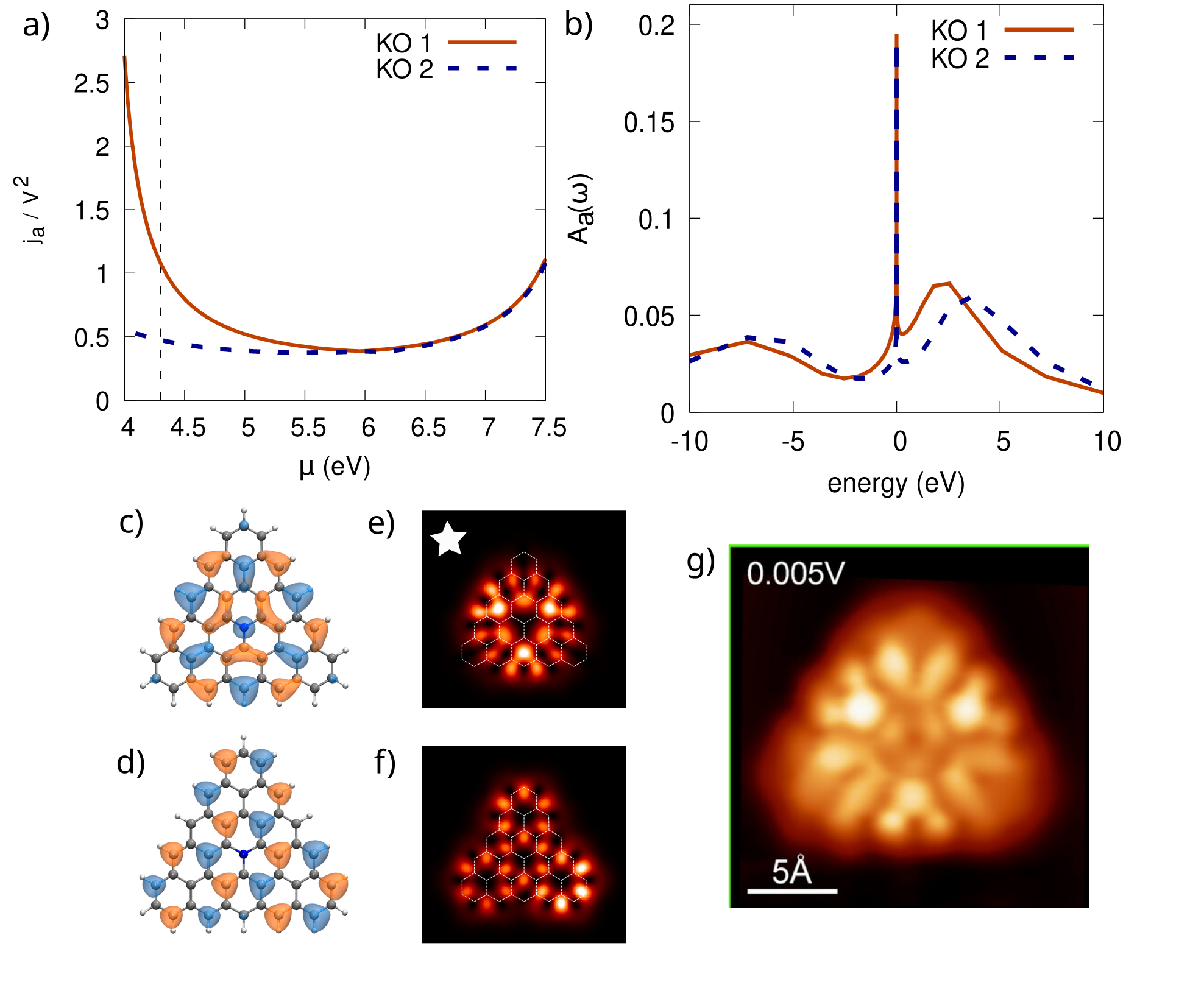"}
\caption{
    Same information as in Fig. \ref{fig:FAT} for
    $\textbf{A5T}$ with parameters $\mu=4.3$\,eV and
    $\Gamma=$~1.42\,eV, with HWHM
    ($T_\K$) values 0.75\,meV
    (8.7\,K) and $0.06$\,meV
    (0.7\,K) for KOs 1 and 2, respectively. Only the
    KOs with the strongest AFM coupling are shown.
    Experimental d$I$/d$V$ map taken at 5 meV adopted
    from Ref. \cite{bib:vilas23}.
}
\label{fig:A5T}
\end{figure}

 
DFT calculations for this molecule give us four degenerate
molecular orbitals, shown in Fig. \ref{fig:
Aza-[5]-Triangulene (A5T)}. For low values of $\mu$, we find
that the strongest KO has the shape of the HOMO-1, which is
identified in Ref. \cite{bib:vilas23} as the orbital
responsible for the screening, while the second strongest KO
resembles the HOMO, which would produce a d$I$/d$V$ signal
not observed in experiment, see Fig. \ref{fig:A5T}
a,c-g). We choose $\mu=4.3$\,eV to obtain a
sufficient separation between $j_1$ and $j_2$ and we fix
$\Gamma=$~1.42\,eV, which gives us for KO 1
$\HWHM_\NRG=$~0.75\,meV
($T_\K=8.7$~K~$>T_\ex$),
close to the temperature-corrected experimental value
0.74~meV, and for KO 2
$T_\K=0.7\,\K<T_\ex$, corresponding to a one-channel
underscreened impurity spin with the predicted d$I$/d$V$
signal given in Fig. \ref{fig:A5T} e)
, which matches well with experimental dI/dV maps at
5 meV shown in Fig. \ref{fig:A5T}g).

\section{Comparison with natural transition orbitals}
The characteristic second-order processes leading to the effective magnetic interaction between the molecule and the substrate involve electron scattering processes where the impurity spin orientation changes from $m_0$ to $m_0\pm 1$. This motivates invoking the NTOs \cite{NaturalOrbs} as the mediators of the spin-flip processes.
\par
In their general definition, the NTOs are defined as the eigenvectors of the matrices $A= T T^\dagger$ and $ B= T^\dagger T$, with the transition matrix elements of $T$ defined in our case as
\begin{equation}
T_{\alpha\alpha'} 
=
\bra{M_0,m_0-1} f^\dagger_{\alpha\downarrow} f_{\alpha'\uparrow} \ket{M_0,m_0}
\label{eq:T}
\end{equation}
for a fixed $m_0$ in the range $-S_0<m_0\leq S_0$. For our spin-flip transitions, it can straighforwardly be shown that $T=T^\dagger$ and, consequently, $A=B$ (see Appendix~\ref{appendix_nto_vs_ko}). Therefore, we obtain a single set of NTOs $\ket{n}$  with eigenvalues $\lambda_n$ fullfiling $1\geq\lambda_n\geq 0$. From these NTOs we can compute the spin-flip d$I$/d$V$ map by summing the intensities obtained from each NTO weighted by their eigenvalues $\lambda_n$.

\begin{figure}
\centering
\includegraphics[width=1.0\linewidth]{"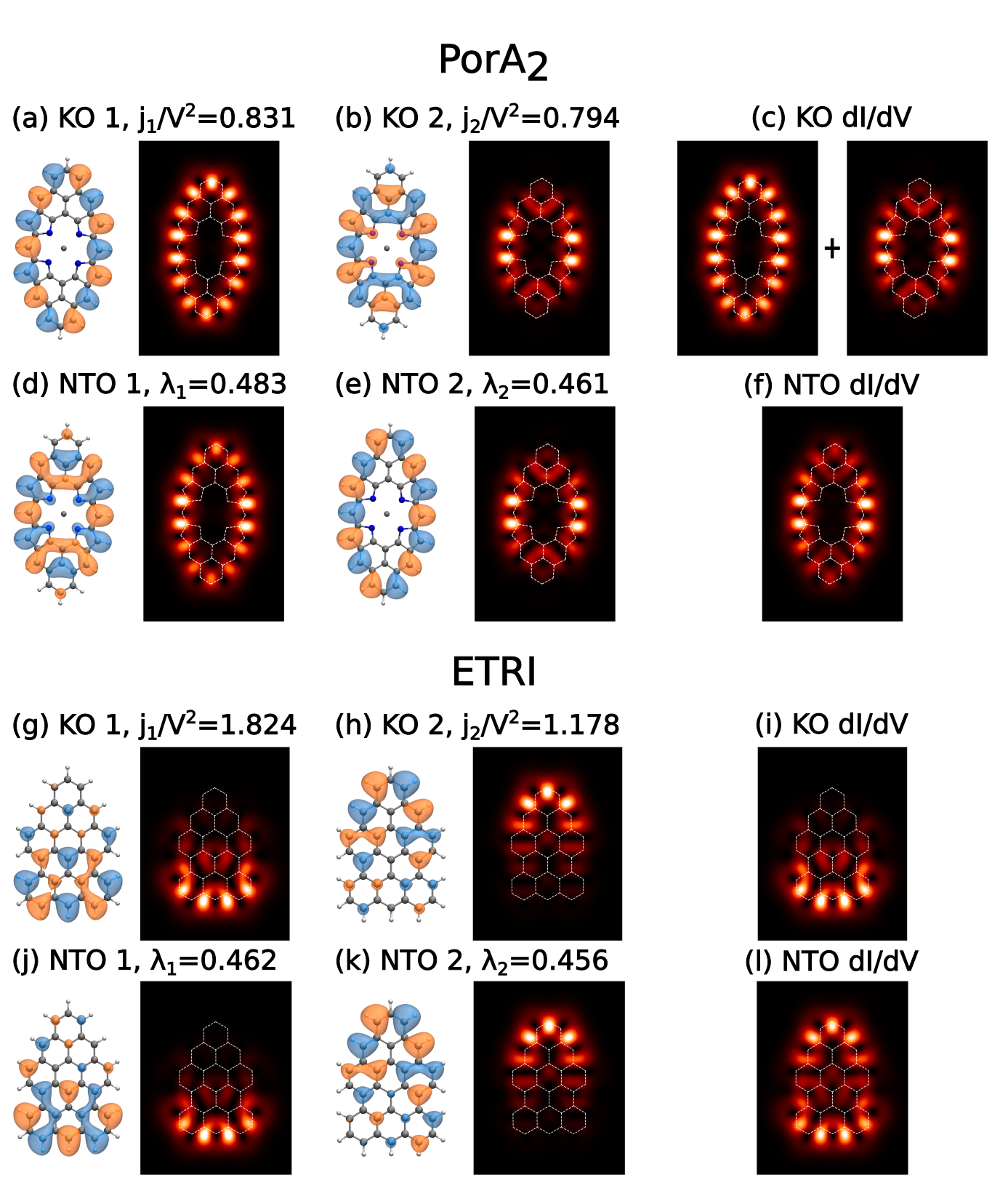"}
\caption{Comparison of KOs and NTOs for $\ZP$ and
$\textbf{ETRI}$. For $\ZP$ [\textbf{ETRI}] we show a,b)
[g,h)] the KOs with the value of their rescaled
couplings $j_a/V^2$ and their d$I$/d$V$ map, c) [i)]
the d$I$/d$V$ map expected from the Kondo orbital
analysis, d,e) [j,k)] the NTOs with their $\lambda_a$
eigenvalue and their d$I$/d$V$ maps, and f) [l)] the 
sum of the NTO d$I$/d$V$ maps weighted by their eigenvalues.}
\label{fig:KO_NTO_zp_rocket}
\end{figure}

\begin{figure}
\centering
\includegraphics[width=1.0\linewidth]{"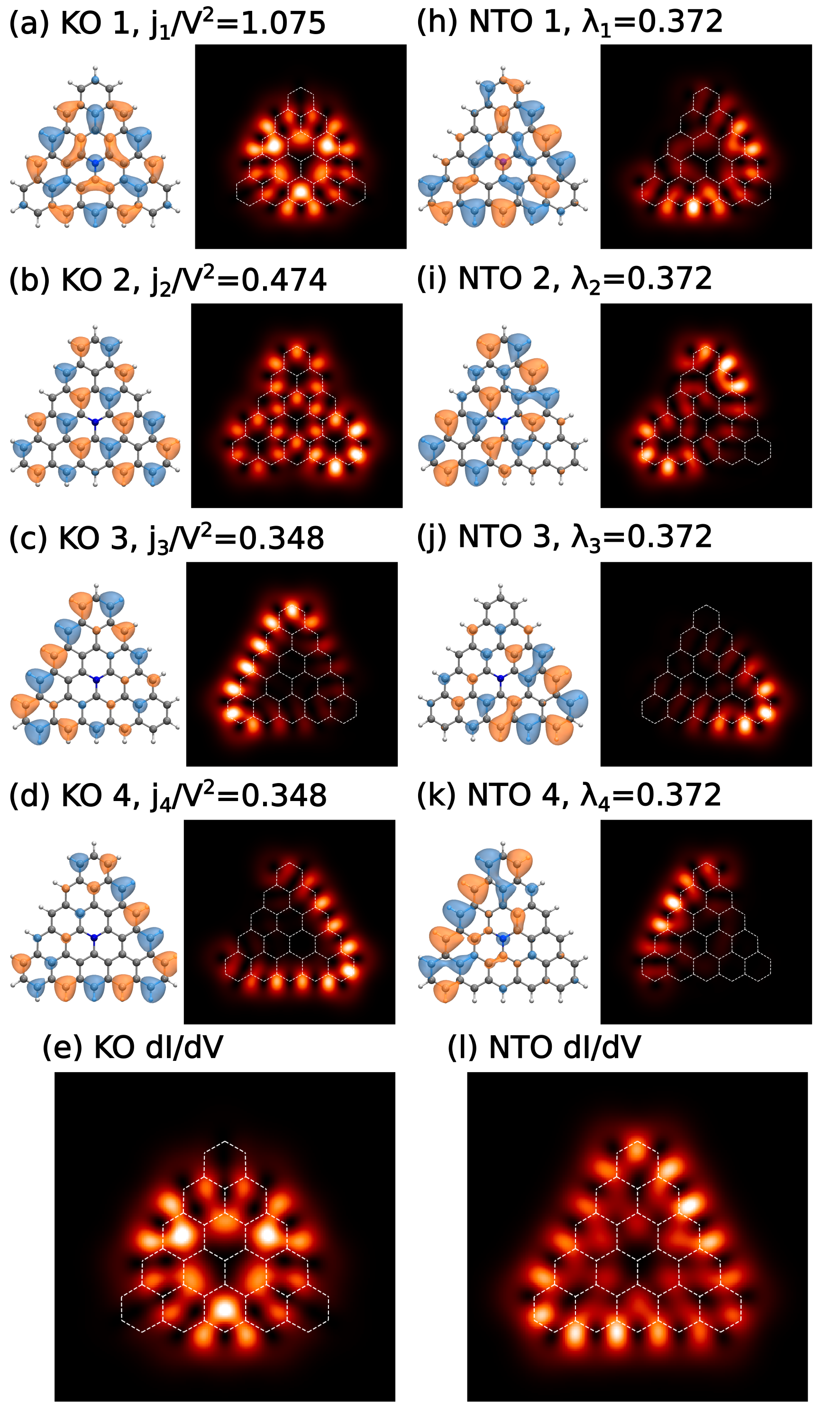"}
\caption{Comparison of KOs and NTOs for \textbf{A5T}. a-d)
KOs with their $j_a/V^2$ values and d$I$/d$V$ maps, e) d$I$/d$V$
map expected from the Kondo orbital analysis, h-k) NTOs
with their $\lambda_a$ eigenvalues and d$I$/d$V$ maps, and l) 
sum of the NTO d$I$/d$V$ maps weighted by their eigenvalues.}
\label{fig:KO_NTO_A5T}
\end{figure}

\par
In Fig. \ref{fig:KO_NTO_zp_rocket} we compare the results of
the KO+NRG analysis in Section \ref{Results} with the
d$I$/d$V$ maps calculated from the NTO for $\ZP$ and
$\textbf{ETRI}$. For $\ZP$, KOs and NTOs are very similar
and have the form of the HOMO and LUMO. The d$I$/d$V$ maps
obtained using both methods contain contributions from both
MOs, resulting in intensity distributions that qualitatively
match the experimental results. For $\textbf{ETRI}$, on the
other hand, NTOs have similar weights for both MOs,
resulting in a d$I$/d$V$ map that extends across the entire
molecule. This does not match the uneven underscreened
ground state and neither the uneven spatial distribution
observed in the experiment, which are well captured in the
KO+NRG results. As in the case of $\textbf{ETRI}$,
$\textbf{A5T}$ (Fig. \ref{fig:A5T}) also shows a discrepancy
between the two approaches. In this case, we obtain four
degenerate NTOs that combine with equal weights to yield a
d$I$/d$V$ map that does not match the experiment.  
\par
The reason for the discrepancies between the NTO and KO+NRG
Kondo maps is that the NTOs only take into account the
composition of the ground state, leaving aside all the
structure of excited states that is incorporated into the KO
analysis through the Schrieffer-Wolff transformation and
naturally into the NRG. This can be clearly seen by
manipulating Eq. \ref{eq:T} to rewrite the matrix elements
of $T$ as 
\begin{equation}
    T_{\alpha\alpha'}
    =
    \frac{S^-_{m_0-1,m_0}}{4}
    \sum_{M_c} C(M_c)_{\alpha\alpha'}.
\label{eq:Trewritten}
\end{equation}
where $S^-_{m_0-1,m_0}$ is the matrix element of the spin lowering operator and
$C(M_c)_{\alpha\alpha'}$ are the same coefficients appearing in Eq. \ref{eq:J}
and given by Eq. \ref{eq:C} (see Appendix~\ref{appendix_nto_vs_ko}). The only
difference between $T$ and $J$, apart from a constant factor, is the energy
denominator in $J$ that acts as a weight in the sum of the contributions. Thus,
we expect $T$ to retain the features of $J$ insomuch as the excitation energies
of the excited multiplets $M_c$ with sizeable Lehmann amplitudes are similar
(see Appendix~\ref{appendix_lehmann}). When this condition is met, $T$ can be
regarded as an approximation of $J$ and the signs of its eigenvalues tell us
whether each associated NTO mediates a FM or AFM interaction (the eigenstates of
$T$ are the same as those of $T^2$ with signs $\pm\sqrt{\lambda_n}$).

\section{Conclusions}

We presented the perturbative approach of spin-flip processes between various molecular multiplets, which enables us to identify the presence of FM and AFM screening channels. 
The latter give rise to multi-orbital Kondo screening effects. We demonstrated that in order to fully capture this physics, several low-energy excited states need to be taken into account. From the diagonalization of the magnetic exchange coupling matrix, we introduce the concept of \emph{Kondo orbital}, which allows us to map a real-space localization of the Kondo resonance observed experimentally.  The KO concept incorporates the excited state electronic structure through the Schrieffer-Wolff transformation. This allows to overcome the limitations of other quantities frequently used to identify Kondo effects in the d$I$/d$V$ spatial features, such as frontier orbitals and natural transition orbitals (NTO), which only account for the ground state.

The procedure has been applied to four different open-shell molecules with several unpaired electrons and NRG calculations have confirmed the existence of multi-orbital Kondo effects. In particular, we have been able to reproduce published experimental d$I$/d$V$ maps and zero-bias spectral features. We anticipate that the presented theoretical framework can be adopted for analysis of the emergence of the Kondo regime in other strongly correlated atomic systems with several unpaired electrons, too.

\appendix

\section{Schrieffer-Wolff transformation}\label{appendix_schrieffer_wolff}

The effective Kondo interaction for the ionic Anderson Hamiltonian to second order in perturbation theory takes the form
\begin{equation}
\label{eq:schriefferwolff}
\begin{aligned}
    \hat{\mathcal H}_\K 
    = 
    \sum_{m_0,m_0'}
    \int_{-D}^D
    d\epsilon 
    \int_{-D}^D 
    d\epsilon'
    T(m_0,\epsilon\alpha i\leftarrow m_0',\epsilon'\alpha' i') &
    \\ 
    \times \hat c^\dagger_{\epsilon\alpha i}
    \ket{M_0,m_0} \bra{M_0,m_0'}
    \hat c_{\epsilon'\alpha' i'},&
\end{aligned}
\end{equation}
where $T(m_0,\alpha\epsilon i\leftarrow m_0',\alpha'\epsilon' i')$ is the second order scattering amplitude  corresponding to virtual processes where an incoming electron with quantum numbers $\epsilon'$ (energy),  $\alpha'$ (channel) and $i'$ (spin) is scattered off the molecule into a state with quantum numbers $\epsilon$, $\alpha$ and $\sigma$, and the molecule spin changes from $m_0'$ to $m_0$. This amplitude can be written  as a sum of contributions from the intermediate excited multiplets,
\begin{equation}
\begin{aligned}
    T(m_0,\epsilon\alpha i& \leftarrow m_0',\epsilon'\alpha' i')
    = \\ &
    \sum_{M_c}
    T(M_c,N_c,S_c;m_0,\epsilon\alpha i\leftarrow
    m_0',\epsilon'\alpha' i'),
    \\
    \qquad
\end{aligned}
\end{equation}
where $T(M_c,N_c,S_c;m_0,\epsilon\alpha i\leftarrow
m_0',\epsilon'\alpha' i')$ is the amplitude for processes
where the intermediate molecular state belongs to the $M_c$
multiplet, which has occupation and spin quantum numbers
$N_c,S_c$. The Feynman diagrams for such processes are given
in Fig. \ref{fig:feynman_schrieffer-wolff}.
\par 
The amplitude for scattering through an intermediate state with one more electron ($N_c=N_0+1$) is
\begin{equation}
\begin{aligned}
    T( & M_c,N_0+1,S_c;  m_0,\epsilon\alpha i\leftarrow m_0',\epsilon'\alpha' i')
    = \\
    & -\rho V^2
    \frac{
        \bra{M_c}|\hat f^\dagger_{\alpha}|\ket{M_0}^*
        \bra{M_c}|\hat f^\dagger_{\alpha'}|\ket{M_0}
    }{
        \epsilon_{M_c}-\epsilon_{M_0}
    }
    \\
    &\times(\tfrac{1}{2},i;S_0, m_0|S_c,m_0+i)^*
    (\tfrac{1}{2},i';S_0, m_0'|S_c, m_0+i),
\label{eq:Tplus}
\end{aligned}
\end{equation}
where we have used the  Wigner-Eckart theorem to decompose the excitation matrix elements as
\begin{equation}
    \bra{M_1,m_1}\hat f^\dagger_{\alpha i}\ket{M_2,m_2}
    =
    \bra{M_1}|\hat f^\dagger_\alpha|\ket{M_2}
    ( \tfrac{1}{2},  i; S_2, m_2 | S_1, m_1)^*,
\end{equation}
where $\bra{M_1}|f^\dagger_\alpha|\ket{M_2}$ is the reduced matrix element with respect to spin symmetry and the $(\dots;\dots|\dots)$ terms are spin Clebsch-Gordan coefficients, which are given by
\begin{align}
    (\tfrac{1}{2},\pm\tfrac{1}{2};S,m|S+\tfrac{1}{2},m\pm\tfrac{1}{2}) 
    & = 
    \sqrt{\frac{1}{2}\left(1\pm\frac{m\pm\frac{1}{2}}{S+\tfrac{1}{2}}\right)},
    \\
    (\tfrac{1}{2},\pm\tfrac{1}{2};S,m|S-\tfrac{1}{2},m\pm\tfrac{1}{2}) 
    & =
    \mp \sqrt{\frac{1}{2}\left(1\mp\frac{m\pm\tfrac{1}{2}}{S+\tfrac{1}{2}}\right)}.
\end{align}
\par
Applying the previous results to $T(M_c,N_0+1,S_0-\frac{1}{2};m_0,\epsilon\alpha i\leftarrow m_0',\epsilon\alpha' i')$, we obtain
\begin{widetext}
\begin{equation}
\begin{aligned}
    T(& M_c,N_0+1,S_0-\frac{1}{2}; m_0,\epsilon\alpha i  \leftarrow m_0',\epsilon'\alpha' i')
    = \\
    & -\rho V^2
    \frac{
        \bra{M_c}|\hat f^\dagger_{\alpha}|\ket{M_0}^*
        \bra{M_c}|\hat f^\dagger_{\alpha'}\ket{M_0}
    }{
        \epsilon_{M_c}-\epsilon_{M_0}
    }
    \times\begin{cases}
        &-\frac{1}{S_0+\frac{1}{2}}
        \frac{1}{2}
        m_0
        +\frac{1}{2}\frac{S_0}{S_0+1/2}
        ,\text{ if } i= i'=\frac{1}{2},m_0=m_0',
        \\
        &+\frac{1}{S_0+\frac{1}{2}}
        \frac{1}{2}
        m_0
        +\frac{1}{2}\frac{S_0}{S_0+1/2}
        ,\text{ if } i= i'=-\frac{1}{2},m_0=m_0',
        \\
        &-\frac{1}{S_0+\frac{1}{2}}
        \frac{1}{2}\sqrt{S_0(S_0+1)-m_0(m_0+1)}
        ,\text{ if } i=\frac{1}{2}, i'=-\frac{1}{2},m_0=m_0'-1,
        \\
        &-\frac{1}{S_0+\frac{1}{2}}
        \frac{1}{2}\sqrt{S_0(S_0+1)-m_0(m_0-1)}
        ,\text{ if } i=-\frac{1}{2}, i'=\frac{1}{2},m_0=m_0'+1
    \end{cases}
    \\
    =& 
    \rho V^2
    \frac{
        \bra{M_c}|\hat f^\dagger_{\alpha}|\ket{M_0}^*
        \bra{M_c}|\hat f^\dagger_{\alpha'}\ket{M_0}
    }{
        (\epsilon_{M_c}-\epsilon_{M_0})(S_0+\frac{1}{2})
    }
    \left(
        \frac{1}{2}
        \mathbf S_{m_0 m_0'}
        \cdot
        \boldsymbol \sigma_{ii'}
        +
        S_0\delta_{ii'}\delta_{m_0 m_0'}
    \right).
\label{eq:T1}
\end{aligned}
\end{equation}
\end{widetext}
The diagonal term proportional to
$S_0\delta_{ii'}\delta_{m_0 m_0'}$ gives rise to a scattering
potential that we will ignore from now on, as its main
contribution is to slightly renormalize the magnetic
coupling parameters \cite{krishna-murthy1980b}. $\mathbf S$
is the matrix representation of the spin (vector) operator,
which we obtain together with the Pauli matrices by
identifying their components using the expressions
\begin{equation}
\begin{aligned}
    &S^z_{m_0 m_0'} = m_0 \delta_{m_0,m_0'},
    \\
    &S^-_{m_0 m_0'} 
    = 
    \sqrt{S_0(S_0+1)-m_0(m_0+1)}
    \delta_{m_0,m_0'-1},
    \\
    &S^+_{m_0 m_0'} 
    = 
    \sqrt{S_0(S_0+1)-m_0(m_0-1)}
    \delta_{m_0,m_0'+1}.
\label{eq:Selements}
\end{aligned}
\end{equation}
Following the same procedure for the remaining $N_c,S_c$ combinations, we obtain the following contributions from second-order scattering to the magnetic coupling:
\begin{widetext}
\begin{align}
    & T(M_c,N_0-1,S_0-\tfrac{1}{2};m_0,\epsilon\alpha i \leftarrow m_0',\epsilon'\alpha' i)
    =
    \rho V^2
    \frac{
        \bra{M_0}|\hat f^\dagger_{\alpha}|\ket{M_c}
        \bra{M_0}|\hat f^\dagger_{\alpha'}|\ket{M_c}^*
    }{
        (\epsilon_{M_c}-\epsilon_{M_0})S_0
    }
    \frac{1}{2}
        \mathbf S_{m_0 m_0'}
        \cdot
        \boldsymbol \sigma_{ii'}, \label{eq:T2}
\\
    & T(M_c,N_0+1,S_0+\tfrac{1}{2};m_0,\epsilon\alpha i \leftarrow m_0',\epsilon'\alpha' i)
    =
    - \rho V^2
    \frac{
        \bra{M_c}|\hat f^\dagger_{\alpha}|\ket{M_0}^*
        \bra{M_c}|\hat f^\dagger_{\alpha'}|\ket{M_0}
    }
    {
        (\epsilon_{M_c}-\epsilon_{M_0})(S_0+\frac{1}{2})
    }
    \frac{1}{2}
    \mathbf S_{m_0 m_0'}
    \cdot
    \boldsymbol \sigma_{ii'}, \label{eq:T3}
\\
    & T(M_c,N_0-1,S_0+\tfrac{1}{2};m_0,\epsilon\alpha i \leftarrow m_0',\epsilon'\alpha' i)
    =
    -
    \rho V^2
    \frac{  
        \bra{M_0}|\hat f^\dagger_{\alpha}|\ket{M_c}
        \bra{M_0}|\hat f^\dagger_{\alpha'}|\ket{M_c}^*
    }{
        (\epsilon_{M_c}-\epsilon_{M_0})(S_0+1)
    }
    \frac{1}{2}
    \mathbf S_{m_0 m_0'}
    \cdot
    \boldsymbol \sigma_{ii'}. \label{eq:T4}
\end{align}
\end{widetext}
\par 
Substituting Eqs. \ref{eq:T1}, \ref{eq:T2}, \ref{eq:T3} and \ref{eq:T4} in Eq. \ref{eq:schriefferwolff}, we obtain the Kondo Hamiltonian
\begin{equation}
    \hat H _\K
    = 
    \frac{J_{\alpha\alpha'}}{2}
    \sum_{i,i'}
    \sum_{\alpha,\alpha'}
    \left(
    \hat{\mathbf S}
    \cdot 
    \boldsymbol \sigma_{ii'}
    \right)
    \hat \psi ^\dagger _{\alpha i}
    \hat \psi _{\alpha' i'},
\end{equation}
where $J_{\alpha\alpha'}$ is given by Eqs. \ref{eq:J} and \ref{eq:C}, $\hat{\mathbf S}=\sum_{m_0 m_0'} \mathbf S_{m_0 m_0'}\ket{M_0 m_0}\bra{M_0 m_0'}$ is the impurity spin operator, and $\hat{\mathbf s}_{\alpha\alpha'}$ is the local spin density operator of the conduction electrons \cite{pustilnik2001},
\begin{equation}
    \hat{\mathbf s}_{\alpha\alpha'} 
    = 
    \sum_{ii'}
    \frac{1}{2}
    \hat \psi ^\dagger _{\alpha i}
    \boldsymbol \sigma_{ii'}
    \hat \psi _{\alpha' i'},
\end{equation}
where the $\hat \psi ^{(\dagger)} _{\alpha i}$ operators are defined as $\hat \psi _{\alpha i} = \int d\epsilon \rho^{\frac{1}{2}} \hat c_{\epsilon\alpha i}$ and fulfill the commutation relation $\{\hat \psi^\dagger_{\alpha i}, \hat \psi_{\alpha'i'}\}=\delta_{\alpha\alpha'}\delta_{ii'}$. 
\par 
Throughout the calculation, we have ignored the renormalization of the excitation energies. Although its effect is usually small for systems close to particle-hole symmetry, in particle-hole asymmetric cases it can result in the renormalization of coupling constants, in driving the system close to the mixed-valence regime, and even in changes in the ground state as the temperature is lowered \cite{krishna-murthy1980b}. In principle, this could be relevant for the \textbf{ETRI} and \textbf{A5T} molecules (Sections \ref{Rocket} and \ref{Aza5Triangulene}). For those systems, we rely on the NRG calculations to verify that the system does indeed exhibit the zero-bias peaks characteristic of the Kondo regime and that the larger peak widths correspond to the largest coupling constants.

\section{Scaling equations for the coupling constants}\label{appendix_scaling}
\begin{figure}[b]
\scalebox{1.3}{
\begin{tikzpicture}
\begin{feynman}
\vertex (i1) {\tiny$m_0$};
\vertex [above=of i1] (e1) {\tiny$\epsilon\alpha i$};
\vertex [dot, right=of i1] (v1) {};
\vertex [dot, right=of v1] (v2) {};
\vertex [right=of v2] (i2) {\tiny$m_0'$};
\vertex [above=of i2] (e2) {\tiny$\epsilon'\alpha' i'$};
\diagram* {
(e2) -- [charged scalar] (v2) -- [charged scalar, half left, edge label=\tiny$\epsilon''\alpha''i''$] (v1) -- [charged scalar] (e1),
(i2) -- [fermion] (v2) -- [fermion, edge label=\tiny$m_0''$] (v1) -- [fermion] (i1),
};
\end{feynman}
\end{tikzpicture} }
\scalebox{1.3}{
\begin{tikzpicture}
\begin{feynman}
\vertex (i1) {\tiny$m_0$};
\vertex [above=of i1] (e1) {\tiny$\epsilon\alpha i$};
\vertex [dot, right=of i1] (v1) {};
\vertex [dot, right=of v1] (v2) {};
\vertex [right=of v2] (i2){\tiny$m_0'$};
\vertex [above=of i2] (e2){\tiny$\epsilon'\alpha' i'$};
\diagram* {
(e2) -- [charged scalar] (v1) -- [charged scalar, half right, edge label'=\tiny$\epsilon''\alpha'' i''$] (v2) -- [charged scalar] (e1),
(i2) -- [fermion] (v2) -- [fermion,edge label=\tiny$m_0''$] (v1) -- [fermion] (i1),
};
\end{feynman}
\end{tikzpicture}
}
\caption{Feynman diagrams for virtual processes with an intermediate electron (left) and hole (left).}
\label{fig:feynman_spin-flip}
\end{figure}
The scaling equations for the coupling constants $J_{\alpha\alpha'}$ and $j_a$ are obtained using the poor man's scaling method \cite{anderson1970} to the Kondo interaction derived in Section \ref{appendix_scaling}. Following a procedure analogous to the Schrieffer-Wolff transformation in Appendix \ref{appendix_schrieffer_wolff}, we sum over the second-order processes depicted in Fig. \ref{fig:feynman_spin-flip} to eliminate electronic degrees of freedom lying in the energy range$|\epsilon|\in[D-|\delta D|,D]$. By doing so, we obtain the differential scaling of the Kondo interaction,
\begin{equation}
\begin{aligned}
    \delta \hat{\mathcal H} _K
    =
    \sum_{m_0,m_0'}
    \int_{-D}^D d\epsilon 
    \int_{-D}^D d\epsilon'
    \delta T(m_0,\epsilon\alpha  i\leftarrow
    m_0',\epsilon'\alpha'i') &
    \\ \times
    \hat c ^\dagger _{ \epsilon\alpha i}
    \ket{M_0, m_0}
    \bra{M_0, m_0'}
    \hat c _{ \epsilon'\alpha' i'} &,
\label{eq:deltaHK}
\end{aligned}
\end{equation}
where $\delta T(m_0,\epsilon\alpha  i\leftarrow m_0',\epsilon'\alpha'i')$ is the sum over second order scattering amplitudes involving the creation of an intermediate particle ($p$) with or hole ($h$) with energy $|\epsilon|\in[D-\delta D,D]$: 
\begin{equation}
\begin{aligned}
    \delta T(m_0,\epsilon\alpha  i\leftarrow m_0',\epsilon'\alpha'i')
    &=
    \delta T_p(m_0,\epsilon\alpha i\leftarrow m_0',\epsilon'\alpha'i')
    \\
    &+
    \delta T_h(m_0,\epsilon\alpha  i\leftarrow m_0',\epsilon'\alpha' i').
\end{aligned}
\end{equation}
To compute the scattering amplitudes, we generalize the equations in \cite{coleman2015} to include many channels. For the contribution from processes with an intermediate particle, we obtain
\begin{widetext}
\begin{equation}
\begin{aligned}
    \delta T_p(m_0,\epsilon\alpha i\leftarrow m_0',\epsilon'\alpha'i')
    &=
    \sum_{\alpha'',m_0''}
    \frac{1}{2}
    \rho
    J_{\alpha\alpha''}
    \left(
        \mathbf S_{m_0 m_0''}
        \cdot
        \boldsymbol \sigma_{ii''}
    \right)
    \frac{\dD }{(-D)}
    \frac{1}{2}
    \rho
    J_{\alpha''\alpha'}
    \left(
        \mathbf S_{m_0'' m_0'}
        \cdot
        \boldsymbol \sigma_{i''i'}
    \right)
    \\
    &=
    -\frac{1}{4}
    \rho^2
    \left[J^2\right]_{\alpha\alpha'}
    \frac{\dD}{D}
    \sum_{a,b}
    [S^a S^b]_{m_0 m_0'}
    [\sigma^a \sigma^b]_{ii'}
\end{aligned}
\end{equation}
for the scattering through intermediate states with an excited electron, where $\rho \dD$ is the number of particle states in the $[D-\dD,D]$ energy range and $S^a$ and $\sigma^a$ are the $a$ components of $\mathbf S$ and $\boldsymbol \sigma$, respectively. For the scattering through an intermediate state with the incoming electron plus an electron-hole pair, we have
\begin{equation}
\begin{aligned}
    \delta T_h(m_0,\epsilon\alpha i\leftarrow m_0',\epsilon'\alpha'i')
    &=
    \sum_{\alpha'',m_0''}
    -
    \frac{1}{2}
    \rho
    J_{\alpha''\alpha'}
    \left(
        \mathbf S_{m_0 m_0''}
        \cdot
        \sigma_{i''i'}
    \right)
    \frac{\dD }{(-D)}
    \frac{1}{2}
    \rho
    J_{\alpha\alpha''}
    \left(
        \mathbf S_{m_0'' m_0'}
        \cdot
        \boldsymbol \sigma_{ii''}
    \right)
    \\
    &=
    \frac{1}{4}
    \rho^2
    \frac{\dD}{D}
    \left[J^2\right]_{\alpha\alpha'}
    \sum_{a,b}
    [S^a S^b]_{m_0 m_0'}
    [\sigma^b \sigma^a]_{ii'}.
\end{aligned}
\end{equation}
\end{widetext}
To obtain the total amplitude, we sum the particle and hole terms and apply the commutation relations of the spin and Pauli matrices to arrive at
\begin{equation}
    \delta T(m_0,\epsilon\alpha i\leftarrow m_0',\epsilon'\alpha'i')
    =
    -
    \frac{1}{2}
    \rho^2 
    \left[ J^2\right]_{\alpha\alpha'}
    \frac{\dD}{D}
    \mathbf S_{m_0 m_0'}
    \cdot
    \boldsymbol \sigma_{ii'}.
\end{equation}
Introducing this in Eq. \ref{eq:deltaHK}, we get
\begin{equation}
\begin{aligned}
    \delta \hat H_\K
    =
    -
    \sum_{m_0,m_0'}
    \sum_{i,i'}
    \sum_{\alpha,\alpha'}
    \int_{-D}^D
    d\epsilon
    \int_{-D}^D
    d\epsilon'
    \frac{1}{2}
    \rho^2 
    \left[J^2\right]_{\alpha\alpha'}
    \frac{\dD}{D} &
    \\  \times
    \left(
    \mathbf S_{m_0 m_0'}
    \cdot
    \boldsymbol \sigma_{ii'}    
    \right)
    \hat c ^\dagger _{\epsilon\alpha  i}
    \ket{M_0, m_0}
    \bra{M_0, m_0'}
    \hat c _{\epsilon'\alpha'  i'} &
    \\ =
    -
    \sum_{i,i'}
    \sum_{\alpha,\alpha'}
    \frac{1}{2}
    \rho
    \left[J^2\right]_{\alpha\alpha'}
    \frac{\dD}{D}
    (
        \hat{\mathbf S}
        \cdot
        \boldsymbol \sigma_{ii'}    
    )
    \hat \psi ^\dagger _{\alpha i}
    \hat \psi _{\alpha' i'}.&
\end{aligned}
\end{equation}
Isolating the coupling constant, we find
\begin{equation}
    \delta J_{\alpha\alpha'}
    =
    \rho [J^2]_{\alpha\alpha'}\frac{|\dD|}{D}.
\end{equation}
In the diagonal channel basis, this reduces to the simple expression
\begin{equation}
    \delta j_a
    =
    \rho j_a^2\frac{|\dD|}{D},
\end{equation}
Therefore, as the energy scale is lowered, the magnitude $|j_a|$ of the coupling will grow in antiferromagnetically coupled channels ($j_a>0$) decrease in the ferromagnetically coupled channels ($j_a<0$). 

\section{Natural transition orbitals and Kondo orbitals}\label{appendix_nto_vs_ko}
To prove that $ T= T^\dagger$, we introduce the resolution of the identity $\hat I =\sum_{M,m} \ket{Mm}\bra{Mm}$ in Eq. \ref{eq:T} and we apply the Wigner-Eckart theorem,
\begin{equation}
\begin{aligned}
T_{\alpha\alpha'} 
= &
\bra{M_0,m_0-1}
\hat f^\dagger_{\alpha \downarrow}
\hat I
\hat f_{\alpha'\uparrow}
\ket{M_0,m_0} 
\\ = &
\sum_{M,m}
\bra{M_0}|\hat f^\dagger_{\alpha}|\ket{M}
(\tfrac{1}{2},-\tfrac{1}{2};S,m|S_0,m_0-1)^*
\\
& \quad \times
\bra{M_0}|\hat f^\dagger_{\alpha'}|\ket{M}^*
(\tfrac{1}{2},\tfrac{1}{2};S,m|S_0,m_0).
\end{aligned}
\end{equation}
Since we can choose the molecular states to be real, it follows that the reduced matrix elements can also be chosen as real and therefore $T_{\alpha\alpha'}$ is invariant under swapping the indices $\alpha$ and $\alpha'$.
\par
To compare NTOs and the KOs analytically, we begin by rewriting the matrix elements of $T$ as
\begin{equation}
\begin{aligned}
2 \cdot T_{\alpha\alpha'}
= &
\bra{M_0,m_0-1} \hat f^\dagger_{\alpha\downarrow} \hat I \hat f_{\alpha'\uparrow} \ket{M_0,m_0}
\\  -&
\bra{M_0,m_0-1} \hat f_{\alpha'\downarrow}\hat I \hat f^\dagger_{\alpha\uparrow} \ket{M_0,m_0},
\end{aligned}
\end{equation}
where we have used the commutation relation $\{ \hat f^\dagger_{\alpha\downarrow} ,\hat f_{\alpha'\uparrow}\}=0$. Following the same matrix element expansion as in Appendix \ref{appendix_schrieffer_wolff}, we arrive at results similar to Eqs. \ref{eq:T1}, \ref{eq:T2}, \ref{eq:T3} and \ref{eq:T4},
\begin{widetext}
\begin{align}
    \bra{M_0,m_0-1}
    \hat f^\dagger_{\alpha\downarrow} 
    \hat I 
    \hat f_{\alpha'\uparrow}
    \ket{M_0,m_0}
    =
    \frac{1}{2}
    S^-_{m_0-1,m_0}
    \sum_{M_c}
    \frac{
        \bra{M_0}|
        \hat f ^\dagger _{\alpha} 
        |\ket{M_c}
        \bra{M_c}|
        \hat f _{\alpha'} 
        |\ket{M_0}
    }{
        F(N_0-1,S_c,S_0) 
    },
    \\
    \bra{M_0,m_0-1}
    \hat f_{\alpha'\uparrow}
    \hat I 
    \hat f^\dagger_{\alpha\downarrow} 
    \ket{M_0,m_0}
    =
    -\frac{1}{2}
    S^-_{m_0-1,m_0}
    \sum_{M_c}
    \frac{
        \bra{M_0}|
        \hat f ^\dagger _{\alpha} 
        |\ket{M_c}
        \bra{M_c}|
        \hat f _{\alpha'} 
        |\ket{M_0}
    }{
        F(N_0+1,S_c,S_0) 
    },
\end{align}
where 
\begin{equation}
    F(N_c,S_c,S_0)=
    \begin{cases}
    (S_0+\frac{1}{2})
    , &\text{if } N_c=N_0+1,\;S_c=S_0-\frac{1}{2},
    \\
    S_0
    , &\text{if } N_c=N_0-1,\;S_c=S_0-\frac{1}{2},
    \\
    -(S_0+\frac{1}{2})
    , & \text{if } N_c=N_0+1,\;S_c=S_0+\frac{1}{2},
    \\
    -(S_0+1)
    , & \text{if } N_c=N_0-1,\;S_c=S_0+\frac{1}{2}
    \end{cases}
\label{eq:F}
\end{equation}
\end{widetext}
is the same factor appearing in Eq. \ref{eq:C}. Therefore, the matrix elements $T_{\alpha\alpha'}$ can be written compactly as in Eq. \ref{eq:Trewritten}.

\section{Electronic structure of the molecules}\label{appendix_multiplet}

Here we describe the many-body wave function of the ground and excited states obtained from CASCI calculations for neutral ($N$) and charged ($N\pm1$) states for all molecules presented in this manuscript. In the case of neutral molecules we employed CAS(12,12), but most of the many-body character in the Slater-determinants arises in the orbitals around the Fermi level. Therefore, to make the many-body representation of the wave-functions we use only six electrons.

\begin{figure}[H]
\centering
\includegraphics[width=1.0\linewidth]{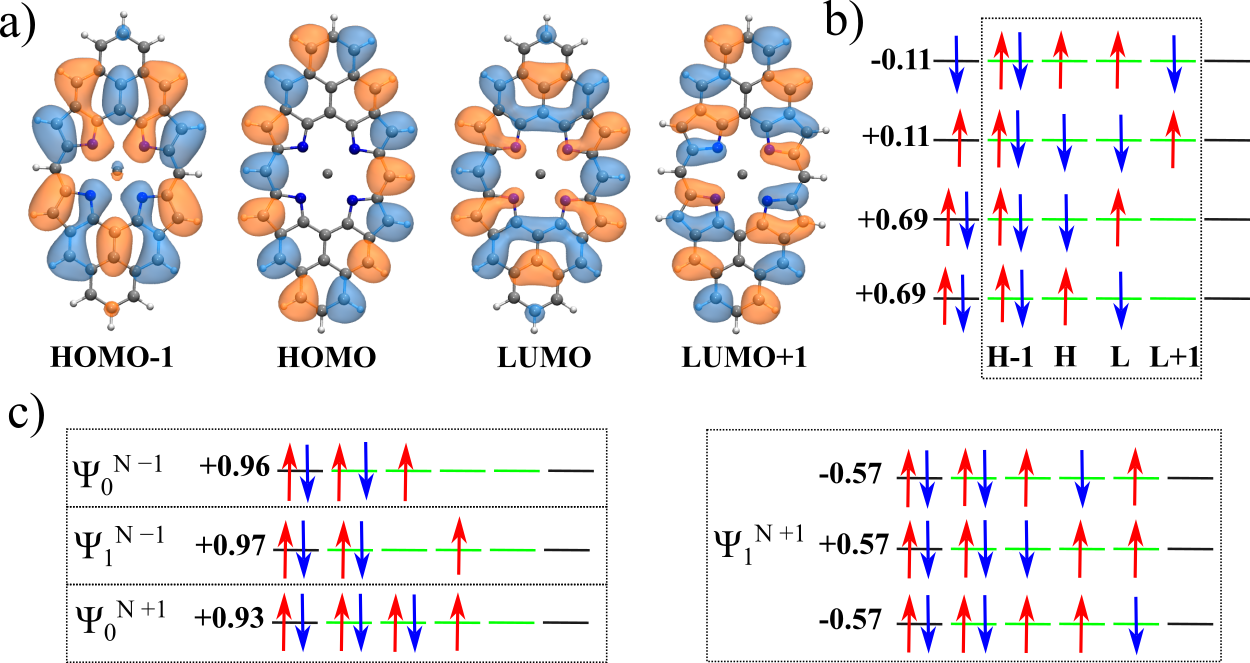}
\caption{
    a) The four main DFT orbitals employed in the active
    space calculation for the $\ZP$ molecule. b) Major
    Slater determinants for the ground states $\ZP$. The
    dashed box contains the four frontier orbitals: HOMO-1,
    HOMO, LUMO, and LUMO+1.  The values are the coefficients
    of those determinants in the orbital wave function. c)
    Major Slater determinants for N-1 and N+1 charge states
    for ground and first multiplet.
}
\label{fig:electronic_structure_ZP}
\end{figure}

\begin{table}[H]
\setlength{\tabcolsep}{10pt}
\renewcommand{\arraystretch}{1.2}
    \centering
\begin{tabular}{l| c| c| c  }
    \hline
   State& N & N-1 & N+1   \\
    \hline
 Ground &0.0000(1) &	7.9298(0.5) &	2.0303(0.5) \\
$1^{st}$ excited &0.0527(0)&	8.2659(0.5) &	2.3651(1.5) \\
$2^{nd}$ excited & 1.1738(1)	&9.0140(0.5)	&2.5542(0.5) \\
$3^{rd}$ excited & 1.6342(0)	&9.2818(1.5)	&2.7697(0.5) \\

\hline
\end{tabular}
\\
\caption{\label{tab:energies_ZP} Calculated energies \added{and spin (in
parentheses)} of the ground state and lowest excited state\added{s} for the
neutral (N) and charged (N$\pm$1) charge states obtained from CASCI calculations
of the gas-phase $\ZP$ molecule. The energies (in eV) \added{include the
chemical potential $\mu=3$~eV and} are aligned to the ground state of the
neutral molecule, which is set to zero.} 
\end{table}

\begin{figure}[H]
\centering
\includegraphics[width=1.0\linewidth]{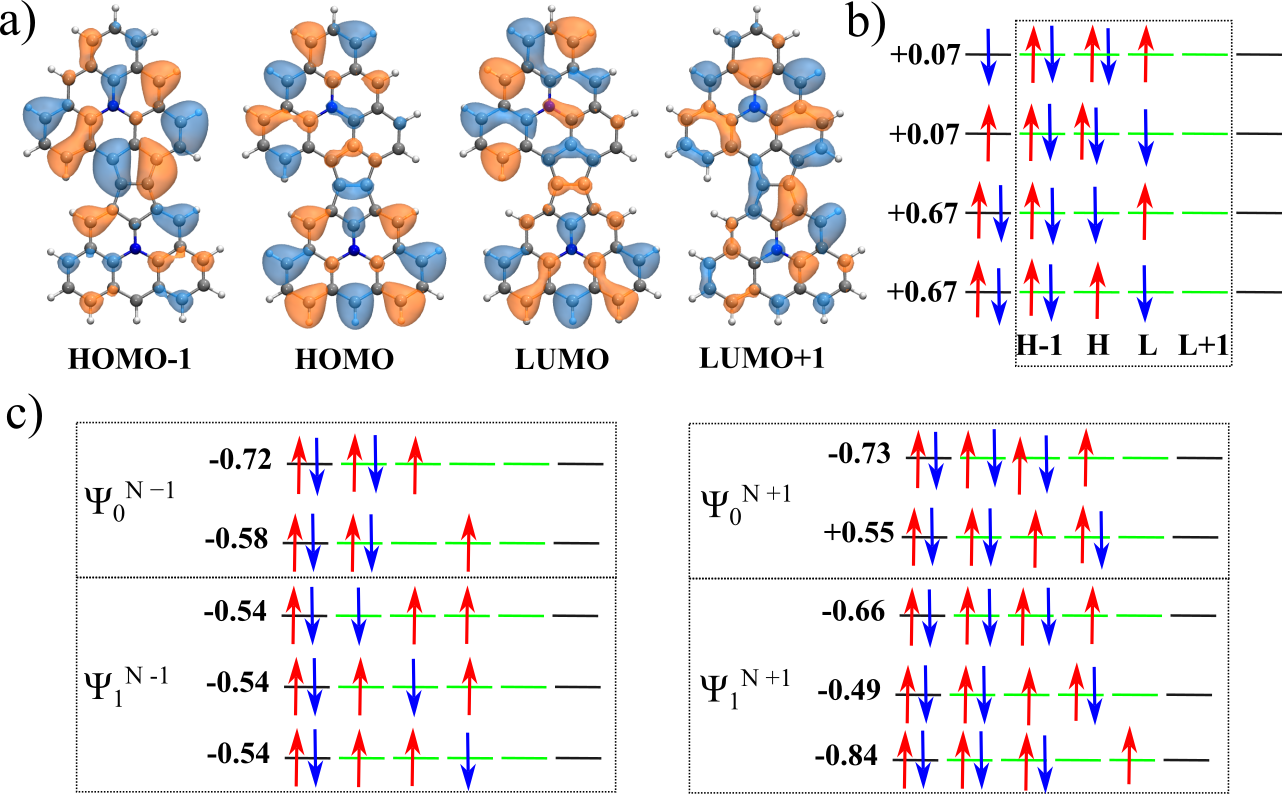}
\caption{a) The four main DFT orbitals employed in the active space calculation for the $\FAT$ molecule. b) Major Slater determinants for the ground states $\FAT$ molecule. The dashed box contains the four frontier orbitals: HOMO-1, HOMO, LUMO, and LUMO+1.  The values are the coefficients of those determinants in the orbital wave function. c) Major Slater determinants for N-1 and N+1 charge states for ground and first multiplet.}
\label{fig:electronic_structure_FAT}
\end{figure}

\begin{table}[H]
\setlength{\tabcolsep}{10pt}
\renewcommand{\arraystretch}{1.2}
    \centering
\begin{tabular}{l| c| c| c  }
    \hline
   State& N & N-1 & N+1   \\
    \hline

 Ground & 0.0000(1)	&12.0506(0.5)&	1.8190(0.5) \\
$1^{st}$ excited & 0.0009(0)&	12.5610(1.5)&	2.4848(0.5) \\
$2^{nd}$ excited & 1.6143(0)&	12.7091(0.5)&	2.8623(1.5) \\
$3^{rd}$ excited & 1.8229(1)&	12.9194(0.5)&	3.0735(0.5) \\

\hline
\end{tabular}
\\
\caption{\added{Same as Table \ref{tab:energies_ZP} for the $\FAT$ molecule
with the chemical potential $\mu=5$~eV.}
\deleted{Calculated energies of the ground state and lowest excited state for the
neutral (N) and charged (N$\pm$1) charge states obtained from CASCI calculations
of gas-phase $\FAT$ molecule. The energies (in eV) are aligned to the ground
state of the neutral molecule, which is set to zero.}} 
\label{tab:energies_FAT}
\end{table}

\begin{figure}[H]
\centering
\includegraphics[width=1.0\linewidth]{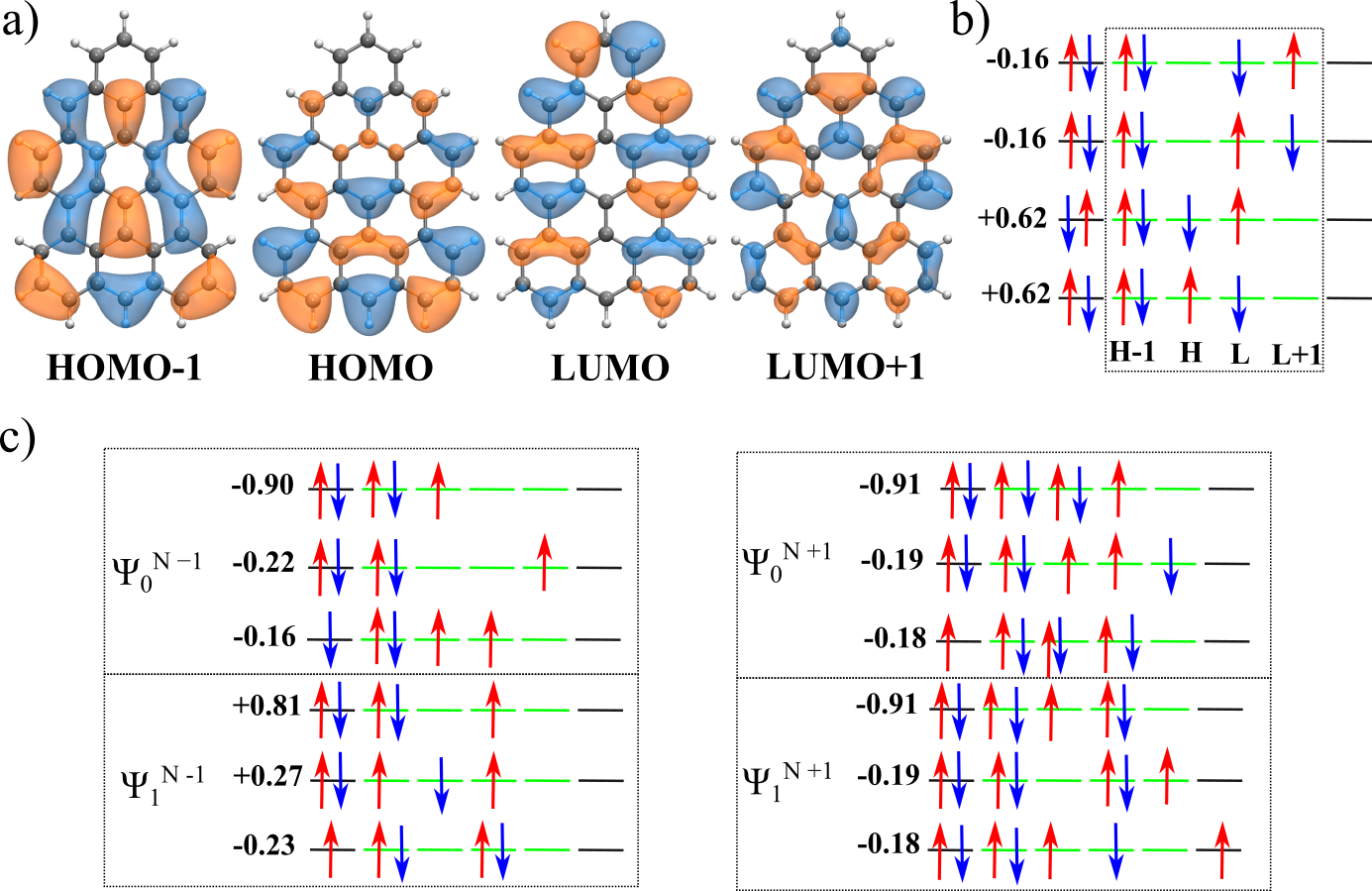}
\caption{a) The four main DFT orbitals employed in the active space calculation for the \textbf{ETRI} molecule. b) Major Slater determinants for the ground states \textbf{ETRI} molecule. The dashed box contains the four frontier orbitals: HOMO-1, HOMO, LUMO, and LUMO+1.  The values are the coefficients of those determinants in the orbital wave function. c) Major Slater determinants for N-1 and N+1 charge states for ground and first multiplet.}
\label{fig:electronic_structure_ROCKET}
\end{figure}

\begin{table}[H]
\setlength{\tabcolsep}{10pt}
\renewcommand{\arraystretch}{1.2}
    \centering
\begin{tabular}{l| c| c| c  }
    \hline
   State& N & N-1 & N+1   \\
    \hline

 Ground & 0.0000(1)	&8.7642(0.5)&	3.8533(0.5) \\
$1^{st}$ excited & 0.3618(0)&	8.9630(0.5)&	4.0815(0.5) \\
$2^{nd}$ excited & 1.2966(0)	&10.6777(1.5) &	5.6539(1.5) \\
$3^{rd}$ excited & 2.1004(0)	&11.1707(1.5)	&6.1963(0.5) \\

\hline
\end{tabular}
\\
\caption{\added{Same as Table \ref{tab:energies_ZP} for the {\bf ETRI} molecule
with the chemical potential $\mu=4$~eV}
\deleted{Calculated energies of the ground state and lowest excited state for the
neutral (N) and charged (N$\pm$1) charge states obtained from CASCI calculations
of gas-phase \textbf{ETRI} molecule. The energies (in eV) are aligned to the
ground state of the neutral molecule, which is set to zero.}} 
\label{tab:energies_ROCKET}
\end{table}

\begin{figure}[H]
\centering
\includegraphics[width=1.0\linewidth]{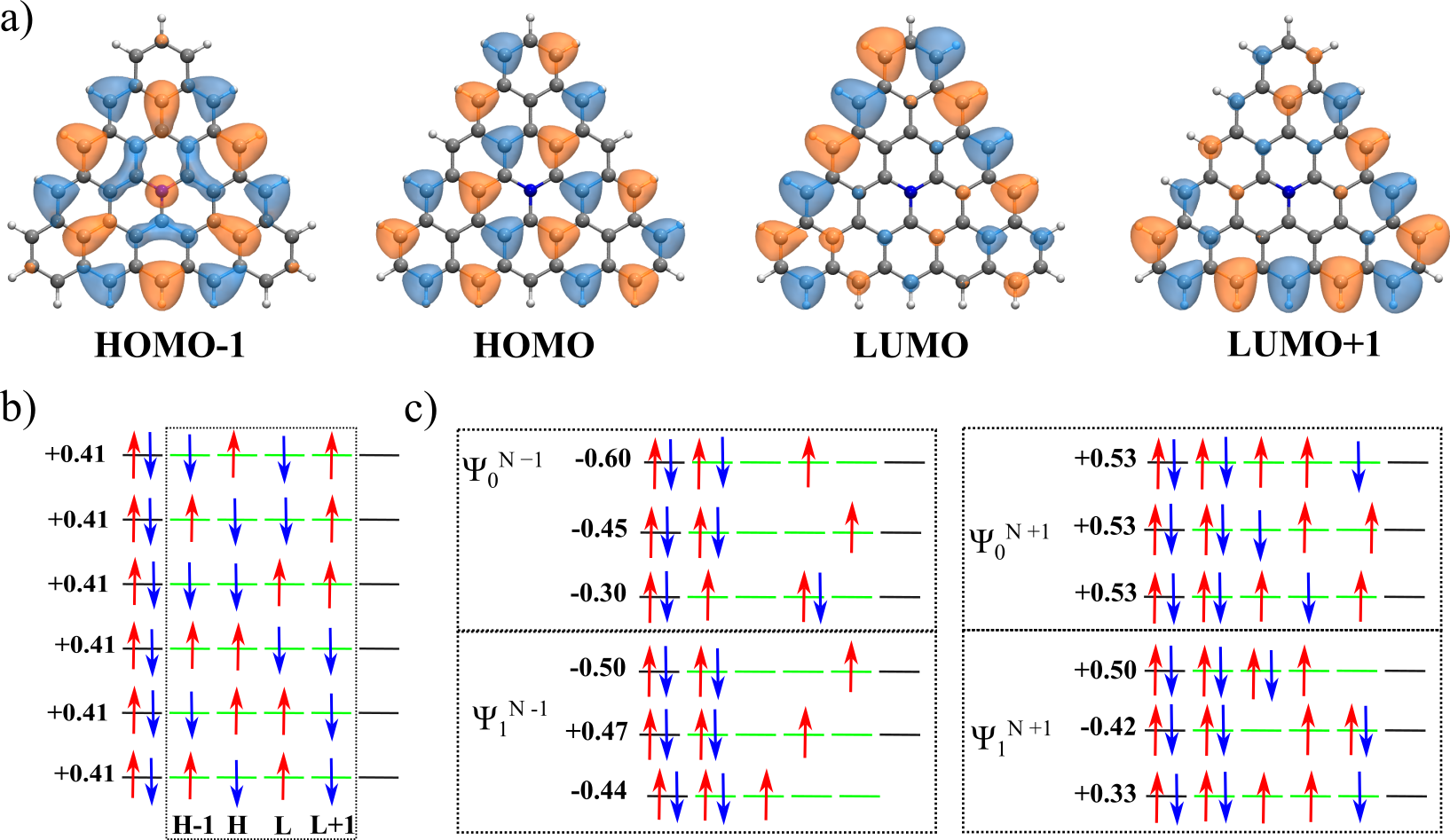}
\caption{a) The four main DFT orbitals employed in the active space calculation for the \textbf{A5T} molecule. b) Major Slater determinants for the ground states \textbf{A5T} molecule. The dashed box contains the four frontier orbitals: HOMO-1, HOMO, LUMO, and LUMO+1.  The values are the coefficients of those determinants in the orbital wave function. c) Major Slater determinants for N-1 and N+1 charge states for ground and first multiplet.}
\label{fig:electronic_structure_A5T}
\end{figure}

\begin{table}[H]
\setlength{\tabcolsep}{10pt}
\renewcommand{\arraystretch}{1.2}
    \centering
\begin{tabular}{l| c| c| c  }
    \hline
   State& N & N-1 & N+1   \\
    \hline

 Ground & 0.0000(2) &	12.0759(0.5)&	0.4809(1.5) \\
$1^{st}$ excited & 0.5491(1)&	12.1114(0.5)&	0.9573(0.5) \\ 
$2^{nd}$ excited & 0.5701(1)&	12.2326(0.5) &	0.9675(0.5) \\
$3^{rd}$ excited & 0.8931(0)&	12.2585(1.5) &	1.4241(1.5) \\
\hline
\end{tabular}
\\
\caption{
\added{Same as Table \ref{tab:energies_ZP} for the {\bf A5T} molecule with the chemical potential $\mu=4.3$~eV.}
\deleted{Calculated energies of the ground state and lowest
excited state for the neutral (N) and charged (N$\pm$1) charge states obtained
from CASCI calculations of gas-phase \textbf{A5T} molecule. The energies (in eV)
are aligned to the ground state of the neutral molecule, which is set to zero.}} 
\label{tab:energies_A5T}
\end{table}

\section{Reduced Lehmann amplitudes}\label{appendix_lehmann}

The reduced matrix elements $\bra{M_c}|\hat{f}^{(\dagger)}_\alpha|\ket{M_0}$ appearing in the Schrieffer-Wolf transformation are shown below for the four studied molecules.

\begin{table}[H]
\begin{tabular}{ r | m{1.5cm} m{1.5cm} m{1.5cm} m{1.5cm} }
    \hline
    $M_c$
    &$M_{0-}$
    &$M_{1-}$
    &$M_{0+}$
    &$M_{1+}$
    \\
    \hline
1 & 0.00 &	0.00	&	0.00&	-0.09 \\
2 & 0.00	&0.00	&	0.00&	0.00 \\
 $\alpha\qquad$ 3 & 0.00	&-0.96	&	-1.23&	0.00 \\
4 & -0.97	&0.00	&	0.00&	0.00 \\
5 & 0.00	&0.00	&	0.00&	0.96 \\
6 & 0.00	&0.04	&	-0.20&	0.00 \\
\hline
\end{tabular}
\\
\caption{Reduced matrix elements $\bra{M_c}|\hat{f}^{(\dagger)}_\alpha|\ket{M_0}$
    for positively (+) and negatively (-) charged excited multiplets $M_c$ for the $\ZP$ molecule.} 
\label{tab:contributions}
\end{table}

\begin{table}[H]
\begin{tabular}{ r | m{1.5cm} m{1.5cm} m{1.5cm} m{1.5cm} }
    \hline
    $M_c$
    &$M_{0-}$
    &$M_{1-}$
    &$M_{0+}$
    &$M_{1+}$
    \\
    \hline
1&-0.03	&	0.06	&	-0.03	&	0.06 \\
2&-0.02	&	0.77	&	0.01	&	0.00 \\
 $\alpha\qquad$3&0.42	&	0.00	&	0.53	&	-0.35 \\
4&-0.50	&	-0.01	&	0.39	&	0.47 \\
5&0.03	&	0.04	&	-0.04	&	0.11 \\
6&-0.01	&	0.01	&	0.07	&	-0.07 \\
\hline
\end{tabular}
\\
\caption{
    Same as Table~\ref{tab:contributions}
    for the $\FAT$ molecule.
} 
\end{table}

\begin{table}[H]
\begin{tabular}{ r | m{1.5cm} m{1.5cm} m{1.5cm} m{1.5cm} }
    \hline
    $M_c$
    &$M_{0-}$
    &$M_{1-}$
    &$M_{0+}$
    &$M_{1+}$
    \\
    \hline
1&0.16	&	0.00	&	0.00	&	-0.21 \\
2&0.00	&	0.19	&	-0.11	&	0.00 \\
 $\alpha\qquad$ 3&0.00	&	-0.89	&	1.14	&	0.00 \\
4&-0.93	&	0.00	&	0.00	&	1.11 \\
5&0.00	&	-0.21	&	0.22	&	0.00 \\
6&0.00	&	-0.03	&	0.01	&	0.00 \\

\hline
\end{tabular}
\\
\caption{Same as Table~\ref{tab:contributions} for the \textbf{ETRI} molecule.} 
\end{table}

\begin{table}[H]
\begin{tabular}{ r | m{1.5cm} m{1.5cm} m{1.5cm} m{1.5cm} }
    \hline
    $M_c$
    &$M_{0-}$
    &$M_{1-}$
    &$M_{0+}$
    &$M_{1+}$
    \\
    \hline
1&-0.02	&	-0.04	&	0.00	&	0.00 \\
2&0.00	&	-0.04	&	-1.03	&	0.00 \\
 $\alpha\qquad$3&0.37	&	-0.03	&	0.00	&	1.11 \\
4&-0.03	&	-0.96	&	0.00	&	0.00 \\
5&-0.89	&	0.02	&	0.00	&	-0.02 \\
6&0.00	&	0.00	&	0.36	&	0.00 \\
\hline
\end{tabular}
\\
\caption{Same as Table~\ref{tab:contributions} for the \textbf{A5T} molecule.} 
\end{table}

\begin{acknowledgments}
A.C.-F., A.E. and M.B.-R. acknowledge grants No. IT-1527-22, funded by the Department of Education, Universities and Research of the Basque Government, and PID2022-137685NB-I00, funded by MCIN/AEI 10.13039/501100011033/ and by “ERDF A way of making Europe”. 
A.C.-F. acknowledges grant No. PRE2020-092046 funded by the Spanish MCIN. D.S., M.K. and P.J. acknowledge financial support from the CzechNanoLab Research Infrastructure supported by MEYS CR (LM2023051) and the GACR project no. 23-05486S.  We acknowledge fruitful discussion with L. Veis and F. Flores. 
\end{acknowledgments}

\bibliography{bibliography3.bib}

\begin{thebibliography}{62}%
\makeatletter
\providecommand \@ifxundefined [1]{%
 \@ifx{#1\undefined}
}%
\providecommand \@ifnum [1]{%
 \ifnum #1\expandafter \@firstoftwo
 \else \expandafter \@secondoftwo
 \fi
}%
\providecommand \@ifx [1]{%
 \ifx #1\expandafter \@firstoftwo
 \else \expandafter \@secondoftwo
 \fi
}%
\providecommand \natexlab [1]{#1}%
\providecommand \enquote  [1]{``#1''}%
\providecommand \bibnamefont  [1]{#1}%
\providecommand \bibfnamefont [1]{#1}%
\providecommand \citenamefont [1]{#1}%
\providecommand \href@noop [0]{\@secondoftwo}%
\providecommand \href [0]{\begingroup \@sanitize@url \@href}%
\providecommand \@href[1]{\@@startlink{#1}\@@href}%
\providecommand \@@href[1]{\endgroup#1\@@endlink}%
\providecommand \@sanitize@url [0]{\catcode `\\12\catcode `\$12\catcode
  `\&12\catcode `\#12\catcode `\^12\catcode `\_12\catcode `\%12\relax}%
\providecommand \@@startlink[1]{}%
\providecommand \@@endlink[0]{}%
\providecommand \url  [0]{\begingroup\@sanitize@url \@url }%
\providecommand \@url [1]{\endgroup\@href {#1}{\urlprefix }}%
\providecommand \urlprefix  [0]{URL }%
\providecommand \Eprint [0]{\href }%
\providecommand \doibase [0]{https://doi.org/}%
\providecommand \selectlanguage [0]{\@gobble}%
\providecommand \bibinfo  [0]{\@secondoftwo}%
\providecommand \bibfield  [0]{\@secondoftwo}%
\providecommand \translation [1]{[#1]}%
\providecommand \BibitemOpen [0]{}%
\providecommand \bibitemStop [0]{}%
\providecommand \bibitemNoStop [0]{.\EOS\space}%
\providecommand \EOS [0]{\spacefactor3000\relax}%
\providecommand \BibitemShut  [1]{\csname bibitem#1\endcsname}%
\let\auto@bib@innerbib\@empty
\bibitem [{\citenamefont {Hewson}(1993)}]{bib:hewson}%
  \BibitemOpen
  \bibfield  {author} {\bibinfo {author} {\bibfnamefont {A.~C.}\ \bibnamefont
  {Hewson}},\ }\href {https://doi.org/10.1017/CBO9780511470752} {\emph
  {\bibinfo {title} {The Kondo Problem to Heavy Fermions}}},\ Cambridge Studies
  in Magnetism\ (\bibinfo  {publisher} {Cambridge University Press},\ \bibinfo
  {year} {1993})\BibitemShut {NoStop}%
\bibitem [{\citenamefont {Kondo}(1964)}]{kondo1964resistance}%
  \BibitemOpen
  \bibfield  {author} {\bibinfo {author} {\bibfnamefont {J.}~\bibnamefont
  {Kondo}},\ }\bibfield  {title} {\bibinfo {title} {Resistance minimum in
  dilute magnetic alloys},\ }\href@noop {} {\bibfield  {journal} {\bibinfo
  {journal} {Progress of theoretical physics}\ }\textbf {\bibinfo {volume}
  {32}},\ \bibinfo {pages} {37} (\bibinfo {year} {1964})}\BibitemShut {NoStop}%
\bibitem [{\citenamefont {de~Haas}\ \emph {et~al.}(1934)\citenamefont
  {de~Haas}, \citenamefont {de~Boer},\ and\ \citenamefont {van~dën
  Berg}}]{deHaas1934}%
  \BibitemOpen
  \bibfield  {author} {\bibinfo {author} {\bibfnamefont {W.}~\bibnamefont
  {de~Haas}}, \bibinfo {author} {\bibfnamefont {J.}~\bibnamefont {de~Boer}},\
  and\ \bibinfo {author} {\bibfnamefont {G.}~\bibnamefont {van~dën Berg}},\
  }\bibfield  {title} {\bibinfo {title} {The electrical resistance of gold,
  copper and lead at low temperatures},\ }\href
  {https://doi.org/10.1016/s0031-8914(34)80310-2} {\bibfield  {journal}
  {\bibinfo  {journal} {Physica}\ }\textbf {\bibinfo {volume} {1}},\ \bibinfo
  {pages} {1115–1124} (\bibinfo {year} {1934})}\BibitemShut {NoStop}%
\bibitem [{NP2(2014)}]{NP2014-Kondo}%
  \BibitemOpen
  \href {https://doi.org/10.1038/nphys2972} {\bibfield  {journal} {\bibinfo
  {journal} {Nature Physics}\ }\textbf {\bibinfo {volume} {10}},\ \bibinfo
  {pages} {329–329} (\bibinfo {year} {2014})}\BibitemShut {NoStop}%
\bibitem [{\citenamefont {Madhavan}\ \emph {et~al.}(1998)\citenamefont
  {Madhavan}, \citenamefont {Chen}, \citenamefont {Jamneala}, \citenamefont
  {Crommie},\ and\ \citenamefont {Wingreen}}]{bib:madhavan98}%
  \BibitemOpen
  \bibfield  {author} {\bibinfo {author} {\bibfnamefont {V.}~\bibnamefont
  {Madhavan}}, \bibinfo {author} {\bibfnamefont {W.}~\bibnamefont {Chen}},
  \bibinfo {author} {\bibfnamefont {T.}~\bibnamefont {Jamneala}}, \bibinfo
  {author} {\bibfnamefont {M.~F.}\ \bibnamefont {Crommie}},\ and\ \bibinfo
  {author} {\bibfnamefont {N.~S.}\ \bibnamefont {Wingreen}},\ }\bibfield
  {title} {\bibinfo {title} {Tunneling into a single magnetic atom:
  Spectroscopic evidence of the {Kondo} resonance},\ }\href
  {https://doi.org/10.1126/science.280.5363.567} {\bibfield  {journal}
  {\bibinfo  {journal} {Science}\ }\textbf {\bibinfo {volume} {280}},\ \bibinfo
  {pages} {567} (\bibinfo {year} {1998})}\BibitemShut {NoStop}%
\bibitem [{\citenamefont {Li}\ \emph {et~al.}(1998)\citenamefont {Li},
  \citenamefont {Schneider}, \citenamefont {Berndt},\ and\ \citenamefont
  {Delley}}]{bib:li98}%
  \BibitemOpen
  \bibfield  {author} {\bibinfo {author} {\bibfnamefont {J.}~\bibnamefont
  {Li}}, \bibinfo {author} {\bibfnamefont {W.-D.}\ \bibnamefont {Schneider}},
  \bibinfo {author} {\bibfnamefont {R.}~\bibnamefont {Berndt}},\ and\ \bibinfo
  {author} {\bibfnamefont {B.}~\bibnamefont {Delley}},\ }\bibfield  {title}
  {\bibinfo {title} {Kondo scattering observed at a single magnetic impurity},\
  }\href {https://doi.org/10.1103/PhysRevLett.80.2893} {\bibfield  {journal}
  {\bibinfo  {journal} {Phys. Rev. Lett.}\ }\textbf {\bibinfo {volume} {80}},\
  \bibinfo {pages} {2893} (\bibinfo {year} {1998})}\BibitemShut {NoStop}%
\bibitem [{\citenamefont {Knorr}\ \emph {et~al.}(2002)\citenamefont {Knorr},
  \citenamefont {Schneider}, \citenamefont {Diekh\"oner}, \citenamefont
  {Wahl},\ and\ \citenamefont {Kern}}]{bib:knorr02}%
  \BibitemOpen
  \bibfield  {author} {\bibinfo {author} {\bibfnamefont {N.}~\bibnamefont
  {Knorr}}, \bibinfo {author} {\bibfnamefont {M.~A.}\ \bibnamefont
  {Schneider}}, \bibinfo {author} {\bibfnamefont {L.}~\bibnamefont
  {Diekh\"oner}}, \bibinfo {author} {\bibfnamefont {P.}~\bibnamefont {Wahl}},\
  and\ \bibinfo {author} {\bibfnamefont {K.}~\bibnamefont {Kern}},\ }\bibfield
  {title} {\bibinfo {title} {Kondo effect of single {Co} adatoms on {Cu}
  surfaces},\ }\href {https://doi.org/10.1103/PhysRevLett.88.096804} {\bibfield
   {journal} {\bibinfo  {journal} {Phys. Rev. Lett.}\ }\textbf {\bibinfo
  {volume} {88}},\ \bibinfo {pages} {096804} (\bibinfo {year}
  {2002})}\BibitemShut {NoStop}%
\bibitem [{\citenamefont {Otte}\ \emph {et~al.}(2008)\citenamefont {Otte},
  \citenamefont {Ternes}, \citenamefont {von Bergmann}, \citenamefont {Loth},
  \citenamefont {Brune}, \citenamefont {Lutz}, \citenamefont {Hirjibehedin},\
  and\ \citenamefont {Heinrich}}]{bib:otte08}%
  \BibitemOpen
  \bibfield  {author} {\bibinfo {author} {\bibfnamefont {A.~F.}\ \bibnamefont
  {Otte}}, \bibinfo {author} {\bibfnamefont {M.}~\bibnamefont {Ternes}},
  \bibinfo {author} {\bibfnamefont {K.}~\bibnamefont {von Bergmann}}, \bibinfo
  {author} {\bibfnamefont {S.}~\bibnamefont {Loth}}, \bibinfo {author}
  {\bibfnamefont {H.}~\bibnamefont {Brune}}, \bibinfo {author} {\bibfnamefont
  {C.~P.}\ \bibnamefont {Lutz}}, \bibinfo {author} {\bibfnamefont {C.~F.}\
  \bibnamefont {Hirjibehedin}},\ and\ \bibinfo {author} {\bibfnamefont {A.~J.}\
  \bibnamefont {Heinrich}},\ }\bibfield  {title} {\bibinfo {title} {The role of
  magnetic anisotropy in the {Kondo} effect},\ }\href
  {https://doi.org/10.1038/nphys1072} {\bibfield  {journal} {\bibinfo
  {journal} {Nature Physics}\ }\textbf {\bibinfo {volume} {4}},\ \bibinfo
  {pages} {847} (\bibinfo {year} {2008})}\BibitemShut {NoStop}%
\bibitem [{\citenamefont {Esat}\ \emph {et~al.}(2015)\citenamefont {Esat},
  \citenamefont {Deilmann}, \citenamefont {Lechtenberg}, \citenamefont
  {Wagner}, \citenamefont {Kr\"uger}, \citenamefont {Temirov}, \citenamefont
  {Anders}, \citenamefont {Rohlfing},\ and\ \citenamefont {Tautz}}]{Esat2015}%
  \BibitemOpen
  \bibfield  {author} {\bibinfo {author} {\bibfnamefont {T.}~\bibnamefont
  {Esat}}, \bibinfo {author} {\bibfnamefont {T.}~\bibnamefont {Deilmann}},
  \bibinfo {author} {\bibfnamefont {B.}~\bibnamefont {Lechtenberg}}, \bibinfo
  {author} {\bibfnamefont {C.}~\bibnamefont {Wagner}}, \bibinfo {author}
  {\bibfnamefont {P.}~\bibnamefont {Kr\"uger}}, \bibinfo {author}
  {\bibfnamefont {R.}~\bibnamefont {Temirov}}, \bibinfo {author} {\bibfnamefont
  {F.~B.}\ \bibnamefont {Anders}}, \bibinfo {author} {\bibfnamefont
  {M.}~\bibnamefont {Rohlfing}},\ and\ \bibinfo {author} {\bibfnamefont
  {F.~S.}\ \bibnamefont {Tautz}},\ }\bibfield  {title} {\bibinfo {title}
  {Transfering spin into an extended $\ensuremath{\pi}$ orbital of a large
  molecule},\ }\href {https://doi.org/10.1103/PhysRevB.91.144415} {\bibfield
  {journal} {\bibinfo  {journal} {Phys. Rev. B}\ }\textbf {\bibinfo {volume}
  {91}},\ \bibinfo {pages} {144415} (\bibinfo {year} {2015})}\BibitemShut
  {NoStop}%
\bibitem [{\citenamefont {Scott}\ and\ \citenamefont
  {Natelson}(2010)}]{bib:scott10}%
  \BibitemOpen
  \bibfield  {author} {\bibinfo {author} {\bibfnamefont {G.~D.}\ \bibnamefont
  {Scott}}\ and\ \bibinfo {author} {\bibfnamefont {D.}~\bibnamefont
  {Natelson}},\ }\bibfield  {title} {\bibinfo {title} {Kondo resonances in
  molecular devices},\ }\href {https://doi.org/10.1021/nn100793s} {\bibfield
  {journal} {\bibinfo  {journal} {ACS Nano}\ }\textbf {\bibinfo {volume} {4}},\
  \bibinfo {pages} {3560} (\bibinfo {year} {2010})}\BibitemShut {NoStop}%
\bibitem [{\citenamefont {Komeda}\ \emph {et~al.}(2011)\citenamefont {Komeda},
  \citenamefont {Isshiki}, \citenamefont {Liu}, \citenamefont {Zhang},
  \citenamefont {Lorente}, \citenamefont {Katoh}, \citenamefont {Breedlove},\
  and\ \citenamefont {Yamashita}}]{bib:komeda11}%
  \BibitemOpen
  \bibfield  {author} {\bibinfo {author} {\bibfnamefont {T.}~\bibnamefont
  {Komeda}}, \bibinfo {author} {\bibfnamefont {H.}~\bibnamefont {Isshiki}},
  \bibinfo {author} {\bibfnamefont {J.}~\bibnamefont {Liu}}, \bibinfo {author}
  {\bibfnamefont {Y.-F.}\ \bibnamefont {Zhang}}, \bibinfo {author}
  {\bibfnamefont {N.}~\bibnamefont {Lorente}}, \bibinfo {author} {\bibfnamefont
  {K.}~\bibnamefont {Katoh}}, \bibinfo {author} {\bibfnamefont {B.~K.}\
  \bibnamefont {Breedlove}},\ and\ \bibinfo {author} {\bibfnamefont
  {M.}~\bibnamefont {Yamashita}},\ }\bibfield  {title} {\bibinfo {title}
  {Observation and electric current control of a local spin in a
  single-molecule magnet},\ }\href {https://doi.org/10.1038/ncomms1210}
  {\bibfield  {journal} {\bibinfo  {journal} {Nature Communications}\ }\textbf
  {\bibinfo {volume} {2}},\ \bibinfo {pages} {217} (\bibinfo {year}
  {2011})}\BibitemShut {NoStop}%
\bibitem [{\citenamefont {Minamitani}\ \emph {et~al.}(2012)\citenamefont
  {Minamitani}, \citenamefont {Tsukahara}, \citenamefont {Matsunaka},
  \citenamefont {Kim}, \citenamefont {Takagi},\ and\ \citenamefont
  {Kawai}}]{bib:minamitani12}%
  \BibitemOpen
  \bibfield  {author} {\bibinfo {author} {\bibfnamefont {E.}~\bibnamefont
  {Minamitani}}, \bibinfo {author} {\bibfnamefont {N.}~\bibnamefont
  {Tsukahara}}, \bibinfo {author} {\bibfnamefont {D.}~\bibnamefont
  {Matsunaka}}, \bibinfo {author} {\bibfnamefont {Y.}~\bibnamefont {Kim}},
  \bibinfo {author} {\bibfnamefont {N.}~\bibnamefont {Takagi}},\ and\ \bibinfo
  {author} {\bibfnamefont {M.}~\bibnamefont {Kawai}},\ }\bibfield  {title}
  {\bibinfo {title} {Symmetry-driven novel {Kondo} effect in a molecule},\
  }\href {https://doi.org/10.1103/PhysRevLett.109.086602} {\bibfield  {journal}
  {\bibinfo  {journal} {Phys. Rev. Lett.}\ }\textbf {\bibinfo {volume} {109}},\
  \bibinfo {pages} {086602} (\bibinfo {year} {2012})}\BibitemShut {NoStop}%
\bibitem [{\citenamefont {Hiraoka}\ \emph {et~al.}(2017)\citenamefont
  {Hiraoka}, \citenamefont {Minamitani}, \citenamefont {Arafune}, \citenamefont
  {Tsukahara}, \citenamefont {Watanabe}, \citenamefont {Kawai},\ and\
  \citenamefont {Takagi}}]{bib:hiraoka17}%
  \BibitemOpen
  \bibfield  {author} {\bibinfo {author} {\bibfnamefont {R.}~\bibnamefont
  {Hiraoka}}, \bibinfo {author} {\bibfnamefont {E.}~\bibnamefont {Minamitani}},
  \bibinfo {author} {\bibfnamefont {R.}~\bibnamefont {Arafune}}, \bibinfo
  {author} {\bibfnamefont {N.}~\bibnamefont {Tsukahara}}, \bibinfo {author}
  {\bibfnamefont {S.}~\bibnamefont {Watanabe}}, \bibinfo {author}
  {\bibfnamefont {M.}~\bibnamefont {Kawai}},\ and\ \bibinfo {author}
  {\bibfnamefont {N.}~\bibnamefont {Takagi}},\ }\bibfield  {title} {\bibinfo
  {title} {Single-molecule quantum dot as a {Kondo} simulator},\ }\href
  {https://doi.org/10.1038/ncomms16012} {\bibfield  {journal} {\bibinfo
  {journal} {Nature Communications}\ }\textbf {\bibinfo {volume} {8}},\
  \bibinfo {pages} {16012} (\bibinfo {year} {2017})}\BibitemShut {NoStop}%
\bibitem [{\citenamefont {{\v{Z}}itko}\ \emph {et~al.}(2021)\citenamefont
  {{\v{Z}}itko}, \citenamefont {Blesio}, \citenamefont {Manuel},\ and\
  \citenamefont {Aligia}}]{bib:zitko21}%
  \BibitemOpen
  \bibfield  {author} {\bibinfo {author} {\bibfnamefont {R.}~\bibnamefont
  {{\v{Z}}itko}}, \bibinfo {author} {\bibfnamefont {G.~G.}\ \bibnamefont
  {Blesio}}, \bibinfo {author} {\bibfnamefont {L.~O.}\ \bibnamefont {Manuel}},\
  and\ \bibinfo {author} {\bibfnamefont {A.~A.}\ \bibnamefont {Aligia}},\
  }\bibfield  {title} {\bibinfo {title} {Iron phthalocyanine on au(111) is a
  {``non-Landau''} {Fermi} liquid},\ }\href
  {https://doi.org/10.1038/s41467-021-26339-z} {\bibfield  {journal} {\bibinfo
  {journal} {Nature Communications}\ }\textbf {\bibinfo {volume} {12}},\
  \bibinfo {pages} {6027} (\bibinfo {year} {2021})}\BibitemShut {NoStop}%
\bibitem [{\citenamefont {Fern\'andez-Torrente}\ \emph
  {et~al.}(2008)\citenamefont {Fern\'andez-Torrente}, \citenamefont {Franke},\
  and\ \citenamefont {Pascual}}]{bib:fernandez08}%
  \BibitemOpen
  \bibfield  {author} {\bibinfo {author} {\bibfnamefont {I.}~\bibnamefont
  {Fern\'andez-Torrente}}, \bibinfo {author} {\bibfnamefont {K.~J.}\
  \bibnamefont {Franke}},\ and\ \bibinfo {author} {\bibfnamefont {J.~I.}\
  \bibnamefont {Pascual}},\ }\bibfield  {title} {\bibinfo {title} {Vibrational
  {Kondo} effect in pure organic charge-transfer assemblies},\ }\href
  {https://doi.org/10.1103/PhysRevLett.101.217203} {\bibfield  {journal}
  {\bibinfo  {journal} {Phys. Rev. Lett.}\ }\textbf {\bibinfo {volume} {101}},\
  \bibinfo {pages} {217203} (\bibinfo {year} {2008})}\BibitemShut {NoStop}%
\bibitem [{\citenamefont {Choi}\ \emph {et~al.}(2010)\citenamefont {Choi},
  \citenamefont {Bedwani}, \citenamefont {Rochefort}, \citenamefont {Chen},
  \citenamefont {Epstein},\ and\ \citenamefont {Gupta}}]{bib:choi10}%
  \BibitemOpen
  \bibfield  {author} {\bibinfo {author} {\bibfnamefont {T.}~\bibnamefont
  {Choi}}, \bibinfo {author} {\bibfnamefont {S.}~\bibnamefont {Bedwani}},
  \bibinfo {author} {\bibfnamefont {A.}~\bibnamefont {Rochefort}}, \bibinfo
  {author} {\bibfnamefont {C.-Y.}\ \bibnamefont {Chen}}, \bibinfo {author}
  {\bibfnamefont {A.~J.}\ \bibnamefont {Epstein}},\ and\ \bibinfo {author}
  {\bibfnamefont {J.~A.}\ \bibnamefont {Gupta}},\ }\bibfield  {title} {\bibinfo
  {title} {A single molecule {Kondo} switch: Multistability of
  tetracyanoethylene on cu(111)},\ }\href {https://doi.org/10.1021/nl1024563}
  {\bibfield  {journal} {\bibinfo  {journal} {Nano Letters}\ }\textbf {\bibinfo
  {volume} {10}},\ \bibinfo {pages} {4175} (\bibinfo {year}
  {2010})}\BibitemShut {NoStop}%
\bibitem [{\citenamefont {Perera}\ \emph {et~al.}(2010)\citenamefont {Perera},
  \citenamefont {Kulik}, \citenamefont {Iancu}, \citenamefont {Dias~da Silva},
  \citenamefont {Ulloa}, \citenamefont {Marzari},\ and\ \citenamefont
  {Hla}}]{bib:perera10}%
  \BibitemOpen
  \bibfield  {author} {\bibinfo {author} {\bibfnamefont {U.~G.~E.}\
  \bibnamefont {Perera}}, \bibinfo {author} {\bibfnamefont {H.~J.}\
  \bibnamefont {Kulik}}, \bibinfo {author} {\bibfnamefont {V.}~\bibnamefont
  {Iancu}}, \bibinfo {author} {\bibfnamefont {L.~G. G.~V.}\ \bibnamefont
  {Dias~da Silva}}, \bibinfo {author} {\bibfnamefont {S.~E.}\ \bibnamefont
  {Ulloa}}, \bibinfo {author} {\bibfnamefont {N.}~\bibnamefont {Marzari}},\
  and\ \bibinfo {author} {\bibfnamefont {S.-W.}\ \bibnamefont {Hla}},\
  }\bibfield  {title} {\bibinfo {title} {Spatially extended {Kondo} state in
  magnetic molecules induced by interfacial charge transfer},\ }\href
  {https://doi.org/10.1103/PhysRevLett.105.106601} {\bibfield  {journal}
  {\bibinfo  {journal} {Phys. Rev. Lett.}\ }\textbf {\bibinfo {volume} {105}},\
  \bibinfo {pages} {106601} (\bibinfo {year} {2010})}\BibitemShut {NoStop}%
\bibitem [{\citenamefont {Requist}\ \emph {et~al.}(2014)\citenamefont
  {Requist}, \citenamefont {Modesti}, \citenamefont {Baruselli}, \citenamefont
  {Smogunov}, \citenamefont {Fabrizio},\ and\ \citenamefont
  {Tosatti}}]{bib:requist14}%
  \BibitemOpen
  \bibfield  {author} {\bibinfo {author} {\bibfnamefont {R.}~\bibnamefont
  {Requist}}, \bibinfo {author} {\bibfnamefont {S.}~\bibnamefont {Modesti}},
  \bibinfo {author} {\bibfnamefont {P.~P.}\ \bibnamefont {Baruselli}}, \bibinfo
  {author} {\bibfnamefont {A.}~\bibnamefont {Smogunov}}, \bibinfo {author}
  {\bibfnamefont {M.}~\bibnamefont {Fabrizio}},\ and\ \bibinfo {author}
  {\bibfnamefont {E.}~\bibnamefont {Tosatti}},\ }\bibfield  {title} {\bibinfo
  {title} {Kondo conductance across the smallest spin 1/2 radical molecule},\
  }\href {https://doi.org/10.1073/pnas.1322239111} {\bibfield  {journal}
  {\bibinfo  {journal} {Proceedings of the National Academy of Sciences}\
  }\textbf {\bibinfo {volume} {111}},\ \bibinfo {pages} {69} (\bibinfo {year}
  {2014})}\BibitemShut {NoStop}%
\bibitem [{\citenamefont {Minamitani}\ \emph {et~al.}(2015)\citenamefont
  {Minamitani}, \citenamefont {Fu}, \citenamefont {Xue}, \citenamefont {Kim},\
  and\ \citenamefont {Watanabe}}]{bib:minamitani15}%
  \BibitemOpen
  \bibfield  {author} {\bibinfo {author} {\bibfnamefont {E.}~\bibnamefont
  {Minamitani}}, \bibinfo {author} {\bibfnamefont {Y.-S.}\ \bibnamefont {Fu}},
  \bibinfo {author} {\bibfnamefont {Q.-K.}\ \bibnamefont {Xue}}, \bibinfo
  {author} {\bibfnamefont {Y.}~\bibnamefont {Kim}},\ and\ \bibinfo {author}
  {\bibfnamefont {S.}~\bibnamefont {Watanabe}},\ }\bibfield  {title} {\bibinfo
  {title} {Spatially extended underscreened {Kondo} state from collective
  molecular spin},\ }\href {https://doi.org/10.1103/PhysRevB.92.075144}
  {\bibfield  {journal} {\bibinfo  {journal} {Phys. Rev. B}\ }\textbf {\bibinfo
  {volume} {92}},\ \bibinfo {pages} {075144} (\bibinfo {year}
  {2015})}\BibitemShut {NoStop}%
\bibitem [{\citenamefont {Patera}\ \emph {et~al.}(2019)\citenamefont {Patera},
  \citenamefont {Sokolov}, \citenamefont {Low}, \citenamefont {Campos},
  \citenamefont {Venkataraman},\ and\ \citenamefont {Repp}}]{bib:patera19}%
  \BibitemOpen
  \bibfield  {author} {\bibinfo {author} {\bibfnamefont {L.~L.}\ \bibnamefont
  {Patera}}, \bibinfo {author} {\bibfnamefont {S.}~\bibnamefont {Sokolov}},
  \bibinfo {author} {\bibfnamefont {J.~Z.}\ \bibnamefont {Low}}, \bibinfo
  {author} {\bibfnamefont {L.~M.}\ \bibnamefont {Campos}}, \bibinfo {author}
  {\bibfnamefont {L.}~\bibnamefont {Venkataraman}},\ and\ \bibinfo {author}
  {\bibfnamefont {J.}~\bibnamefont {Repp}},\ }\bibfield  {title} {\bibinfo
  {title} {Resolving the unpaired-electron orbital distribution in a stable
  organic radical by {Kondo} resonance mapping},\ }\href
  {https://doi.org/https://doi.org/10.1002/anie.201904851} {\bibfield
  {journal} {\bibinfo  {journal} {Angewandte Chemie International Edition}\
  }\textbf {\bibinfo {volume} {58}},\ \bibinfo {pages} {11063} (\bibinfo {year}
  {2019})}\BibitemShut {NoStop}%
\bibitem [{\citenamefont {Koshida}\ \emph {et~al.}(2021)\citenamefont
  {Koshida}, \citenamefont {Okuyama}, \citenamefont {Hatta}, \citenamefont
  {Aruga},\ and\ \citenamefont {Minamitani}}]{bib:koshida21}%
  \BibitemOpen
  \bibfield  {author} {\bibinfo {author} {\bibfnamefont {H.}~\bibnamefont
  {Koshida}}, \bibinfo {author} {\bibfnamefont {H.}~\bibnamefont {Okuyama}},
  \bibinfo {author} {\bibfnamefont {S.}~\bibnamefont {Hatta}}, \bibinfo
  {author} {\bibfnamefont {T.}~\bibnamefont {Aruga}},\ and\ \bibinfo {author}
  {\bibfnamefont {E.}~\bibnamefont {Minamitani}},\ }\bibfield  {title}
  {\bibinfo {title} {Effect of local geometry on magnetic property of nitric
  oxide on
  $\mathrm{Au}(110)\text{\ensuremath{-}}(1\ifmmode\times\else\texttimes\fi{}2)$},\
  }\href {https://doi.org/10.1103/PhysRevB.103.155412} {\bibfield  {journal}
  {\bibinfo  {journal} {Phys. Rev. B}\ }\textbf {\bibinfo {volume} {103}},\
  \bibinfo {pages} {155412} (\bibinfo {year} {2021})}\BibitemShut {NoStop}%
\bibitem [{\citenamefont {Lu}\ \emph {et~al.}(2021)\citenamefont {Lu},
  \citenamefont {Nam}, \citenamefont {Xiao}, \citenamefont {Liu}, \citenamefont
  {Guo}, \citenamefont {Bai}, \citenamefont {Cheng}, \citenamefont {Deng},
  \citenamefont {Li}, \citenamefont {Zhou}, \citenamefont {Henkelman},
  \citenamefont {Fiete}, \citenamefont {Gao}, \citenamefont {MacDonald},
  \citenamefont {Zhang},\ and\ \citenamefont {Shih}}]{bib:lu21}%
  \BibitemOpen
  \bibfield  {author} {\bibinfo {author} {\bibfnamefont {S.}~\bibnamefont
  {Lu}}, \bibinfo {author} {\bibfnamefont {H.}~\bibnamefont {Nam}}, \bibinfo
  {author} {\bibfnamefont {P.}~\bibnamefont {Xiao}}, \bibinfo {author}
  {\bibfnamefont {M.}~\bibnamefont {Liu}}, \bibinfo {author} {\bibfnamefont
  {Y.}~\bibnamefont {Guo}}, \bibinfo {author} {\bibfnamefont {Y.}~\bibnamefont
  {Bai}}, \bibinfo {author} {\bibfnamefont {Z.}~\bibnamefont {Cheng}}, \bibinfo
  {author} {\bibfnamefont {J.}~\bibnamefont {Deng}}, \bibinfo {author}
  {\bibfnamefont {Y.}~\bibnamefont {Li}}, \bibinfo {author} {\bibfnamefont
  {H.}~\bibnamefont {Zhou}}, \bibinfo {author} {\bibfnamefont {G.}~\bibnamefont
  {Henkelman}}, \bibinfo {author} {\bibfnamefont {G.~A.}\ \bibnamefont
  {Fiete}}, \bibinfo {author} {\bibfnamefont {H.-J.}\ \bibnamefont {Gao}},
  \bibinfo {author} {\bibfnamefont {A.~H.}\ \bibnamefont {MacDonald}}, \bibinfo
  {author} {\bibfnamefont {C.}~\bibnamefont {Zhang}},\ and\ \bibinfo {author}
  {\bibfnamefont {C.-K.}\ \bibnamefont {Shih}},\ }\bibfield  {title} {\bibinfo
  {title} {{PTCDA} molecular monolayer on {Pb} thin films: An unusual
  $\ensuremath{\pi}$-electron {Kondo} system and its interplay with a
  quantum-confined superconductor},\ }\href
  {https://doi.org/10.1103/PhysRevLett.127.186805} {\bibfield  {journal}
  {\bibinfo  {journal} {Phys. Rev. Lett.}\ }\textbf {\bibinfo {volume} {127}},\
  \bibinfo {pages} {186805} (\bibinfo {year} {2021})}\BibitemShut {NoStop}%
\bibitem [{\citenamefont {Li}\ \emph {et~al.}(2019)\citenamefont {Li},
  \citenamefont {Sanz}, \citenamefont {Corso}, \citenamefont {Choi},
  \citenamefont {Pe{\~{n}}a}, \citenamefont {Frederiksen},\ and\ \citenamefont
  {Pascual}}]{Li2019}%
  \BibitemOpen
  \bibfield  {author} {\bibinfo {author} {\bibfnamefont {J.}~\bibnamefont
  {Li}}, \bibinfo {author} {\bibfnamefont {S.}~\bibnamefont {Sanz}}, \bibinfo
  {author} {\bibfnamefont {M.}~\bibnamefont {Corso}}, \bibinfo {author}
  {\bibfnamefont {D.~J.}\ \bibnamefont {Choi}}, \bibinfo {author}
  {\bibfnamefont {D.}~\bibnamefont {Pe{\~{n}}a}}, \bibinfo {author}
  {\bibfnamefont {T.}~\bibnamefont {Frederiksen}},\ and\ \bibinfo {author}
  {\bibfnamefont {J.~I.}\ \bibnamefont {Pascual}},\ }\bibfield  {title}
  {\bibinfo {title} {Single spin localization and manipulation in graphene
  open-shell nanostructures},\ }\href@noop {} {\bibfield  {journal} {\bibinfo
  {journal} {Nature Communications}\ }\textbf {\bibinfo {volume} {10}}
  (\bibinfo {year} {2019})}\BibitemShut {NoStop}%
\bibitem [{\citenamefont {Li}\ \emph {et~al.}(2020{\natexlab{a}})\citenamefont
  {Li}, \citenamefont {Sanz}, \citenamefont {Castro-Esteban}, \citenamefont
  {Vilas-Varela}, \citenamefont {Friedrich}, \citenamefont {Frederiksen},
  \citenamefont {Pe\~na},\ and\ \citenamefont {Pascual}}]{bib:li20}%
  \BibitemOpen
  \bibfield  {author} {\bibinfo {author} {\bibfnamefont {J.}~\bibnamefont
  {Li}}, \bibinfo {author} {\bibfnamefont {S.}~\bibnamefont {Sanz}}, \bibinfo
  {author} {\bibfnamefont {J.}~\bibnamefont {Castro-Esteban}}, \bibinfo
  {author} {\bibfnamefont {M.}~\bibnamefont {Vilas-Varela}}, \bibinfo {author}
  {\bibfnamefont {N.}~\bibnamefont {Friedrich}}, \bibinfo {author}
  {\bibfnamefont {T.}~\bibnamefont {Frederiksen}}, \bibinfo {author}
  {\bibfnamefont {D.}~\bibnamefont {Pe\~na}},\ and\ \bibinfo {author}
  {\bibfnamefont {J.~I.}\ \bibnamefont {Pascual}},\ }\bibfield  {title}
  {\bibinfo {title} {Uncovering the triplet ground state of triangular graphene
  nanoflakes engineered with atomic precision on a metal surface},\ }\href
  {https://doi.org/10.1103/PhysRevLett.124.177201} {\bibfield  {journal}
  {\bibinfo  {journal} {Phys. Rev. Lett.}\ }\textbf {\bibinfo {volume} {124}},\
  \bibinfo {pages} {177201} (\bibinfo {year} {2020}{\natexlab{a}})}\BibitemShut
  {NoStop}%
\bibitem [{\citenamefont {Mishra}\ \emph {et~al.}(2020)\citenamefont {Mishra},
  \citenamefont {Beyer}, \citenamefont {Eimre}, \citenamefont {Kezilebieke},
  \citenamefont {Berger}, \citenamefont {Gr{\"o}ning}, \citenamefont
  {Pignedoli}, \citenamefont {M{\"u}llen}, \citenamefont {Liljeroth},
  \citenamefont {Ruffieux}, \citenamefont {Feng},\ and\ \citenamefont
  {Fasel}}]{bib:mishra20}%
  \BibitemOpen
  \bibfield  {author} {\bibinfo {author} {\bibfnamefont {S.}~\bibnamefont
  {Mishra}}, \bibinfo {author} {\bibfnamefont {D.}~\bibnamefont {Beyer}},
  \bibinfo {author} {\bibfnamefont {K.}~\bibnamefont {Eimre}}, \bibinfo
  {author} {\bibfnamefont {S.}~\bibnamefont {Kezilebieke}}, \bibinfo {author}
  {\bibfnamefont {R.}~\bibnamefont {Berger}}, \bibinfo {author} {\bibfnamefont
  {O.}~\bibnamefont {Gr{\"o}ning}}, \bibinfo {author} {\bibfnamefont {C.~A.}\
  \bibnamefont {Pignedoli}}, \bibinfo {author} {\bibfnamefont {K.}~\bibnamefont
  {M{\"u}llen}}, \bibinfo {author} {\bibfnamefont {P.}~\bibnamefont
  {Liljeroth}}, \bibinfo {author} {\bibfnamefont {P.}~\bibnamefont {Ruffieux}},
  \bibinfo {author} {\bibfnamefont {X.}~\bibnamefont {Feng}},\ and\ \bibinfo
  {author} {\bibfnamefont {R.}~\bibnamefont {Fasel}},\ }\bibfield  {title}
  {\bibinfo {title} {Topological frustration induces unconventional magnetism
  in a nanographene},\ }\href {https://doi.org/10.1038/s41565-019-0577-9}
  {\bibfield  {journal} {\bibinfo  {journal} {Nature Nanotechnology}\ }\textbf
  {\bibinfo {volume} {15}},\ \bibinfo {pages} {22} (\bibinfo {year}
  {2020})}\BibitemShut {NoStop}%
\bibitem [{\citenamefont {Jacob}\ \emph {et~al.}(2021)\citenamefont {Jacob},
  \citenamefont {Ortiz},\ and\ \citenamefont
  {Fern\'andez-Rossier}}]{bib:jacob21}%
  \BibitemOpen
  \bibfield  {author} {\bibinfo {author} {\bibfnamefont {D.}~\bibnamefont
  {Jacob}}, \bibinfo {author} {\bibfnamefont {R.}~\bibnamefont {Ortiz}},\ and\
  \bibinfo {author} {\bibfnamefont {J.}~\bibnamefont {Fern\'andez-Rossier}},\
  }\bibfield  {title} {\bibinfo {title} {Renormalization of spin excitations
  and {Kondo} effect in open-shell nanographenes},\ }\href
  {https://doi.org/10.1103/PhysRevB.104.075404} {\bibfield  {journal} {\bibinfo
   {journal} {Phys. Rev. B}\ }\textbf {\bibinfo {volume} {104}},\ \bibinfo
  {pages} {075404} (\bibinfo {year} {2021})}\BibitemShut {NoStop}%
\bibitem [{\citenamefont {Turco}\ \emph {et~al.}(2023)\citenamefont {Turco},
  \citenamefont {Bernhardt}, \citenamefont {Krane}, \citenamefont {Valenta},
  \citenamefont {Fasel}, \citenamefont {Juríček},\ and\ \citenamefont
  {Ruffieux}}]{bib:turco23}%
  \BibitemOpen
  \bibfield  {author} {\bibinfo {author} {\bibfnamefont {E.}~\bibnamefont
  {Turco}}, \bibinfo {author} {\bibfnamefont {A.}~\bibnamefont {Bernhardt}},
  \bibinfo {author} {\bibfnamefont {N.}~\bibnamefont {Krane}}, \bibinfo
  {author} {\bibfnamefont {L.}~\bibnamefont {Valenta}}, \bibinfo {author}
  {\bibfnamefont {R.}~\bibnamefont {Fasel}}, \bibinfo {author} {\bibfnamefont
  {M.}~\bibnamefont {Juríček}},\ and\ \bibinfo {author} {\bibfnamefont
  {P.}~\bibnamefont {Ruffieux}},\ }\bibfield  {title} {\bibinfo {title}
  {Observation of the magnetic ground state of the two smallest triangular
  nanographenes},\ }\href {https://doi.org/10.1021/jacsau.2c00666} {\bibfield
  {journal} {\bibinfo  {journal} {JACS Au}\ }\textbf {\bibinfo {volume} {3}},\
  \bibinfo {pages} {1358} (\bibinfo {year} {2023})}\BibitemShut {NoStop}%
\bibitem [{\citenamefont {{Nozi\`eres, Ph.}}\ and\ \citenamefont {{Blandin,
  A.}}(1980)}]{bib:nozieres80}%
  \BibitemOpen
  \bibfield  {author} {\bibinfo {author} {\bibnamefont {{Nozi\`eres, Ph.}}}\
  and\ \bibinfo {author} {\bibnamefont {{Blandin, A.}}},\ }\bibfield  {title}
  {\bibinfo {title} {Kondo effect in real metals},\ }\href
  {https://doi.org/10.1051/jphys:01980004103019300} {\bibfield  {journal}
  {\bibinfo  {journal} {J. Phys. France}\ }\textbf {\bibinfo {volume} {41}},\
  \bibinfo {pages} {193} (\bibinfo {year} {1980})}\BibitemShut {NoStop}%
\bibitem [{\citenamefont {Roch}\ \emph {et~al.}(2009)\citenamefont {Roch},
  \citenamefont {Florens}, \citenamefont {Costi}, \citenamefont {Wernsdorfer},\
  and\ \citenamefont {Balestro}}]{bib:roch09}%
  \BibitemOpen
  \bibfield  {author} {\bibinfo {author} {\bibfnamefont {N.}~\bibnamefont
  {Roch}}, \bibinfo {author} {\bibfnamefont {S.}~\bibnamefont {Florens}},
  \bibinfo {author} {\bibfnamefont {T.~A.}\ \bibnamefont {Costi}}, \bibinfo
  {author} {\bibfnamefont {W.}~\bibnamefont {Wernsdorfer}},\ and\ \bibinfo
  {author} {\bibfnamefont {F.}~\bibnamefont {Balestro}},\ }\bibfield  {title}
  {\bibinfo {title} {Observation of the underscreened {Kondo} effect in a
  molecular transistor},\ }\href
  {https://doi.org/10.1103/PhysRevLett.103.197202} {\bibfield  {journal}
  {\bibinfo  {journal} {Phys. Rev. Lett.}\ }\textbf {\bibinfo {volume} {103}},\
  \bibinfo {pages} {197202} (\bibinfo {year} {2009})}\BibitemShut {NoStop}%
\bibitem [{\citenamefont {Parks}\ \emph {et~al.}(2010)\citenamefont {Parks},
  \citenamefont {Champagne}, \citenamefont {Costi}, \citenamefont {Shum},
  \citenamefont {Pasupathy}, \citenamefont {Neuscamman}, \citenamefont
  {Flores-Torres}, \citenamefont {Cornaglia}, \citenamefont {Aligia},
  \citenamefont {Balseiro}, \citenamefont {Chan}, \citenamefont {na},\ and\
  \citenamefont {Ralph}}]{bib:parks10}%
  \BibitemOpen
  \bibfield  {author} {\bibinfo {author} {\bibfnamefont {J.~J.}\ \bibnamefont
  {Parks}}, \bibinfo {author} {\bibfnamefont {A.~R.}\ \bibnamefont
  {Champagne}}, \bibinfo {author} {\bibfnamefont {T.~A.}\ \bibnamefont
  {Costi}}, \bibinfo {author} {\bibfnamefont {W.~W.}\ \bibnamefont {Shum}},
  \bibinfo {author} {\bibfnamefont {A.~N.}\ \bibnamefont {Pasupathy}}, \bibinfo
  {author} {\bibfnamefont {E.}~\bibnamefont {Neuscamman}}, \bibinfo {author}
  {\bibfnamefont {S.}~\bibnamefont {Flores-Torres}}, \bibinfo {author}
  {\bibfnamefont {P.~S.}\ \bibnamefont {Cornaglia}}, \bibinfo {author}
  {\bibfnamefont {A.~A.}\ \bibnamefont {Aligia}}, \bibinfo {author}
  {\bibfnamefont {C.~A.}\ \bibnamefont {Balseiro}}, \bibinfo {author}
  {\bibfnamefont {G.~K.-L.}\ \bibnamefont {Chan}}, \bibinfo {author}
  {\bibfnamefont {H.~D.~A.}\ \bibnamefont {na}},\ and\ \bibinfo {author}
  {\bibfnamefont {D.~C.}\ \bibnamefont {Ralph}},\ }\bibfield  {title} {\bibinfo
  {title} {Mechanical control of spin states in spin-1 molecules and the
  underscreened {Kondo} effect},\ }\href
  {https://doi.org/10.1126/science.1186874} {\bibfield  {journal} {\bibinfo
  {journal} {Science}\ }\textbf {\bibinfo {volume} {328}},\ \bibinfo {pages}
  {1370} (\bibinfo {year} {2010})}\BibitemShut {NoStop}%
\bibitem [{\citenamefont {Pustilnik}\ and\ \citenamefont
  {Glazman}(2001)}]{pustilnik2001}%
  \BibitemOpen
  \bibfield  {author} {\bibinfo {author} {\bibfnamefont {M.}~\bibnamefont
  {Pustilnik}}\ and\ \bibinfo {author} {\bibfnamefont {L.~I.}\ \bibnamefont
  {Glazman}},\ }\bibfield  {title} {\bibinfo {title} {Kondo effect in real
  quantum dots},\ }\href@noop {} {\bibfield  {journal} {\bibinfo  {journal}
  {Physical Review Letters}\ }\textbf {\bibinfo {volume} {87}},\ \bibinfo
  {pages} {216601} (\bibinfo {year} {2001})}\BibitemShut {NoStop}%
\bibitem [{\citenamefont {Mitchell}\ \emph {et~al.}(2017)\citenamefont
  {Mitchell}, \citenamefont {Pedersen}, \citenamefont {Hedeg{\aa}rd},\ and\
  \citenamefont {Paaske}}]{mitchell2017}%
  \BibitemOpen
  \bibfield  {author} {\bibinfo {author} {\bibfnamefont {A.}~\bibnamefont
  {Mitchell}}, \bibinfo {author} {\bibfnamefont {K.}~\bibnamefont {Pedersen}},
  \bibinfo {author} {\bibfnamefont {P.}~\bibnamefont {Hedeg{\aa}rd}},\ and\
  \bibinfo {author} {\bibfnamefont {J.}~\bibnamefont {Paaske}},\ }\bibfield
  {title} {\bibinfo {title} {Kondo blockade due to quantum interference in
  single-molecule junctions},\ }\href
  {https://doi.org/https://doi.org/10.1038/ncomms15210} {\bibfield  {journal}
  {\bibinfo  {journal} {Nature Communications}\ }\textbf {\bibinfo {volume}
  {8}},\ \bibinfo {pages} {15210} (\bibinfo {year} {2017})}\BibitemShut
  {NoStop}%
\bibitem [{\citenamefont {Girovsky}\ \emph {et~al.}(2017)\citenamefont
  {Girovsky}, \citenamefont {Nowakowski}, \citenamefont {Ali}, \citenamefont
  {Baljozovic}, \citenamefont {Rossmann}, \citenamefont {Nijs}, \citenamefont
  {Aeby}, \citenamefont {Nowakowska}, \citenamefont {Siewert}, \citenamefont
  {Srivastava}, \citenamefont {W{\"a}ckerlin}, \citenamefont {Dreiser},
  \citenamefont {Decurtins}, \citenamefont {Liu}, \citenamefont {Oppeneer},
  \citenamefont {Jung},\ and\ \citenamefont {Ballav}}]{bib:girovsky2017}%
  \BibitemOpen
  \bibfield  {author} {\bibinfo {author} {\bibfnamefont {J.}~\bibnamefont
  {Girovsky}}, \bibinfo {author} {\bibfnamefont {J.}~\bibnamefont
  {Nowakowski}}, \bibinfo {author} {\bibfnamefont {M.~E.}\ \bibnamefont {Ali}},
  \bibinfo {author} {\bibfnamefont {M.}~\bibnamefont {Baljozovic}}, \bibinfo
  {author} {\bibfnamefont {H.~R.}\ \bibnamefont {Rossmann}}, \bibinfo {author}
  {\bibfnamefont {T.}~\bibnamefont {Nijs}}, \bibinfo {author} {\bibfnamefont
  {E.~A.}\ \bibnamefont {Aeby}}, \bibinfo {author} {\bibfnamefont
  {S.}~\bibnamefont {Nowakowska}}, \bibinfo {author} {\bibfnamefont
  {D.}~\bibnamefont {Siewert}}, \bibinfo {author} {\bibfnamefont
  {G.}~\bibnamefont {Srivastava}}, \bibinfo {author} {\bibfnamefont
  {C.}~\bibnamefont {W{\"a}ckerlin}}, \bibinfo {author} {\bibfnamefont
  {J.}~\bibnamefont {Dreiser}}, \bibinfo {author} {\bibfnamefont
  {S.}~\bibnamefont {Decurtins}}, \bibinfo {author} {\bibfnamefont {S.-X.}\
  \bibnamefont {Liu}}, \bibinfo {author} {\bibfnamefont {P.~M.}\ \bibnamefont
  {Oppeneer}}, \bibinfo {author} {\bibfnamefont {T.~A.}\ \bibnamefont {Jung}},\
  and\ \bibinfo {author} {\bibfnamefont {N.}~\bibnamefont {Ballav}},\
  }\bibfield  {title} {\bibinfo {title} {Long-range ferrimagnetic order in a
  two-dimensional supramolecular {Kondo} lattice},\ }\href
  {https://doi.org/10.1038/ncomms15388} {\bibfield  {journal} {\bibinfo
  {journal} {Nature Communications}\ }\textbf {\bibinfo {volume} {8}},\
  \bibinfo {pages} {15388} (\bibinfo {year} {2017})}\BibitemShut {NoStop}%
\bibitem [{\citenamefont {Biswas}\ \emph {et~al.}(2022)\citenamefont {Biswas},
  \citenamefont {Urbani}, \citenamefont {Sánchez-Grande}, \citenamefont
  {Soler-Polo}, \citenamefont {Lauwaet}, \citenamefont {Matěj}, \citenamefont
  {Mutombo}, \citenamefont {Veis}, \citenamefont {Brabec}, \citenamefont
  {Pernal}, \citenamefont {Gallego}, \citenamefont {Miranda}, \citenamefont
  {Écija}, \citenamefont {Jelínek}, \citenamefont {Torres},\ and\
  \citenamefont {Urgel}}]{yo8}%
  \BibitemOpen
  \bibfield  {author} {\bibinfo {author} {\bibfnamefont {K.}~\bibnamefont
  {Biswas}}, \bibinfo {author} {\bibfnamefont {M.}~\bibnamefont {Urbani}},
  \bibinfo {author} {\bibfnamefont {A.}~\bibnamefont {Sánchez-Grande}},
  \bibinfo {author} {\bibfnamefont {D.}~\bibnamefont {Soler-Polo}}, \bibinfo
  {author} {\bibfnamefont {K.}~\bibnamefont {Lauwaet}}, \bibinfo {author}
  {\bibfnamefont {A.}~\bibnamefont {Matěj}}, \bibinfo {author} {\bibfnamefont
  {P.}~\bibnamefont {Mutombo}}, \bibinfo {author} {\bibfnamefont
  {L.}~\bibnamefont {Veis}}, \bibinfo {author} {\bibfnamefont {J.}~\bibnamefont
  {Brabec}}, \bibinfo {author} {\bibfnamefont {K.}~\bibnamefont {Pernal}},
  \bibinfo {author} {\bibfnamefont {J.~M.}\ \bibnamefont {Gallego}}, \bibinfo
  {author} {\bibfnamefont {R.}~\bibnamefont {Miranda}}, \bibinfo {author}
  {\bibfnamefont {D.}~\bibnamefont {Écija}}, \bibinfo {author} {\bibfnamefont
  {P.}~\bibnamefont {Jelínek}}, \bibinfo {author} {\bibfnamefont
  {T.}~\bibnamefont {Torres}},\ and\ \bibinfo {author} {\bibfnamefont {J.~I.}\
  \bibnamefont {Urgel}},\ }\bibfield  {title} {\bibinfo {title} {Interplay
  between $\pi$-conjugation and exchange magnetism in one-dimensional
  porphyrinoid polymers},\ }\href {https://doi.org/10.1021/jacs.2c02700}
  {\bibfield  {journal} {\bibinfo  {journal} {Journal of the American Chemical
  Society}\ }\textbf {\bibinfo {volume} {144}},\ \bibinfo {pages} {12725}
  (\bibinfo {year} {2022})}\BibitemShut {NoStop}%
\bibitem [{\citenamefont {Sun}\ \emph {et~al.}(2020)\citenamefont {Sun},
  \citenamefont {Mateo}, \citenamefont {Robles}, \citenamefont {Ruffieux},
  \citenamefont {Lorente}, \citenamefont {Bottari}, \citenamefont {Torres},\
  and\ \citenamefont {Fasel}}]{Sun2020}%
  \BibitemOpen
  \bibfield  {author} {\bibinfo {author} {\bibfnamefont {Q.}~\bibnamefont
  {Sun}}, \bibinfo {author} {\bibfnamefont {L.~M.}\ \bibnamefont {Mateo}},
  \bibinfo {author} {\bibfnamefont {R.}~\bibnamefont {Robles}}, \bibinfo
  {author} {\bibfnamefont {P.}~\bibnamefont {Ruffieux}}, \bibinfo {author}
  {\bibfnamefont {N.}~\bibnamefont {Lorente}}, \bibinfo {author} {\bibfnamefont
  {G.}~\bibnamefont {Bottari}}, \bibinfo {author} {\bibfnamefont
  {T.}~\bibnamefont {Torres}},\ and\ \bibinfo {author} {\bibfnamefont
  {R.}~\bibnamefont {Fasel}},\ }\bibfield  {title} {\bibinfo {title} {Inducing
  open-shell character in porphyrins through surface-assisted phenalenyl
  $\pi$-extension},\ }\href {https://doi.org/10.1021/jacs.0c07781} {\bibfield
  {journal} {\bibinfo  {journal} {Journal of the American Chemical Society}\
  }\textbf {\bibinfo {volume} {142}},\ \bibinfo {pages} {18109} (\bibinfo
  {year} {2020})}\BibitemShut {NoStop}%
\bibitem [{\citenamefont {Roos}\ \emph {et~al.}(1980)\citenamefont {Roos},
  \citenamefont {Taylor},\ and\ \citenamefont {Sigbahn}}]{CAS}%
  \BibitemOpen
  \bibfield  {author} {\bibinfo {author} {\bibfnamefont {B.~O.}\ \bibnamefont
  {Roos}}, \bibinfo {author} {\bibfnamefont {P.~R.}\ \bibnamefont {Taylor}},\
  and\ \bibinfo {author} {\bibfnamefont {P.~E.}\ \bibnamefont {Sigbahn}},\
  }\bibfield  {title} {\bibinfo {title} {A complete active space {SCF} method
  ({CASSCF}) using a density matrix formulated super-{CI} approach},\ }\href
  {https://doi.org/https://doi.org/10.1016/0301-0104(80)80045-0} {\bibfield
  {journal} {\bibinfo  {journal} {Chemical Physics}\ }\textbf {\bibinfo
  {volume} {48}},\ \bibinfo {pages} {157} (\bibinfo {year} {1980})}\BibitemShut
  {NoStop}%
\bibitem [{\citenamefont {Krishna-murthy}\ \emph
  {et~al.}(1980{\natexlab{a}})\citenamefont {Krishna-murthy}, \citenamefont
  {Wilkins},\ and\ \citenamefont {Wilson}}]{krishna-murthy1980a}%
  \BibitemOpen
  \bibfield  {author} {\bibinfo {author} {\bibfnamefont {H.~R.}\ \bibnamefont
  {Krishna-murthy}}, \bibinfo {author} {\bibfnamefont {J.~W.}\ \bibnamefont
  {Wilkins}},\ and\ \bibinfo {author} {\bibfnamefont {K.~G.}\ \bibnamefont
  {Wilson}},\ }\bibfield  {title} {\bibinfo {title} {Renormalization-group
  approach to the {Anderson} model of dilute magnetic alloys. {I. Static}
  properties for the symmetric case},\ }\href@noop {} {\bibfield  {journal}
  {\bibinfo  {journal} {Physical Review B}\ }\textbf {\bibinfo {volume} {21}},\
  \bibinfo {pages} {1003} (\bibinfo {year} {1980}{\natexlab{a}})}\BibitemShut
  {NoStop}%
\bibitem [{\citenamefont {Calupitan}\ \emph {et~al.}(2023)\citenamefont
  {Calupitan}, \citenamefont {Berdonces-Layunta}, \citenamefont
  {Aguilar-Galindo}, \citenamefont {Vilas-Varela}, \citenamefont {Peña},
  \citenamefont {Casanova}, \citenamefont {Corso}, \citenamefont {de~Oteyza},\
  and\ \citenamefont {Wang}}]{bib:calupitan23}%
  \BibitemOpen
  \bibfield  {author} {\bibinfo {author} {\bibfnamefont {J.~P.}\ \bibnamefont
  {Calupitan}}, \bibinfo {author} {\bibfnamefont {A.}~\bibnamefont
  {Berdonces-Layunta}}, \bibinfo {author} {\bibfnamefont {F.}~\bibnamefont
  {Aguilar-Galindo}}, \bibinfo {author} {\bibfnamefont {M.}~\bibnamefont
  {Vilas-Varela}}, \bibinfo {author} {\bibfnamefont {D.}~\bibnamefont {Peña}},
  \bibinfo {author} {\bibfnamefont {D.}~\bibnamefont {Casanova}}, \bibinfo
  {author} {\bibfnamefont {M.}~\bibnamefont {Corso}}, \bibinfo {author}
  {\bibfnamefont {D.~G.}\ \bibnamefont {de~Oteyza}},\ and\ \bibinfo {author}
  {\bibfnamefont {T.}~\bibnamefont {Wang}},\ }\bibfield  {title} {\bibinfo
  {title} {Emergence of $\pi$-magnetism in fused {Aza-Triangulenes}: Symmetry
  and charge transfer effects},\ }\href
  {https://doi.org/10.1021/acs.nanolett.3c02586} {\bibfield  {journal}
  {\bibinfo  {journal} {Nano Letters}\ }\textbf {\bibinfo {volume} {23}},\
  \bibinfo {pages} {9832} (\bibinfo {year} {2023})}\BibitemShut {NoStop}%
\bibitem [{\citenamefont {Vilas-Varela}\ \emph {et~al.}(2023)\citenamefont
  {Vilas-Varela}, \citenamefont {Romero-Lara}, \citenamefont {Vegliante},
  \citenamefont {Calupitan}, \citenamefont {Martínez}, \citenamefont {Meyer},
  \citenamefont {Uriarte-Amiano}, \citenamefont {Friedrich}, \citenamefont
  {Wang}, \citenamefont {Schulz}, \citenamefont {Koval}, \citenamefont
  {Sandoval-Salinas}, \citenamefont {Casanova}, \citenamefont {Corso},
  \citenamefont {Artacho}, \citenamefont {Peña},\ and\ \citenamefont
  {Pascual}}]{bib:vilas23}%
  \BibitemOpen
  \bibfield  {author} {\bibinfo {author} {\bibfnamefont {M.}~\bibnamefont
  {Vilas-Varela}}, \bibinfo {author} {\bibfnamefont {F.}~\bibnamefont
  {Romero-Lara}}, \bibinfo {author} {\bibfnamefont {A.}~\bibnamefont
  {Vegliante}}, \bibinfo {author} {\bibfnamefont {J.~P.}\ \bibnamefont
  {Calupitan}}, \bibinfo {author} {\bibfnamefont {A.}~\bibnamefont
  {Martínez}}, \bibinfo {author} {\bibfnamefont {L.}~\bibnamefont {Meyer}},
  \bibinfo {author} {\bibfnamefont {U.}~\bibnamefont {Uriarte-Amiano}},
  \bibinfo {author} {\bibfnamefont {N.}~\bibnamefont {Friedrich}}, \bibinfo
  {author} {\bibfnamefont {D.}~\bibnamefont {Wang}}, \bibinfo {author}
  {\bibfnamefont {F.}~\bibnamefont {Schulz}}, \bibinfo {author} {\bibfnamefont
  {N.~E.}\ \bibnamefont {Koval}}, \bibinfo {author} {\bibfnamefont {M.~E.}\
  \bibnamefont {Sandoval-Salinas}}, \bibinfo {author} {\bibfnamefont
  {D.}~\bibnamefont {Casanova}}, \bibinfo {author} {\bibfnamefont
  {M.}~\bibnamefont {Corso}}, \bibinfo {author} {\bibfnamefont
  {E.}~\bibnamefont {Artacho}}, \bibinfo {author} {\bibfnamefont
  {D.}~\bibnamefont {Peña}},\ and\ \bibinfo {author} {\bibfnamefont {J.~I.}\
  \bibnamefont {Pascual}},\ }\bibfield  {title} {\bibinfo {title} {On-surface
  synthesis and characterization of a high-spin {Aza-[5]-Triangulene}},\ }\href
  {https://doi.org/https://doi.org/10.1002/anie.202307884} {\bibfield
  {journal} {\bibinfo  {journal} {Angewandte Chemie International Edition}\
  }\textbf {\bibinfo {volume} {62}},\ \bibinfo {pages} {e202307884} (\bibinfo
  {year} {2023})}\BibitemShut {NoStop}%
\bibitem [{\citenamefont {Ovchinnikov}(1978)}]{bib:ovchinnikov78}%
  \BibitemOpen
  \bibfield  {author} {\bibinfo {author} {\bibfnamefont {A.~A.}\ \bibnamefont
  {Ovchinnikov}},\ }\bibfield  {title} {\bibinfo {title} {Multiplicity of the
  ground state of large alternant organic molecules with conjugated bonds},\
  }\href {https://doi.org/10.1007/BF00549259} {\bibfield  {journal} {\bibinfo
  {journal} {Theoretica chimica acta}\ }\textbf {\bibinfo {volume} {47}},\
  \bibinfo {pages} {297} (\bibinfo {year} {1978})}\BibitemShut {NoStop}%
\bibitem [{\citenamefont {Ternes}(2015)}]{bib:ternes15}%
  \BibitemOpen
  \bibfield  {author} {\bibinfo {author} {\bibfnamefont {M.}~\bibnamefont
  {Ternes}},\ }\bibfield  {title} {\bibinfo {title} {Spin excitations and
  correlations in scanning tunneling spectroscopy},\ }\href
  {https://doi.org/10.1088/1367-2630/17/6/063016} {\bibfield  {journal}
  {\bibinfo  {journal} {New Journal of Physics}\ }\textbf {\bibinfo {volume}
  {17}},\ \bibinfo {pages} {063016} (\bibinfo {year} {2015})}\BibitemShut
  {NoStop}%
\bibitem [{\citenamefont {Anderson}(1970)}]{anderson1970}%
  \BibitemOpen
  \bibfield  {author} {\bibinfo {author} {\bibfnamefont {P.~W.}\ \bibnamefont
  {Anderson}},\ }\bibfield  {title} {\bibinfo {title} {A poor man's derivation
  of scaling laws for the {Kondo} problem},\ }\href@noop {} {\bibfield
  {journal} {\bibinfo  {journal} {J. Phys. C: Solid St. Phys}\ }\textbf
  {\bibinfo {volume} {3}} (\bibinfo {year} {1970})}\BibitemShut {NoStop}%
\bibitem [{\citenamefont {Hirst}(1978)}]{hirst1978}%
  \BibitemOpen
  \bibfield  {author} {\bibinfo {author} {\bibfnamefont {L.}~\bibnamefont
  {Hirst}},\ }\bibfield  {title} {\bibinfo {title} {Theory of the coupling
  between conduction electrons and moments of 3d and 4f ions in metals},\
  }\href {https://doi.org/10.1080/00018737800101374} {\bibfield  {journal}
  {\bibinfo  {journal} {Advances in Physics}\ }\textbf {\bibinfo {volume}
  {27}},\ \bibinfo {pages} {231} (\bibinfo {year} {1978})}\BibitemShut
  {NoStop}%
\bibitem [{\citenamefont {Schrieffer}\ and\ \citenamefont
  {Wolff}(1966)}]{schrieffer1966}%
  \BibitemOpen
  \bibfield  {author} {\bibinfo {author} {\bibfnamefont {J.~R.}\ \bibnamefont
  {Schrieffer}}\ and\ \bibinfo {author} {\bibfnamefont {P.~A.}\ \bibnamefont
  {Wolff}},\ }\bibfield  {title} {\bibinfo {title} {Relation between the
  anderson and kondo hamiltonians},\ }\href@noop {} {\bibfield  {journal}
  {\bibinfo  {journal} {J. Kondo, Progr. Theoret. Phys. (Kyoto)}\ }\textbf
  {\bibinfo {volume} {14}},\ \bibinfo {pages} {61} (\bibinfo {year}
  {1966})}\BibitemShut {NoStop}%
\bibitem [{\citenamefont {Flores}\ and\ \citenamefont
  {Goldberg}(2016)}]{flores2017}%
  \BibitemOpen
  \bibfield  {author} {\bibinfo {author} {\bibfnamefont {F.}~\bibnamefont
  {Flores}}\ and\ \bibinfo {author} {\bibfnamefont {E.~C.}\ \bibnamefont
  {Goldberg}},\ }\bibfield  {title} {\bibinfo {title} {Ionic {Hamiltonians} for
  transition metal atoms: effective exchange coupling and {Kondo}
  temperature},\ }\href {https://doi.org/10.1088/1361-648X/aa4dda} {\bibfield
  {journal} {\bibinfo  {journal} {Journal of Physics: Condensed Matter}\
  }\textbf {\bibinfo {volume} {29}},\ \bibinfo {pages} {055602} (\bibinfo
  {year} {2016})}\BibitemShut {NoStop}%
\bibitem [{\citenamefont {Sen}\ and\ \citenamefont
  {Mitchell}(2024)}]{mitchell2023}%
  \BibitemOpen
  \bibfield  {author} {\bibinfo {author} {\bibfnamefont {S.}~\bibnamefont
  {Sen}}\ and\ \bibinfo {author} {\bibfnamefont {A.~K.}\ \bibnamefont
  {Mitchell}},\ }\bibfield  {title} {\bibinfo {title} {Many-body quantum
  interference route to the two-channel kondo effect: Inverse design for
  molecular junctions and quantum dot devices},\ }\href
  {https://doi.org/10.1103/PhysRevLett.133.076501} {\bibfield  {journal}
  {\bibinfo  {journal} {Phys. Rev. Lett.}\ }\textbf {\bibinfo {volume} {133}},\
  \bibinfo {pages} {076501} (\bibinfo {year} {2024})}\BibitemShut {NoStop}%
\bibitem [{\citenamefont {Calvo-Fernández}\ \emph {et~al.}(2024)\citenamefont
  {Calvo-Fernández}, \citenamefont {Blanco-Rey},\ and\ \citenamefont
  {Eiguren}}]{calvo-fernandez2023}%
  \BibitemOpen
  \bibfield  {author} {\bibinfo {author} {\bibfnamefont {A.}~\bibnamefont
  {Calvo-Fernández}}, \bibinfo {author} {\bibfnamefont {M.}~\bibnamefont
  {Blanco-Rey}},\ and\ \bibinfo {author} {\bibfnamefont {A.}~\bibnamefont
  {Eiguren}},\ }\bibfield  {title} {\bibinfo {title} {{The PointGroupNRG code
  for numerical renormalization group calculations with discrete point-group
  symmetries}},\ }\href
  {https://doi.org/https://doi.org/10.1016/j.cpc.2023.109032} {\bibfield
  {journal} {\bibinfo  {journal} {Computer Physics Communications}\ }\textbf
  {\bibinfo {volume} {296}},\ \bibinfo {pages} {109032} (\bibinfo {year}
  {2024})}\BibitemShut {NoStop}%
\bibitem [{\citenamefont {Calvo-Fernández}(2024)}]{PointGroupNRG}%
  \BibitemOpen
  \bibfield  {author} {\bibinfo {author} {\bibfnamefont {A.}~\bibnamefont
  {Calvo-Fernández}},\ }\href@noop {} {\bibinfo {title} {{PointGroupNRG}}},\
  \bibinfo {howpublished} {\url{https://github.com/aitorcf/PointGroupNRG}}
  (\bibinfo {year} {2024})\BibitemShut {NoStop}%
\bibitem [{\citenamefont {Gruber}\ \emph {et~al.}(2018)\citenamefont {Gruber},
  \citenamefont {Weismann},\ and\ \citenamefont {Berndt}}]{gruber2018kondo}%
  \BibitemOpen
  \bibfield  {author} {\bibinfo {author} {\bibfnamefont {M.}~\bibnamefont
  {Gruber}}, \bibinfo {author} {\bibfnamefont {A.}~\bibnamefont {Weismann}},\
  and\ \bibinfo {author} {\bibfnamefont {R.}~\bibnamefont {Berndt}},\
  }\bibfield  {title} {\bibinfo {title} {The {Kondo} resonance line shape in
  scanning tunnelling spectroscopy: instrumental aspects},\ }\href@noop {}
  {\bibfield  {journal} {\bibinfo  {journal} {Journal of Physics: Condensed
  Matter}\ }\textbf {\bibinfo {volume} {30}},\ \bibinfo {pages} {424001}
  (\bibinfo {year} {2018})}\BibitemShut {NoStop}%
\bibitem [{\citenamefont {Costi}\ and\ \citenamefont
  {Hewson}(1992)}]{costi1992}%
  \BibitemOpen
  \bibfield  {author} {\bibinfo {author} {\bibfnamefont {T.~A.}\ \bibnamefont
  {Costi}}\ and\ \bibinfo {author} {\bibfnamefont {A.~C.}\ \bibnamefont
  {Hewson}},\ }\bibfield  {title} {\bibinfo {title} {Resistivity cross-over for
  the non-degenerate {Anderson} model},\ }\href
  {https://doi.org/10.1080/13642819208215080} {\bibfield  {journal} {\bibinfo
  {journal} {Philosophical Magazine B}\ }\textbf {\bibinfo {volume} {65}},\
  \bibinfo {pages} {1165} (\bibinfo {year} {1992})}\BibitemShut {NoStop}%
\bibitem [{\citenamefont {Campo}\ and\ \citenamefont
  {Oliveira}(2005)}]{campo2005}%
  \BibitemOpen
  \bibfield  {author} {\bibinfo {author} {\bibfnamefont {V.~L.}\ \bibnamefont
  {Campo}}\ and\ \bibinfo {author} {\bibfnamefont {L.~N.}\ \bibnamefont
  {Oliveira}},\ }\bibfield  {title} {\bibinfo {title} {Alternative
  discretization in the numerical renormalization-group method},\ }\href
  {https://doi.org/10.1103/PhysRevB.72.104432} {\bibfield  {journal} {\bibinfo
  {journal} {Phys. Rev. B}\ }\textbf {\bibinfo {volume} {72}},\ \bibinfo
  {pages} {104432} (\bibinfo {year} {2005})}\BibitemShut {NoStop}%
\bibitem [{\citenamefont {Perdew}\ \emph
  {et~al.}(1996{\natexlab{a}})\citenamefont {Perdew}, \citenamefont
  {Ernzerhof},\ and\ \citenamefont {Burke}}]{pbe0}%
  \BibitemOpen
  \bibfield  {author} {\bibinfo {author} {\bibfnamefont {J.~P.}\ \bibnamefont
  {Perdew}}, \bibinfo {author} {\bibfnamefont {M.}~\bibnamefont {Ernzerhof}},\
  and\ \bibinfo {author} {\bibfnamefont {K.}~\bibnamefont {Burke}},\ }\bibfield
   {title} {\bibinfo {title} {{Rationale for mixing exact exchange with density
  functional approximations}},\ }\href {https://doi.org/10.1063/1.472933}
  {\bibfield  {journal} {\bibinfo  {journal} {The Journal of Chemical Physics}\
  }\textbf {\bibinfo {volume} {105}},\ \bibinfo {pages} {9982} (\bibinfo {year}
  {1996}{\natexlab{a}})}\BibitemShut {NoStop}%
\bibitem [{\citenamefont {Blum}\ \emph {et~al.}(2009)\citenamefont {Blum},
  \citenamefont {Gehrke}, \citenamefont {Hanke}, \citenamefont {Havu},
  \citenamefont {Havu}, \citenamefont {Ren}, \citenamefont {Reuter},\ and\
  \citenamefont {Scheffler}}]{AIMS}%
  \BibitemOpen
  \bibfield  {author} {\bibinfo {author} {\bibfnamefont {V.}~\bibnamefont
  {Blum}}, \bibinfo {author} {\bibfnamefont {R.}~\bibnamefont {Gehrke}},
  \bibinfo {author} {\bibfnamefont {F.}~\bibnamefont {Hanke}}, \bibinfo
  {author} {\bibfnamefont {P.}~\bibnamefont {Havu}}, \bibinfo {author}
  {\bibfnamefont {V.}~\bibnamefont {Havu}}, \bibinfo {author} {\bibfnamefont
  {X.}~\bibnamefont {Ren}}, \bibinfo {author} {\bibfnamefont {K.}~\bibnamefont
  {Reuter}},\ and\ \bibinfo {author} {\bibfnamefont {M.}~\bibnamefont
  {Scheffler}},\ }\bibfield  {title} {\bibinfo {title} {Ab initio molecular
  simulations with numeric atom-centered orbitals},\ }\href
  {https://doi.org/https://doi.org/10.1016/j.cpc.2009.06.022} {\bibfield
  {journal} {\bibinfo  {journal} {Computer Physics Communications}\ }\textbf
  {\bibinfo {volume} {180}},\ \bibinfo {pages} {2175} (\bibinfo {year}
  {2009})}\BibitemShut {NoStop}%
\bibitem [{\citenamefont {Perdew}\ \emph
  {et~al.}(1996{\natexlab{b}})\citenamefont {Perdew}, \citenamefont {Burke},\
  and\ \citenamefont {Ernzerhof}}]{pbe}%
  \BibitemOpen
  \bibfield  {author} {\bibinfo {author} {\bibfnamefont {J.~P.}\ \bibnamefont
  {Perdew}}, \bibinfo {author} {\bibfnamefont {K.}~\bibnamefont {Burke}},\ and\
  \bibinfo {author} {\bibfnamefont {M.}~\bibnamefont {Ernzerhof}},\ }\bibfield
  {title} {\bibinfo {title} {Generalized gradient approximation made simple},\
  }\href {https://doi.org/10.1103/PhysRevLett.77.3865} {\bibfield  {journal}
  {\bibinfo  {journal} {Phys. Rev. Lett.}\ }\textbf {\bibinfo {volume} {77}},\
  \bibinfo {pages} {3865} (\bibinfo {year} {1996}{\natexlab{b}})}\BibitemShut
  {NoStop}%
\bibitem [{\citenamefont {L\"owdin}(1955)}]{NaturalOrbs}%
  \BibitemOpen
  \bibfield  {author} {\bibinfo {author} {\bibfnamefont {P.-O.}\ \bibnamefont
  {L\"owdin}},\ }\bibfield  {title} {\bibinfo {title} {Quantum theory of
  many-particle systems. {I. Physical} interpretations by means of density
  matrices, natural spin-orbitals, and convergence problems in the method of
  configurational interaction},\ }\href
  {https://doi.org/10.1103/PhysRev.97.1474} {\bibfield  {journal} {\bibinfo
  {journal} {Phys. Rev.}\ }\textbf {\bibinfo {volume} {97}},\ \bibinfo {pages}
  {1474} (\bibinfo {year} {1955})}\BibitemShut {NoStop}%
\bibitem [{\citenamefont {Neese}(2012)}]{orca}%
  \BibitemOpen
  \bibfield  {author} {\bibinfo {author} {\bibfnamefont {F.}~\bibnamefont
  {Neese}},\ }\bibfield  {title} {\bibinfo {title} {The {ORCA} program
  system},\ }\href@noop {} {\bibfield  {journal} {\bibinfo  {journal} {WIREs
  Computational Molecular Science}\ }\textbf {\bibinfo {volume} {2}},\ \bibinfo
  {pages} {73} (\bibinfo {year} {2012})}\BibitemShut {NoStop}%
\bibitem [{\citenamefont {Neese}(2018)}]{orca4.0}%
  \BibitemOpen
  \bibfield  {author} {\bibinfo {author} {\bibfnamefont {F.}~\bibnamefont
  {Neese}},\ }\bibfield  {title} {\bibinfo {title} {Software update: the {ORCA}
  program system, version 4.0},\ }\href@noop {} {\bibfield  {journal} {\bibinfo
   {journal} {WIREs Computational Molecular Science}\ }\textbf {\bibinfo
  {volume} {8}} (\bibinfo {year} {2018})}\BibitemShut {NoStop}%
\bibitem [{\citenamefont {Krej{\v{c}}{\'{\i}}}\ \emph
  {et~al.}(2017)\citenamefont {Krej{\v{c}}{\'{\i}}}, \citenamefont {Hapala},
  \citenamefont {Ondr{\'{a}}{\v{c}}ek},\ and\ \citenamefont
  {Jel{\'{\i}}nek}}]{Krej2017}%
  \BibitemOpen
  \bibfield  {author} {\bibinfo {author} {\bibfnamefont {O.}~\bibnamefont
  {Krej{\v{c}}{\'{\i}}}}, \bibinfo {author} {\bibfnamefont {P.}~\bibnamefont
  {Hapala}}, \bibinfo {author} {\bibfnamefont {M.}~\bibnamefont
  {Ondr{\'{a}}{\v{c}}ek}},\ and\ \bibinfo {author} {\bibfnamefont
  {P.}~\bibnamefont {Jel{\'{\i}}nek}},\ }\bibfield  {title} {\bibinfo {title}
  {Principles and simulations of high-resolution {STM} imaging with a flexible
  tip apex},\ }\href@noop {} {\bibfield  {journal} {\bibinfo  {journal}
  {Physical Review B}\ }\textbf {\bibinfo {volume} {95}} (\bibinfo {year}
  {2017})}\BibitemShut {NoStop}%
\bibitem [{\citenamefont {Frota}(1992)}]{frota1992}%
  \BibitemOpen
  \bibfield  {author} {\bibinfo {author} {\bibfnamefont {H.~O.}\ \bibnamefont
  {Frota}},\ }\bibfield  {title} {\bibinfo {title} {Shape of the {Kondo}
  resonance},\ }\href {https://doi.org/10.1103/PhysRevB.45.1096} {\bibfield
  {journal} {\bibinfo  {journal} {Phys. Rev. B}\ }\textbf {\bibinfo {volume}
  {45}},\ \bibinfo {pages} {1096} (\bibinfo {year} {1992})}\BibitemShut
  {NoStop}%
\bibitem [{\citenamefont {Li}\ \emph {et~al.}(2020{\natexlab{b}})\citenamefont
  {Li}, \citenamefont {Sanz}, \citenamefont {Castro-Esteban}, \citenamefont
  {Vilas-Varela}, \citenamefont {Friedrich}, \citenamefont {Frederiksen},
  \citenamefont {Pe\~na},\ and\ \citenamefont {Pascual}}]{li2020}%
  \BibitemOpen
  \bibfield  {author} {\bibinfo {author} {\bibfnamefont {J.}~\bibnamefont
  {Li}}, \bibinfo {author} {\bibfnamefont {S.}~\bibnamefont {Sanz}}, \bibinfo
  {author} {\bibfnamefont {J.}~\bibnamefont {Castro-Esteban}}, \bibinfo
  {author} {\bibfnamefont {M.}~\bibnamefont {Vilas-Varela}}, \bibinfo {author}
  {\bibfnamefont {N.}~\bibnamefont {Friedrich}}, \bibinfo {author}
  {\bibfnamefont {T.}~\bibnamefont {Frederiksen}}, \bibinfo {author}
  {\bibfnamefont {D.}~\bibnamefont {Pe\~na}},\ and\ \bibinfo {author}
  {\bibfnamefont {J.~I.}\ \bibnamefont {Pascual}},\ }\bibfield  {title}
  {\bibinfo {title} {Uncovering the triplet ground state of triangular graphene
  nanoflakes engineered with atomic precision on a metal surface},\ }\href
  {https://doi.org/10.1103/PhysRevLett.124.177201} {\bibfield  {journal}
  {\bibinfo  {journal} {Phys. Rev. Lett.}\ }\textbf {\bibinfo {volume} {124}},\
  \bibinfo {pages} {177201} (\bibinfo {year} {2020}{\natexlab{b}})}\BibitemShut
  {NoStop}%
\bibitem [{\citenamefont {Krishna-murthy}\ \emph
  {et~al.}(1980{\natexlab{b}})\citenamefont {Krishna-murthy}, \citenamefont
  {Wilkins},\ and\ \citenamefont {Wilson}}]{krishna-murthy1980b}%
  \BibitemOpen
  \bibfield  {author} {\bibinfo {author} {\bibfnamefont {H.~R.}\ \bibnamefont
  {Krishna-murthy}}, \bibinfo {author} {\bibfnamefont {J.~W.}\ \bibnamefont
  {Wilkins}},\ and\ \bibinfo {author} {\bibfnamefont {K.~G.}\ \bibnamefont
  {Wilson}},\ }\bibfield  {title} {\bibinfo {title} {Renormalization-group
  approach to the {Anderson} model of dilute magnetic alloys. {II. Static}
  properties for the asymmetric case},\ }\href
  {https://doi.org/10.1103/PhysRevB.21.1044} {\bibfield  {journal} {\bibinfo
  {journal} {Phys. Rev. B}\ }\textbf {\bibinfo {volume} {21}},\ \bibinfo
  {pages} {1044} (\bibinfo {year} {1980}{\natexlab{b}})}\BibitemShut {NoStop}%
\bibitem [{\citenamefont {Coleman}(2015)}]{coleman2015}%
  \BibitemOpen
  \bibfield  {author} {\bibinfo {author} {\bibfnamefont {P.}~\bibnamefont
  {Coleman}},\ }\href@noop {} {\emph {\bibinfo {title} {Introduction to
  Many-Body Physics}}}\ (\bibinfo  {publisher} {Cambridge University Press},\
  \bibinfo {year} {2015})\BibitemShut {NoStop}%
\end{thebibliography}%


\end{document}